\documentstyle[11pt]{article}
\input epsf
\ifx\epsffile\undefined
\newlength{\epsfysize}
\def\epsffile#1#2#3#4]#5{}
\fi

\renewcommand{\thefootnote}{\fnsymbol{footnote}}
\def\roughly#1{\raise.3ex\hbox{$#1$\kern-.75em\lower1ex\hbox{$\sim$}}}

\begin{document}

\begin{flushright}
{\small
SLAC--PUB--7180\\
JHU--TIPAC--96011\\
June 1996\\}
\end{flushright}

\vspace{2cm}

\centerline{PRECISION CORRECTIONS IN THE}
\baselineskip=15pt \centerline{MINIMAL SUPERSYMMETRIC STANDARD MODEL}

\baselineskip=32pt

\centerline{\footnotesize DAMIEN M. PIERCE}

\baselineskip=22pt

\centerline{\footnotesize\it Stanford Linear Accelerator Center}
\baselineskip=13pt
\centerline{\footnotesize\it Stanford University}
\centerline{\footnotesize\it Stanford, California 94309, USA}

\baselineskip=32pt

\centerline{\footnotesize JONATHAN A. BAGGER,}
\baselineskip=13pt
\centerline{\footnotesize KONSTANTIN T. MATCHEV and}
\centerline{\footnotesize REN-JIE ZHANG}

\baselineskip=22pt

\centerline{\footnotesize\it Department of Physics and Astronomy}
\baselineskip=13pt
\centerline{\footnotesize\it Johns Hopkins University}
\centerline{\footnotesize\it 3400 N. Charles Street}
\centerline{\footnotesize\it Baltimore, Maryland 21218, USA}

\vspace{1cm}

\begin{abstract}
In this paper we compute one-loop corrections to masses and couplings
in the minimal supersymmetric standard model.  We present explicit
formulae for the complete corrections and a set of compact
approximations which hold over the unified parameter space associated
with radiative electroweak symmetry breaking.  We illustrate the
importance of the corrections and the accuracy of our approximations
by scanning over the parameter space.  We calculate the supersymmetric
one-loop corrections to the $W$-boson mass, the effective weak mixing
angle, and the quark and lepton masses, and discuss implications for
gauge and Yukawa coupling unification.  We also compute the one-loop
corrections to the entire superpartner and Higgs-boson mass spectrum.
We find significant corrections over much of the parameter space, and
illustrate that our approximations are good to ${\cal O}(1\%)$ for
many of the superparticle masses.
\end{abstract}

\vspace{2cm}

\vfill

{\noindent\em Work supported by Department of Energy contract
DE--AC03--76SF00515 and by the U.S. National Science Foundation,
grant NSF-PHY-9404057.}

\pagebreak

\normalsize\baselineskip=15pt
\setcounter{footnote}{0}
\renewcommand{\thefootnote}{\arabic{footnote}}
\section{Introduction}

Most precision measurements of electroweak parameters agree quite
well with the predictions of the standard model \cite{LWG}.
These experiments rule out many possibilities for physics beyond
the standard model, but they have not touched supersymmetry,
which evades precision constraints because it decouples from
standard-model physics if the scale of supersymmetry breaking
is more than a few hundred GeV.  Definitive tests of supersymmetry
will probably have to wait for direct searches at future colliders.

Once supersymmetry is discovered, a host of new questions arise.
For example, one would like to test the low energy supersymmetric
relations between the particle masses and couplings by making
precision measurements of the supersymmetric parameters.  One
would like to measure the supersymmetric masses as accurately
as possible to shed light on the origin of supersymmetry breaking.
Furthermore, one would also like to know whether weak-scale
supersymmetry sheds any light on physics at even higher energies.
Indeed, the successful unification of gauge couplings encourages
hope that other supersymmetric parameters might unify as well.  It
is important to measure these parameters precisely at low energies
so that they can be extrapolated with confidence to higher energies.

It is in this spirit that we present our calculation of one-loop
corrections to the minimal supersymmetric standard model (MSSM).  We
define the MSSM to be the minimally supersymmetrized standard model,
with no right handed neutrinos, and all possible soft-breaking terms.
We believe that this minimal model provides an appropriate framework
for analyzing the phenomenology of supersymmetry and supersymmetric
unification.

We approach our calculation in the standard fashion associated
with precision electroweak measurements.  We take as inputs
the electromagnetic coupling at zero momentum, $\alpha_{\rm em}
= 1/137.036$, the Fermi constant, $G_\mu = 1.16639\times 10^{-5}$
GeV$^{-2}$, the $Z$-boson pole mass, $M_Z = 91.188$ GeV, the
strong coupling in the \mbox{\footnotesize$\overline{\rm MS}~$}
scheme at the scale $M_Z$, $\alpha_s(M_Z) = 0.118$, as well as
the quark and lepton masses, $m_t = 175$ GeV, $m_b = 4.9$ GeV,
and $m_\tau = 1.777$ GeV \cite{PDG}.

{}From these inputs, for any tree-level supersymmetric spectrum,
we compute the one-loop $W$-boson pole mass, $M_W$, as well as
the one-loop values of the effective weak mixing angle,
$\sin^2\theta^{\rm lept}_{\rm eff}$, and the \mbox{\footnotesize
$\overline{\rm DR}~$} \cite{dr} weak mixing angle, $\hat s^2$.
We also compute the one-loop corrections to the quark and lepton
Yukawa couplings, as well as the masses of all the supersymmetric
and Higgs particles.

We work in the \mbox{\footnotesize$\overline{\rm DR}~$} scheme,
and take the tree-level masses to be given in terms of the running
\mbox{\footnotesize$\overline{\rm DR}~$} parameters.  For each
(bosonic) particle, we determine the one-loop pole mass,
\begin{equation}
M^2 \ =\ \hat M^2(Q)\ -\ {\cal R}e\,\Pi(M^2)\ ,
\label{pole mass}
\end{equation}
where $\hat M(Q)$ is the tree-level \mbox{\footnotesize$\overline{\rm
DR}~$} mass, evaluated at the \mbox{\footnotesize$\overline{\rm DR}~$}
scale $Q$, and $\Pi(p^2)$ is the one-loop self-energy.  (As usual,
$\Pi(p^2)$ depends on $Q$ and on the masses and couplings of the
particles in the loop.  There is a similar expression for the fermion
pole mass.)

In all our computations we include the full self-energies, which
contain both logarithmic and finite contributions.  The logarithmic
corrections can be absorbed by changes in the scale $Q$. Therefore we
checked our logarithmic results against the one-loop supersymmetric
renormalization group equations.  Since we write our results using
Passarino-Veltman functions \cite{PV}, some of our finite terms are
automatically correct.  As a further check, we verified that our
corrections decouple from electroweak observables.

We present our complete calculations in a series of Appendices.  These
appendices include the full one-loop corrections to the gauge and
Yukawa couplings, as well as the complete one-loop corrections to the
entire MSSM mass spectrum. While some of these results are not new
(the gauge-boson \cite{Grifols,CPR}, Higgs-boson \cite{CPR} and
gluino \cite{MV,PP2,Kras,Donini} self-energies and the gauge-coupling
corrections \cite{Grifols,deltar} already appear in the literature),
we include the full set of corrections in order to provide a complete,
self-contained and more useful reference.

In Appendix A we write the tree-level masses in terms of the
parameters of the MSSM, and in Appendix B we define the
Passarino-Veltman functions that we use to present our one-loop
results.  In Appendix C we compute the one-loop radiative corrections
to the gauge couplings of the MSSM, and in Appendix D we write the
one-loop corrections to the masses.  Where appropriate, we evaluate
the corrections to the mass {\em matrices} to account for full
one-loop superparticle mixing.  This allows for an accurate
determination of the masses and mixing through the entire parameter
space.  Finally, in Appendix E we discuss the radiative corrections to
the Higgs boson masses.

The results in the Appendices hold for the MSSM with the most general
pattern of (flavor diagonal) soft supersymmetry breaking.\footnote{
Our results can be readily extended to include inter-generational
mixing at the cost of additional mixing matrices.}\ The parameter
space is huge because of the large number of operators that softly
break supersymmetry.  Therefore in the body of the paper we illustrate
our results in a reduced parameter space, obtained by assuming that
the soft breaking parameters unify at some high scale.

The unification assumption is useful because it reduces the size
of the parameter space.  Moreover, it implies certain mass
relations that can be tested once supersymmetry is discovered.  In
addition, for any set of parameters, it allows us to determine the
unification scale thresholds that are necessary to achieve unification.
As we will see, the present set of precision measurements is sufficient
to begin to constrain the physics at the unification scale.

We implement the unification assumption by solving
the two-loop supersymmetric renormalization group equations
subject to two-sided boundary conditions.  At the weak scale, we
assume a supersymmetric spectrum, and for a given value of the
ratio of vacuum expectation values, $\tan\beta$, we use our one-loop
corrections to extract the \mbox{\footnotesize$\overline{\rm DR}~$}
couplings $g_1,\ g_2, \ g_3,\ \lambda_t,\ \lambda_b$, and $\lambda_\tau$
at the scale $M_Z$.  We then use the two-loop \mbox{\footnotesize$
\overline{\rm DR}~$} renormalization group equations \cite{RGEs} to
run these six parameters to the scale $M_{\rm GUT}$, which we define
to be the scale where $g_1$ and $g_2$ meet.

We require that the soft breaking parameters unify at the scale
$M_{\rm GUT}$.  Therefore at $M_{\rm GUT}$ we fix a universal scalar
mass, $M_0$, a universal gaugino mass, $M_{1/2}$, and a universal
trilinear scalar coupling, $A_0$.  We then run all the soft parameters
back down to the scale $M^2_{\tilde q} = M^2_0 + 4 M^2_{1/2}$, where
we calculate the supersymmetric spectrum using the full one-loop
threshold corrections that we present in this paper.  In section 4
we show that this scale is essentially the scale of the squark masses,
and that the other scalar masses and the Higgsino mass are correlated
with it as well.

We require radiative electroweak symmetry breaking \cite{EWSB}, so
the CP-odd Higgs mass, $m_A$, and the supersymmetric Higgs mass
parameter, $|\mu|$, are determined in a full one-loop calculation at
the scale $M_{\tilde q}$.  The sign of $\mu$ is left undetermined.  We
then iterate the entire procedure to determine a self-consistent
solution.  Typically, convergence to an accuracy of better than
$10^{-4}$ is achieved after four iterations.

Once we have a consistent solution, we use the results of the
Appendices to
illustrate the one-loop corrections in the reduced parameter space
associated with unification.  We display results for a randomly chosen
sample of 4000 points.  Our sample is chosen with a logarithmic
measure in the range $1 < \tan\beta < 60$, $50 < M_{1/2} < 500$ GeV,
$10 < M_0 < 1000$ GeV, and with a linear measure in the range $-3
M_{\tilde q} < A_0 < 3 M_{\tilde q}.$ (The upper limits on $M_0$ and
$M_{1/2}$ are chosen so that the squark masses are less than about 1
TeV.  While larger squark masses are certainly possible, they
reintroduce the fine tunings that supersymmetry is designed to avoid.)

Each of these points corresponds to a (local) minimum of the one-loop
scalar potential with the correct electroweak symmetry breaking, and
each passes a series of phenomenological constraints: We require the
first- and second-generation squark masses to be larger than 220 GeV
\cite{CDF/D0}, the gluino mass to be greater than 170 GeV
\cite{CDF/D0}, the light Higgs mass\footnote{The light Higgs boson is
similar to that of the standard model in almost all of our parameter
space, so we apply the standard model bound.} to be greater than 60
GeV \cite{PDG}, the slepton masses to be greater than 45 GeV
\cite{PDG}, and the chargino masses to be greater than 65 GeV
\cite{ALEPH}.  We also require all the Yukawa couplings to remain
perturbative ($\lambda<3.5$) up to the unification scale, and since we
assume that $R$-parity is unbroken, we enforce the cosmological
requirement that the lightest supersymmetric particle be neutral.

We derive approximations to the radiative corrections that hold with
reasonable accuracy over the unified parameter space.  Where
appropriate, we use scatter plots to illustrate the effectiveness of
our approximations.  The approximations consist of two parts.  First
we identify the most important contributions to the one-loop
corrections.  In most cases these are the loops that involve the
strong and/or third generation Yukawa couplings.  Then we derive
approximations to the loop expressions that hold over the unified
parameter space.

In the next section, we discuss the radiative corrections to the
effective weak mixing angle, $\sin^2\theta^{\rm lept}_{\rm eff}$, and
the $W$-boson pole mass, $M_W$.  We illustrate the magnitudes of the
different supersymmetric contributions to these observables.  We also
discuss the renormalization of the \mbox{\footnotesize$\overline{\rm
DR}~$} weak mixing angle, $\hat s^2$, and comment on the way that it
affects the gauge thresholds at the unification scale.  In section 3
we examine radiative corrections to the third generation quark and
lepton masses.  We illustrate the different contributions and present
approximations which hold to a few percent.  We also examine Yukawa
unification and demonstrate the size of the unification-scale Yukawa
thresholds.  In section 4 we present our results for the radiative
corrections to the supersymmetric and Higgs boson particle masses.  We
find large corrections to the masses of the light superparticles.  We
compare our results with those of the leading logarithmic
approximation and find significant improvements over much of the
unified parameter space.  These corrections are important for
unraveling the underlying supersymmetric structure from the
supersymmetric mass spectrum.

\section{The weak mixing angle and the $W$-boson mass}

\begin{figure}[t]
\epsfysize=2.5in \epsffile[-140 210 0 525]{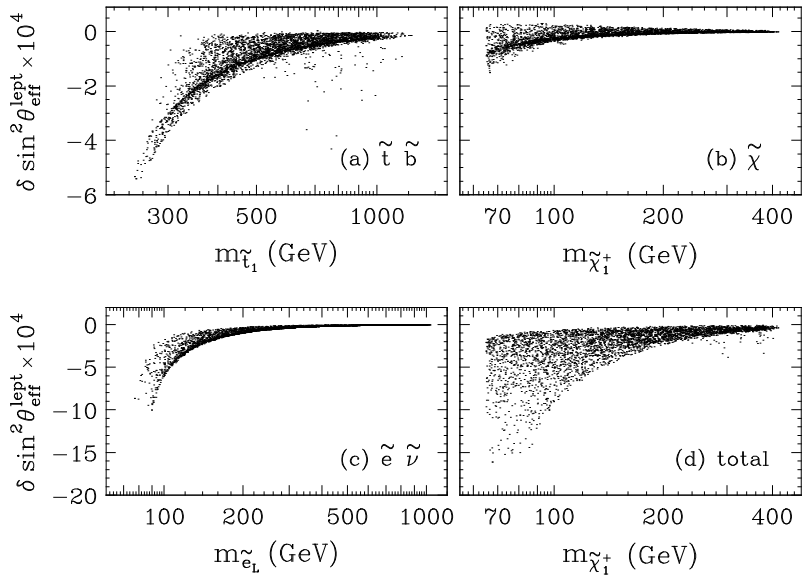}
\begin{center}
\parbox{5.5in}{
\caption[]{\small Supersymmetric corrections to the effective weak
mixing angle, $\sin^2\theta^{\rm lept}_{\rm eff}$.  Figure (a) shows
the corrections from top and bottom squark loops versus the heavy top
squark mass $m_{\tilde t_1}$; (b) shows the neutralino/chargino
contribution versus the light chargino mass; (c) shows the slepton
correction from all three generations against $m_{\tilde e_L}$;
and (d) shows the complete
supersymmetric correction plotted against $m_{{\tilde\chi}_1^+}$.
\label{sw}}}
\end{center}
\end{figure}

The calculation of supersymmetric contributions to electroweak
observables began in 1984 \cite{Grifols}.  Since then, the precise
confrontation of electroweak data with theoretical predictions of the
MSSM has been an active area of study \cite{precision studies,deltar}.
Global fits to precision data in the MSSM have been performed by
several groups \cite{global}.  In this section we display our results
for two electroweak observables over the parameter space associated
with radiative electroweak symmetry breaking and universal
unification-scale boundary conditions.  We extract the contributions
from the various superpartners, and illustrate the manner in which the
different contributions decouple from the low energy observables.

\subsection{Effective weak mixing angle}

We start by considering the effective weak mixing angle, $s^2_\ell
\equiv \sin^2\theta^{\rm lept}_{\rm eff}$.  The full one-loop
calculation is presented\footnote{We do not include the
supersymmetric nonuniversal $Z$-vertex contribution in the
Appendix.  It is a negligible correction in the parameter
space we consider.}
for completeness in Appendix C.  The complete result is rather
involved; for now we
simply say that the calculation follows the outline presented above:
We take $\alpha_{\rm em}$, $G_\mu$, $M_Z$, $\alpha_s(M_Z)$, and the
fermion masses as inputs, and compute $s^2_\ell$ as a function of the
supersymmetric masses.

Because we compute the experimental observable $s^2_\ell$ in terms of
other low-energy observables, its one-loop supersymmetric corrections
decouple as the supersymmetric masses become larger than $M_Z$.  From
Fig.~\ref{sw} we see that $s^2_\ell$ is especially sensitive to light
sleptons, and that the sum of the supersymmetric corrections is always
negative.  We did not plot the Higgs boson and first two generation
squark contributions. They are negligible, less than $1\times10^{-4}$
and $4\times10^{-5}$ in magnitude, respectively.  The corrections to
$\mu$-decay and the corrections to the $Z$-$\ell^+$-$\ell^-$ vertex
comprise the nonuniversal corrections to $s_\ell^2$. The former
contributes between $-3$ and $1.5\times10^{-4}$, the latter between
$\pm1.5\times10^{-4}$, and their sum is in the range $-4$ to
$1\times10^{-4}$.

With $m_t=175$ GeV, we find the standard model value of $s^2_\ell$
varies between 0.2311 to 0.2315 for Higgs masses in the range $60 <
m_h < 130$ GeV.  This is subject to an error of $2.5 \times 10^{-4}$
from the experimental uncertainty in the electromagnetic coupling
evaluated at $M_Z$, and to corrections of this same order from higher
loop effects \cite{h.o.}.  Furthermore, increasing $m_t$ by $10$ GeV
decreases $s^2_\ell$ by $3.3\times 10^{-4}.$

These predictions for $s^2_\ell$ should be compared with the LEP and
SLD average\footnote{The number quoted here assumes lepton
universality.} \cite{LWG} of 0.23165 $\pm$ 0.00024.  Clearly, the
standard model calculation agrees quite well with experiment.  The
additional contribution from supersymmetry can lower the value of
$s_\ell^2$ slightly below 0.2300, or about 6$\sigma$ below the
experimental central value.  However, we note that higher-order
standard-model corrections, changes in $\alpha_{\rm em}$ and $m_t$,
and other precision observables should all be
systematically taken into account to delineate which regions of
parameter space are ruled out by these measurements.  We do not
attempt such a study here.

The corrections to $s^2_\ell$ diminish rapidly as the superpartner
masses become heavy. For example, if we require $m_{\tilde\chi^+},
\ m_{\tilde\ell^+},\ m_h\ > 90$ GeV, we find that $s^2_\ell$ is shifted
by at most $-8\times10^{-4}$ relative to the standard model value.

\begin{figure}[t]
\epsfysize=2.5in \epsffile[-140 220 0 535]{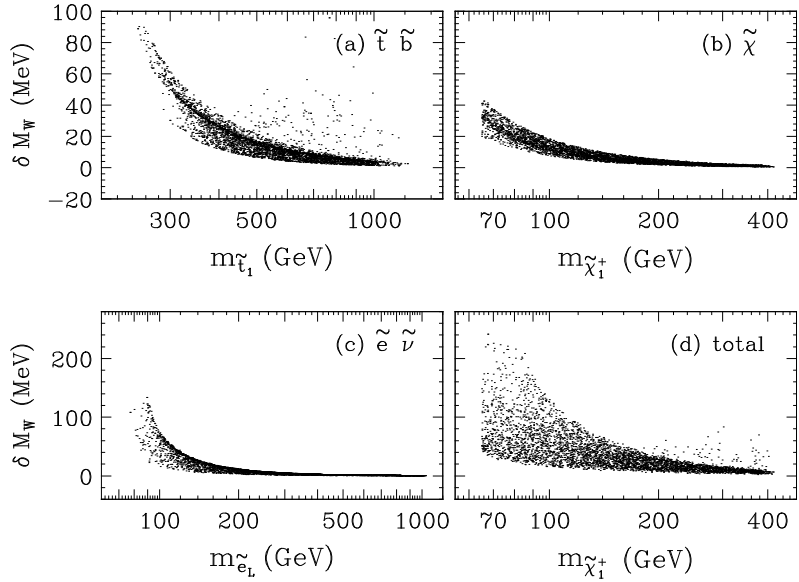}
\begin{center}
\parbox{5.5in}{
\caption[]{\small Finite corrections to $M_W$, in MeV.  Figures (a-d)
are as in Fig.~\ref{sw}.
\label{mw}}}
\end{center}
\end{figure}

\subsection{$W$-boson mass}

We now turn to our second precision electroweak observable,
and compute the one-loop correction to the $W$-boson pole mass, $M_W$.
In Fig.~\ref{mw} we illustrate some of the finite corrections.
The full finite correction increases
the prediction for $M_W$ by up to 250 MeV.  As with $s_\ell^2$,
the contributions from Higgs bosons and the first two generations of
squarks are small, less than 12 and 8 MeV, respectively.  The
nonuniversal correction is also small, between $-6$ and 15 MeV.  For
large supersymmetric masses, the prediction reduces to that of the
standard model because of decoupling.

For $m_t=175$ GeV we find the standard-model value of $M_W$ in the
range 80.39 to 80.43 GeV, with an error of $\pm$13 MeV from the
experimental uncertainty of the electromagnetic coupling. This is
subject to additional corrections of the same order from higher-loop
effects \cite{h.o.}.  If we increase $m_t$ by 10 GeV, we find a 65 MeV
increase in $M_W$.

{}From our calculations we find that the MSSM value for $M_W$ ranges
from 80.39 to 80.64 GeV.  These numbers can be compared to the current
world average, 80.33$\pm$0.15 GeV \cite{PDG}.  With the current
experimental error, all of the supersymmetric parameter space lies
within 2$\sigma$ of the central value.  By the end of the decade, the
error on $M_W$ is expected to be about 50 MeV.  If supersymmetry is
not discovered by that time, one might think that a much more exacting
test could be performed.  However, the limits on the superpartner
spectrum will also have increased to the point where the effects on
weak-scale observables from virtual supersymmetry will be greatly
diminished.  For example, imposing the limits $m_{\tilde\chi^+},
\ m_{\tilde\ell^+},\ m_h > 90$ GeV, we find that typically $\delta M_W <
50$ MeV, and at most $\delta M_W=100$ MeV.

\begin{figure}[t]
\epsfysize=2.5in \epsffile[-140 220 0 535]{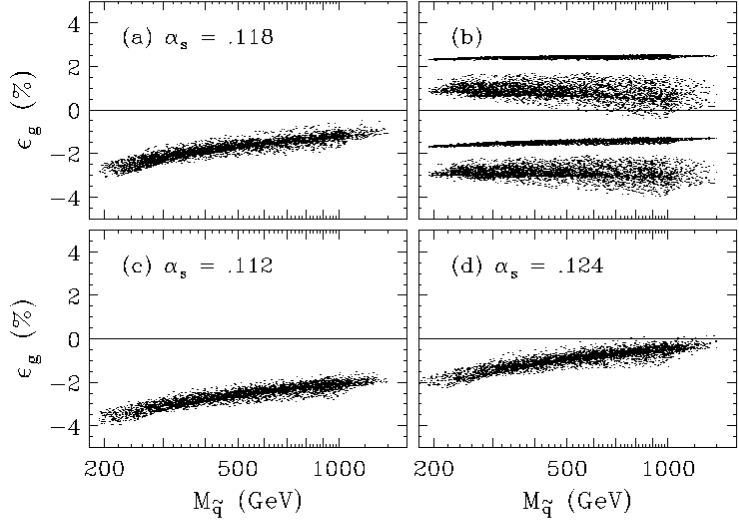}
\begin{center}
\parbox{5.5in}{
\caption[]{\small (a) The unification-scale correction,
$\varepsilon_g$, necessary to obtain $\alpha_s(M_Z) = 0.118$, plotted
versus $M_{\tilde q}$. (b) The maximum and minimum $\varepsilon_g$
allowed in minimal SU(5) (top two regions) and missing doublet SU(5)
(bottom two regions), against $M_{\tilde q}$.  (c) Same as (a) with
$\alpha_s(M_Z) = 0.112$.  (d) Same as (a) with $\alpha_s(M_Z) =
0.124$.
\label{epsg}}}
\end{center}
\end{figure}

\subsection{Gauge coupling unification}

We are now ready to study gauge coupling unification \cite{Lopez}.  We
start by computing the \mbox{\footnotesize$\overline{\rm DR}~$}
electromagnetic coupling constant, $\hat \alpha$, and the
\mbox{\footnotesize$\overline{\rm DR}~$} weak mixing angle, $\hat
s^2$, as described in Appendix C.  The
\mbox{\footnotesize$\overline{\rm DR}~$} weak mixing angle is closely
related to the effective weak mixing angle, $s^2_\ell$.  The main
difference is that $\hat s^2$ is not an experimental observable, so
its radiative corrections involve nondecoupling logarithms of
supersymmetric masses.  However, once we subtract these logarithms we
find finite corrections to $\hat s^2$ which are quantitatively similar
to the corrections to $s^2_\ell$ shown in Fig.~\ref{sw}.

In the context of gauge coupling unification, the finite corrections
to $\hat s^2$ are very important \cite{F&G,CPP,BMP}.  They play a
significant role in determining the required unification-scale
thresholds, which are formally of the same order in perturbation
theory.  As we will see, precision measurements already limit the size
of these thresholds and place constraints on unified models.

We determine the unification thresholds as follows: we calculate the
full one-loop corrections to $\hat\alpha$ and $\hat s^2$, and use them
to determine the \mbox{\footnotesize$\overline{\rm DR}~$} couplings
$g_1$ and $g_2$.  We take $\alpha_s(M_Z) = 0.118$ from experiment
\cite{PDG}, and apply the supersymmetric threshold corrections to fix
the \mbox{\footnotesize$\overline{\rm DR}~$} coupling $g_3$ at the
scale $M_Z$,
\begin{equation}
{g_3^2(M_Z)\over4\pi}\ =\ {\alpha_s(M_Z)\over 1-\Delta\alpha_s}\ ,
\end{equation}
where
\begin{equation}
\Delta\alpha_s \ =\ {\alpha_s(M_Z)\over2\pi}\ \biggl[ {1\over2}\ -
\ {2\over3}\ln\left({m_t\over M_Z}\right) \ -\ 2\ln\left({m_{\tilde
g}\over M_Z}\right) \ -\ {1\over6}\sum_{\tilde
q}\sum_{i=1}^2\ln\left({m_{\tilde q_i} \over M_Z}\right)\biggr]\ .
\label{alpha3}
\end{equation}
The sum indexed by $\tilde q$ runs over the six squark flavors.  The
constant in (\ref{alpha3}) transforms the
\mbox{\footnotesize$\overline{\rm MS}~$} coupling $\alpha_s(M_Z)$ into
the \mbox{\footnotesize$\overline{\rm DR}~$} coupling $g_3$.  We also
calculate the one-loop \mbox{\footnotesize$\overline{\rm DR}~$} Yukawa
couplings $\lambda_t(M_Z), \ \lambda_b(M_Z)$, and $\lambda_\tau(M_Z)$,
as described in the next section.

We then run the six coupled two-loop renormalization group equations
\cite{RGEs} up to the unification scale, $M_{\rm GUT}$, which we
define to be the point where $g_1$ and $g_2$ meet.  At that scale we
define the unification threshold, $\varepsilon_g$, to be the
discrepancy between $g_3$ and the electroweak couplings $g_1$ and
$g_2$,
\begin{equation}
g_3(M_{\rm GUT})\ =\ g_1(M_{\rm GUT})\,\left(1+\varepsilon_g\right)\ .
\end{equation}

In Fig.~\ref{epsg}(a) we plot the threshold, $\varepsilon_g$, versus
the squark mass scale, $M_{\tilde q}$, which we define to be
$M_{\tilde q}^2 \equiv M_0^2 + 4M_{1/2}^2$.  From the figure we see
that for $\alpha_s(M_Z)=0.118$, unification requires a negative
unification-scale threshold correction of between $-1\%$ and $-3\%$,
depending on the weak-scale supersymmetric spectrum.  For small
$M_{\tilde q}$, the finite corrections to the gauge couplings are
comparable in size to the logarithmic corrections; they both decrease
$\varepsilon_g$ at small $M_{\tilde q}$.

In order for a unified model to be consistent with gauge coupling
unification, it must be able to accommodate values of $\varepsilon_g
\simeq -2\%$.  Different unified models give rise to different
unification-scale threshold corrections.  In the minimal \cite{minSU5}
and missing doublet SU(5) \cite{MDSU5} models, $\varepsilon_g$ depends
only on the triplet Higgs mass \cite{BMP}, the same mass that enters
the nucleon decay rate formulae.  The maximum and minimum values of
$\varepsilon_g$ in these two models are shown in Fig.~\ref{epsg}(b).
The two thin regions correspond to the maximum values of
$\varepsilon_g$, obtained by setting the triplet Higgs mass to
$10^{19}$ GeV.  The two larger regions show the minimum values of
$\varepsilon_g$ in each model.  These values are found by setting the
triplet Higgs mass as small as possible, consistent with the bounds
from nucleon decay \cite{HMY}.  From the figure we see that it is
difficult to achieve the necessary thresholds in minimal SU(5), but
that missing doublet SU(5) has unification-scale thresholds in the
right range.

The unification-scale threshold $\varepsilon_g$ necessary for
unification is directly correlated with $\alpha_s(M_Z)$.  Increasing
$\alpha_s(M_Z)$ by 5\% increases $\varepsilon_g$ by about 1\%, as
expected from the one-loop relation
\begin{equation}
{\delta\alpha_s(M_Z)\over \alpha_s(M_Z)} \ =\ 2
\ {\alpha_s(M_Z)\over\alpha_{\rm GUT}}\ \delta\varepsilon_g\ .
\end{equation}
We illustrate this in Figs.~\ref{epsg}(c-d), where we plot
$\varepsilon_g$ for $\alpha_s(M_Z)=0.112$ and 0.124.

\section{Quark and lepton masses}

The full set of radiative corrections to the quark and lepton masses
is presented in Appendix D.  In this section we derive approximations
to these formulae, valid for the third generation.

\subsection{Top quark mass}

The top quark mass provides an important input for radiative
electroweak symmetry breaking.  It receives strong and electroweak
radiative corrections \cite{SUSY94&Wright,Donini}.  Our approximation
begins by eliminating the small electroweak corrections, setting $g
= g' = 0$ and $\lambda_t = \lambda_b = 0$.  We then simplify the
resulting expressions by setting $p^2 = 0$ because $m_t$ is much
smaller than a typical squark or gluino mass.

In this limit, the physical top quark mass is given by
\begin{equation}
m_t\ =\ \hat m_t(Q) \left[\, 1 +\,{\Delta m_t \over m_t}\,\right]
\label{mt phys}\ ,
\end{equation}
where the one-loop correction receives two important contributions.
The first is the well-known gluon correction,\footnote{Here and in the
following, we implicitly perform \mbox{\footnotesize$\overline{\rm
DR}~$} renormalization, so the $1/\hat\epsilon$ poles are subtracted.}
\begin{eqnarray}
\left({\Delta m_t \over m_t}\right)^{tg}\ &=& \ {g^2_3 \over 6\pi^2}
\ \Bigg[\,2B_0(m_t,m_t,0) - B_1(m_t,m_t,0)\,\Bigg] \nonumber \\ &=&
\ {g^2_3\over12\pi^2}\ \left[\,3\ln\left({Q^2\over m_t^2}\right) +
5\,\right]
\label{mt corr}\ .
\end{eqnarray}
Note that this contains a logarithmic and a finite piece; the latter
gives a 6.6\% contribution.\footnote{We have included the two-loop
\mbox{\footnotesize$\overline{\rm MS}~$} contribution $\Delta
m_t=1.11\alpha_s^2m_t$ \cite{mq 2-loop}, which we assume is close to
the \mbox{\footnotesize$\overline{\rm DR}~$} value.}  The second
correction comes from the top squark/gluino loops,
\begin{eqnarray}
\left({\Delta m_t\over m_t}\right)^{\tilde t\tilde g}\ &=& \ -
\ {g^2_3\over12\pi^2}\ \Bigg\{ \,B_{1}(0,m_{\tilde g},m_{\tilde t_1}) +
B_{1}(0,m_{\tilde g},m_{\tilde t_2})\, \nonumber\\ &&\ -\ \sin
(2\theta_t) \ \left({m_{\tilde g} \over m_t}\right)
\ \Bigg[\,B_{0}(0,m_{\tilde g},m_{\tilde t_1}) - B_{0}(0,m_{\tilde
g},m_{\tilde t_2})\,\Bigg] \Bigg\}
\label{mt gluino}
\end{eqnarray}
where $\theta_t$ is the top-squark mixing angle, and
\begin{equation}
B_{0}(0,m_1,m_2)\ =\ - \ \ln\left(M^2\over Q^2\right) + 1 + {m^2\over
m^2-M^2}\ln\left(M^2\over m^2\right)\ ,
\label{b0(0)}
\end{equation}
\begin{equation}
B_{1}(0,m_1,m_2)\ =\ {1\over2}\left[\, -\ \ln\left(M^2\over Q^2\right)
+ {1\over2} + {1\over 1-x} + {\ln x \over (1-x)^2} - \theta(1-x)\ln x
\,\right]\ ,
\label{b1(0)}
\end{equation}
with $M=\max(m_1,m_2),$ $m=\min(m_1,m_2),$ and $x=m_2^2/m_1^2$.  The
full $B$ functions are written in Appendix B; the formulae presented
here are simplifications that hold when the first argument is zero.

\begin{figure}[t]
\epsfysize=2.5in \epsffile[-140 210 0 525]{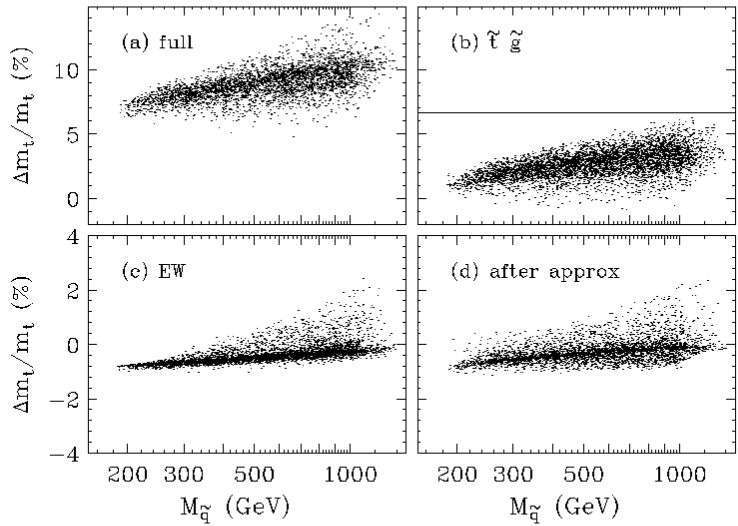}
\begin{center}
\parbox{5.5in}{
\caption[]{\small Corrections to the top quark mass, versus $M_{\tilde
q}$.  Figure (a) shows the full one-loop correction.  Figure (b)
illustrates the correction from the squark/gluino loop; the solid line
shows the gluon contribution for comparison.  Figure (c) shows the
electroweak corrections.  In Fig.~(d) we plot the difference between
the full one-loop result and the approximation given in the text.
\label{mt}}}
\end{center}
\end{figure}

In Fig.~\ref{mt} we show the complete correction to the top quark mass
as well as the contributions from the squark/gluino and electroweak
loops. The tree-level mass is defined to be $\hat m_t(m_t)$. We see
that the squark/gluino loop contribution can be as large as the gluon
contribution for TeV-scale gluino and squark masses. The electroweak
corrections are small because of cancellations.  In the figure we also
plot the difference between the full correction and our approximation.
We see that our approximation is good to typically $\pm1\%$.

\begin{figure}[t]
\epsfysize=2.5in \epsffile[-140 220 0 535]{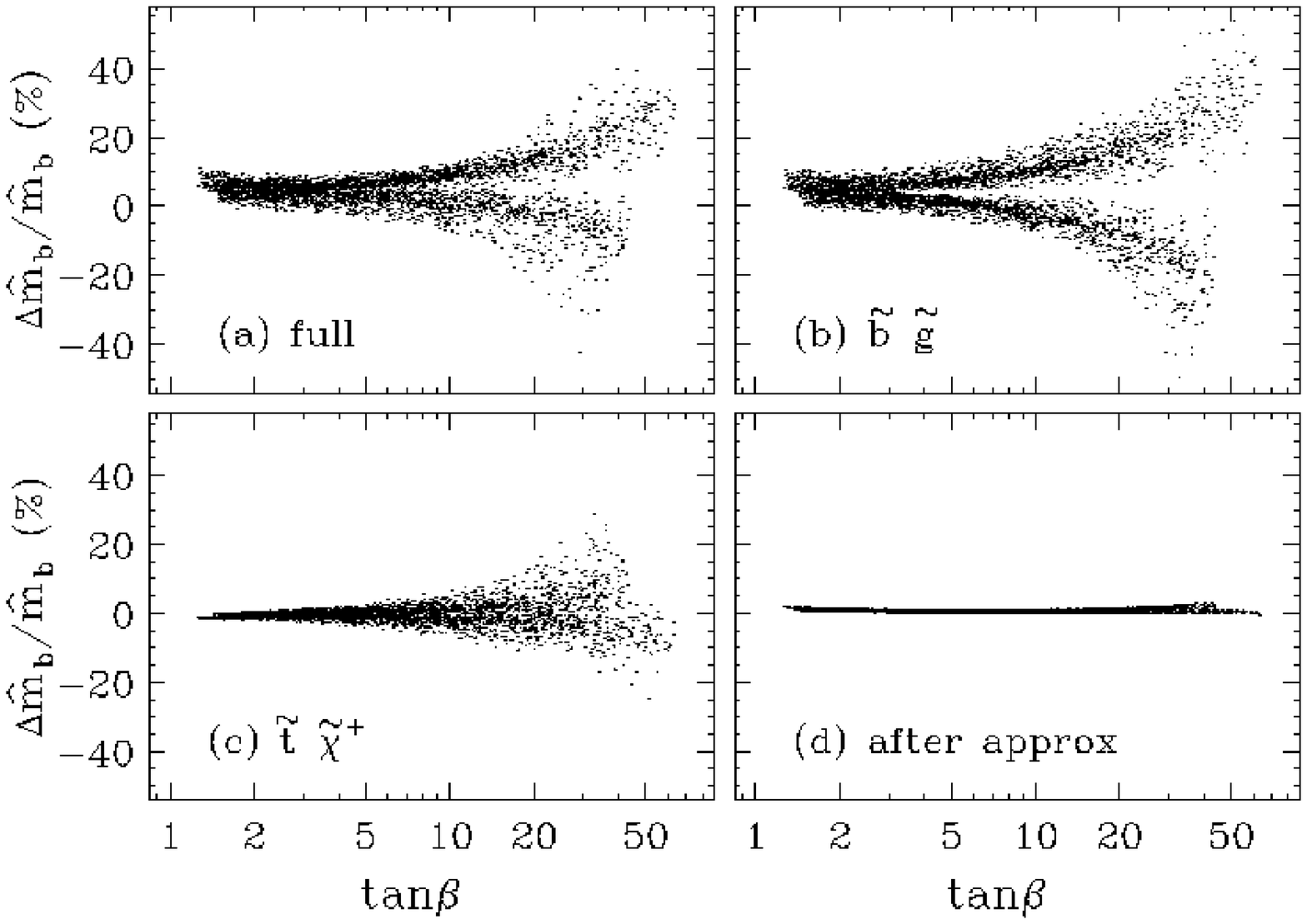}
\begin{center}
\parbox{5.5in}{
\caption[]{\small Corrections to the \mbox{\footnotesize$\overline{\rm
DR}~$} bottom quark mass $\hat m_b(M_Z)$, plotted versus $\tan\beta$.
Figure (a) shows the full one-loop correction; (b) illustrates the
full correction from the bottom squark/gluino loops; (c) shows the
correction from the top squark/chargino loops.  Figure (d) plots the
difference between the full one-loop result and the approximation
given in the text.
\label{bmass}}}
\end{center}
\end{figure}

\subsection{Bottom quark mass}

Corrections to the bottom quark mass in the MSSM have received much
attention because they can contain significant enhanced supersymmetric
contributions \cite{mb,Carena}.  These large contributions play an
important role in Yukawa coupling unification.  Previous studies have
included only the enhanced contributions.  In this paper we present
our results for the full one-loop correction.  Moreover, we
systematically develop approximations to the supersymmetric
corrections.  In this way we can see the importance of the enhanced
contributions relative to the full result.

The corrections to the bottom quark mass are found as follows.
Because the bottom quark is light, $\alpha_s(m_b)$ is large and we
must resum the gluon contribution.  We start with the bottom-quark
pole mass, $m_b$.  We find the standard-model
\mbox{\footnotesize$\overline{\rm DR}~$} bottom quark mass at the
scale $m_b$ using the two-loop QCD correction,
\begin{equation}
\hat m_b(m_b)^{\rm SM}\ =\ m_b\ \left[\,1 - \left({\Delta m_b\over
m_b}\right) ^{bg}\,\right]\ ,
\end{equation}
where\footnote{We do not know the two- and three-loop corrections to
$m_b$ in the \mbox{\footnotesize$\overline{\rm DR}~$} scheme.
Similarly, we do not know the \mbox{\footnotesize$\overline{\rm DR}~$}
three-loop QCD contribution to the running of the strong coupling.  In
both cases we use the \mbox{\footnotesize$\overline{\rm MS}~$} values.
Alternatively, we could have used \mbox{\footnotesize$\overline{\rm
MS}~$} equations to run up to $M_Z$, then convert to
\mbox{\footnotesize$\overline{\rm DR}~$}. The difference between the
two approaches, $\Delta\hat m_b(M_Z)<0.05$ GeV, is nearly an order of
magnitude smaller than the experimental uncertainty in the bottom
quark mass.} \cite{mq 2-loop}
\begin{equation}
\left({\Delta m_b\over m_b}\right)^{bg}\ = \ {5\over3}
{\alpha_s(m_b)\over\pi} \ +\ 12.4 \,\left( {\alpha_s(m_b)\over\pi}
\right)^2\ ,
\end{equation}
and $\alpha_s(m_b)$ is the five-flavor three-loop running
\mbox{\footnotesize$\overline{\rm DR}~$} coupling.  We then evolve
this mass to the scale $M_Z$ using a numerical solution to the
two-loop (plus three-loop ${\cal O}(\alpha_s^3)$) standard-model
renormalization group equations \cite{Ramond}.  Taking the bottom
quark pole mass $m_b=4.9$ GeV, and $\alpha_s(M_Z)=0.118$, we find the
standard-model \mbox{\footnotesize$\overline{\rm DR}~$} value $\hat
m_b(M_Z)^{\rm SM}=2.92$ GeV.

The final step is to add the one-loop corrections from massive
particles,
\begin{equation}
\hat m_b(M_Z)\ =\ \hat m_b(M_Z)^{\rm SM}\ \left[\, 1 - \left( {\Delta
m_b \over m_b}\right)^{\rm massive}\,\right]\ .
\end{equation}
We approximate these corrections as follows.  We ignore the small $W$,
$Z$, Higgs, and neutralino contributions.  This leaves the
squark/gluino and squark/chargino loops,
\begin{equation}
\left({\Delta m_b\over m_b}\right)^{\rm massive}\ = \ \left({\Delta
m_b\over m_b}\right)^{\tilde b\tilde g}\ + \ \left({\Delta m_b\over
m_b}\right)^{\tilde t\tilde\chi^+} \ .\label{mb app}
\end{equation}
We then set $p^2 = 0$.  The squark/gluino contribution is again
given by (\ref{mt gluino}), with the obvious substitution
$t\rightarrow b$.  To approximate the squark/chargino contribution, we
set $g = g' = \lambda_b = \lambda_t = 0$, except for terms that are
enhanced by the Higgsino mass parameter $\mu$ or by $\tan\beta$.  We
simplify our expressions by setting the chargino masses to $M_2$ and
$\mu$, respectively.  In this case the squark/chargino loops give rise
to the following terms,
\begin{eqnarray}
&&\left({\Delta m_b\over m_b}\right)^{\tilde t\tilde\chi^+}\ =
\ {\lambda_t^2\over16\pi^2}\ \mu\ {A_t\tan\beta+\mu \over m_{{\tilde
t}_1}^2-m_{{\tilde t}_2}^2} \ \Bigg[\, B_0(0,\mu,m_{{\tilde
t}_1})-B_0(0,\mu,m_{{\tilde t}_2})\, \Bigg] \nonumber \\\qquad &&+
\ {g^2\over16\pi^2}\,\Bigg\{ {\mu M_2\tan\beta \over
\mu^2-M_2^2}\, \Bigg[\, c_t^2 B_0(0,M_2,m_{{\tilde t}_1}) + s_t^2
B_0(0,M_2,m_{{\tilde t}_2}) \Bigg]\ +\ (\mu\leftrightarrow M_2)
\ \Bigg\}\ ,
\end{eqnarray}
where $B_{0}(0,m_1,m_2)$ is defined in (\ref{b0(0)}), and $c_t\ (s_t)$
is $\cos\theta_t\ (\sin\theta_t)$.

In Fig.~\ref{bmass} we show the corrections to the
\mbox{\footnotesize$\overline{\rm DR}~$} bottom quark mass, $\hat
m_b(M_Z)$, plotted against $\tan\beta$.  Figure \ref{bmass}(a) shows
the full one-loop correction, while (b) and (c) illustrate the
predominately finite corrections from squark/gluino and
squark/chargino loops.  At large $\tan\beta$, the top branches in
Figs.~(a) and (b) correspond to $\mu<0$, while for Fig.~(c) the bottom
branch corresponds to $\mu<0$. The contributions from Figs.~(b) and
(c) tend to cancel.  Because of the cancellations and large
corrections, previous approximations to the bottom quark mass
appearing in the literature can be substantially different from the
full one-loop result.  In Fig.~\ref{bmass}(d) we see that the
approximation (\ref{mb app}) typically agrees with the full one-loop
result to within a few percent.

\begin{figure}[t]
\epsfysize=1.5in \epsffile[-60 375 590 540]{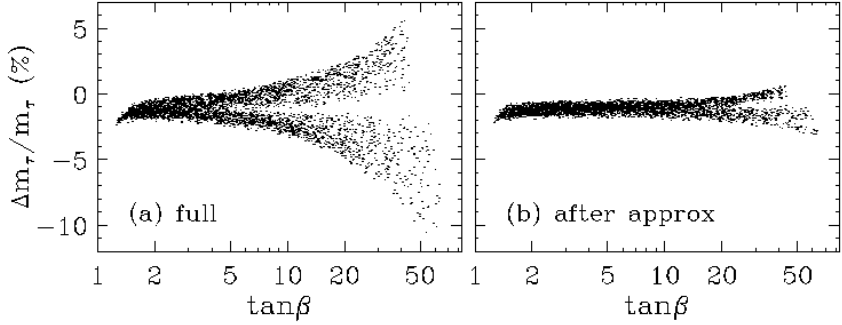}
\begin{center}
\parbox{5.5in}{
\caption[]{\small Supersymmetric corrections to the 
\mbox{\footnotesize$\overline{\rm DR}~$} tau lepton 
mass $\hat m_\tau(M_Z)$, plotted against $\tan\beta$.
Figure (a) gives the full one-loop correction; (b) illustrates the 
difference between the full one-loop result and the approximation 
given in the text.
\label{tau}}}
\end{center}
\end{figure}

\subsection{Tau lepton mass}

The corrections to the tau lepton mass are of course much smaller than
those of the quarks.  After resumming the two-loop QED corrections
which relate the tau pole mass to the
\mbox{\footnotesize$\overline{\rm DR}~$} running mass at $M_Z$
\cite{Ramond}, we obtain the \mbox{\footnotesize$\overline{\rm DR}~$}
mass $\hat m_\tau(M_Z) = 1.7463$ GeV.  We approximate the remaining
corrections by setting $p^2=0$ and keeping only those terms
proportional to $g^2$ and enhanced by $\mu$ or $\tan\beta$.  The only
such terms arise from the chargino loops. They give
\begin{equation}
\left({\Delta m_{\tau}\over m_{\tau}}\right)\ = \ {g^2\over16\pi^2}
\ {\mu M_2 \tan\beta \over \mu^2 - M_2^2} \ \Bigg[\,
B_0(0,M_2,m_{\tilde\nu_{\tau}}) - B_0(0,\mu,m_{\tilde\nu_{\tau}})\,
\Bigg]\ ,
\end{equation}
where $B_{0}(0,m_1,m_2)$ is given in (\ref{b0(0)}).  We illustrate the
tau corrections in Fig.~\ref{tau}.  The full correction ranges from
$-10\%$ to $+6\%$, while the approximation is good to within a few
$\%$.  For large $\tan\beta$, the top branch corresponds to $\mu>0$.

\subsection{Yukawa coupling unification}

In many supersymmetric unified theories, the
\mbox{\footnotesize$\overline{\rm DR}~$} bottom and tau Yukawa
couplings are predicted to unify at the scale $M_{\rm GUT}$ \cite{Lopez}.
To test this hypothesis, one must first extract the running
\mbox{\footnotesize$\overline{\rm DR}~$} Yukawa couplings $\lambda_b$
and $\lambda_\tau$ from the \mbox{\footnotesize$\overline{\rm DR}~$}
bottom and tau masses.  Our procedure is as follows.  We first use the
formulae in Appendix D to find \mbox{\footnotesize$\overline{\rm
DR}~$} masses for the bottom and tau.  We then use the following
relations to find the \mbox{\footnotesize$\overline{\rm DR}~$} Yukawa
couplings at the scale $M_Z$,
\begin{eqnarray}
\hat m_b(M_Z)\ &=&\ {1\over \sqrt{2}}\, \lambda_b(M_Z)\ v(M_Z)
\cos\beta(M_Z) \nonumber\\ \hat m_\tau(M_Z)\ &=&\ {1\over \sqrt{2}}\,
\lambda_\tau(M_Z)\ v(M_Z) \cos\beta(M_Z)\ .
\label{mt yuk}
\end{eqnarray}

We determine the full one-loop \mbox{\footnotesize$\overline{\rm
DR}~$} vev $v(M_Z)$ from the relation
\begin{equation}
M_Z^2 + {\cal R}e\,\Pi_{ZZ}^T(M_Z^2) \ =\ {1\over4}\biggl(g^2(M_Z) +
g'^2(M_Z)\biggr) v^2(M_Z)
\end{equation}
where $g$ and $g'$ are the \mbox{\footnotesize$\overline{\rm DR}~$}
couplings, and $\Pi_{ZZ}^T$ is the transverse $Z$-boson
\mbox{\footnotesize$\overline{\rm DR}~$} self-energy.  Alternatively,
the \mbox{\footnotesize$\overline{\rm DR}~$} vev $v(M_Z)$ can be taken
from the following empirical fit,
\begin{equation}
v(M_Z)\ =\ \left[\,248.6\ +\ 0.9\,\ln\left({M_{\tilde q}\over
M_Z}\right) \right]\ {\rm GeV}\ .
\end{equation}
This expression gives the correct one-loop vev to an accuracy of
better than 1\%.

\begin{figure}[t]
\epsfysize=2.5in \epsffile[-140 220 0 535]{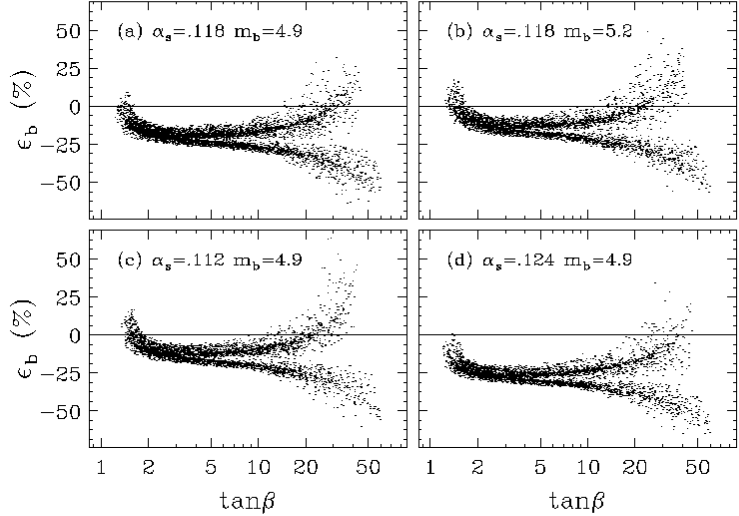}
\begin{center}
\parbox{5.5in}{
\caption[]{\small The unification-scale correction, $\varepsilon_b,$
that is necessary to obtain bottom-tau unification with the given
values of $\alpha_s(M_Z)$ and $m_b$, plotted versus $\tan\beta$. The
bottom quark pole mass is labeled in GeV.
\label{epsb}}}
\end{center}
\end{figure}

Once we have the \mbox{\footnotesize$\overline{\rm DR}~$} Yukawa
couplings at $M_Z$, we run them to the unification scale $M_{\rm
GUT}$.  At that scale we define the unification threshold
$\varepsilon_b$ to be the discrepancy between the couplings,
\begin{equation}
\lambda_b(M_{\rm GUT})\ =\ \lambda_\tau(M_{\rm GUT})\,
\left(1+\varepsilon_b\right) \ .
\end{equation}
Of course, the relation between the bottom quark mass and the
\mbox{\footnotesize$\overline{\rm DR}~$} Yukawa coupling $\lambda_b$
depends strongly on the QCD coupling $\alpha_s(M_Z)$.  Therefore we
compute $\varepsilon_b$ assuming that $\varepsilon_g$ has already been
chosen so that $\alpha_s(M_Z)$ is some fixed value.  Such an analysis
is illustrated in Fig.~\ref{epsb}(a), where we plot $\varepsilon_b$
versus $\tan\beta$ for $\alpha_s(M_Z)=0.118$.  {}From the figure we
see the well-known feature that Yukawa unification is possible with
$\varepsilon_b \simeq0$ for small $\tan\beta$
($1.2\,\roughly{<}\,\tan\beta\,\roughly{<}\,1.7$) and large
$\tan\beta$ ($15\,\roughly{<}\,\tan\beta\,\roughly{<}\,40$).  In the
large $\tan\beta$ region we distinguish the two cases, $\mu<0$ and
$\mu>0$.  For $\mu<0$ we see that $\varepsilon_b$ is always
far from zero, in the range $-24$ to $-60\%$.  For $\mu>0$
we find points which permit bottom-tau unification with
$\varepsilon_b\simeq0$. However, they are not generic;
$\varepsilon_b$ depends sensitively on the parameter space,
and varies between $-20$ to $+30\%$.

The discrepancy $\varepsilon_b$ is sensitive to the value of
$\alpha_s(M_Z)$, as well as the input value for the bottom quark pole
mass.  We illustrate this in Figs.~\ref{epsb}(b-d), for the
$(\alpha_s,\ m_b)$ values (0.118,\ 5.2), (0.112,\ 4.9), and (0.124,
\ 4.9), with $m_b$ in GeV.  We note that setting $\alpha_s(M_Z)= 0.112$
and $m_b=5.2$ GeV, we find solutions with $|\varepsilon_b|<0.05$ over
the whole range $1.3<\tan\beta<30$.

\section{Supersymmetric and Higgs boson masses}

\begin{figure}[t]
\epsfysize=2.5in \epsffile[-40 280 600 535]{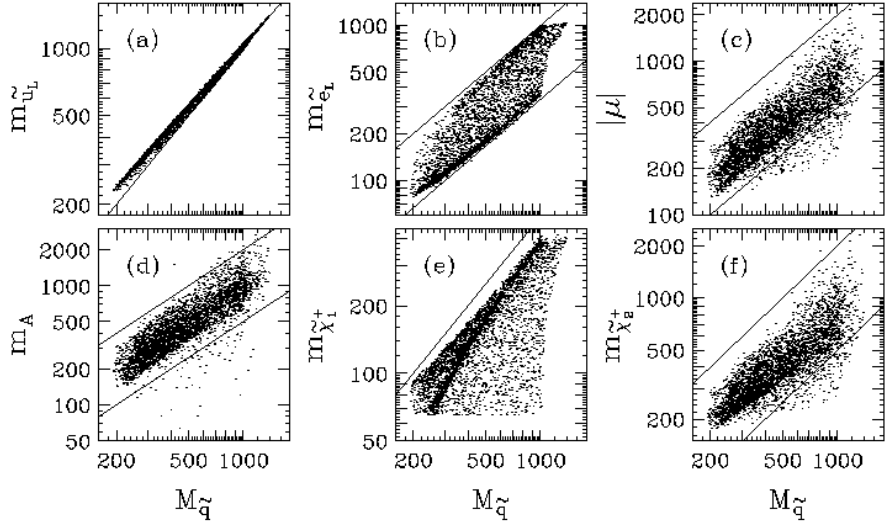}
\begin{center}
\parbox{5.5in}{
\caption[]{\small The masses (a) $m_{\tilde u_L}$, (b) $m_{\tilde
e_L}$, (c) $|\mu|$, (d) $m_A$, (e) $m_{\tilde\chi_1^+}$, and (f)
$m_{\tilde\chi_2^+}$, versus $M_{\tilde q}$.  The lines indicate (a)
$M_{\tilde q}$, (b) $M_{\tilde q}/3$ and $M_{\tilde q}$, (c)
$M_{\tilde q}/2$ and $2M_{\tilde q}$, (d) $M_{\tilde q}/2$ and
$2M_{\tilde q}$, (e) $M_{\tilde q}/2$, and (f) $M_{\tilde q}/2$ and
$2M_{\tilde q}$.  The units for both axes are in GeV.
\label{m vs msq}}}
\end{center}
\end{figure}

We will begin our analysis of the supersymmetric spectrum by
discussing some of its general features.  We will use some of these
features when we derive our approximations for the radiative
corrections.  We will be careful to note when we do, so that one can
assess the validity of our approximations in other scenarios.

\begin{figure}[t]
\epsfysize=2.5in \epsffile[-140 220 0 535]{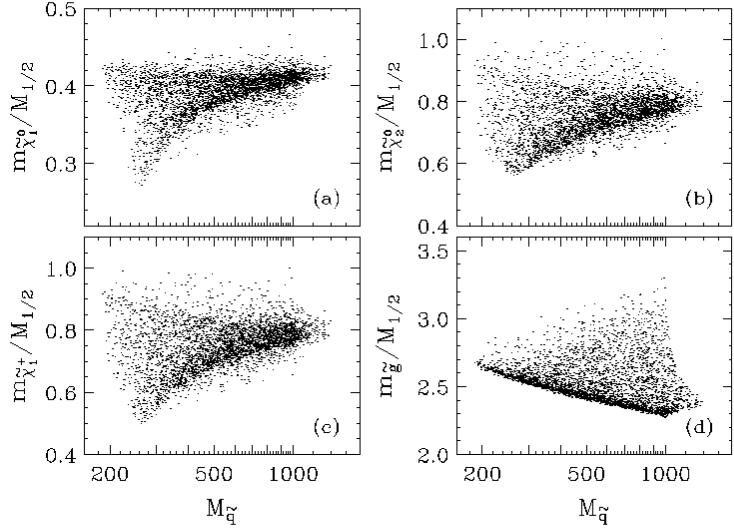}
\begin{center}
\parbox{5.5in}{
\caption[]{\small The ratios (a) $m_{\tilde\chi_1^0}/M_{1/2}$, (b)
$m_{\tilde\chi_2^0}/M_{1/2}$, (c) $m_{\tilde\chi_1^+}/M_{1/2}$, and
(d) $m_{\tilde g}/M_{1/2}$, versus $M_{\tilde q}$.
\label{inos}}}
\end{center}
\end{figure}

Perhaps the most striking consequence of the universal boundary
conditions is the fact that they produce a low energy spectrum which
is tightly correlated with the magnitude of the squark mass scale
$M^2_{\tilde q} = M^2_0 + 4 M^2_{1/2}$.  In Fig.~\ref{m vs msq} we
show the masses $m_{\tilde u_L}$, $m_{\tilde e_L}$, $|\mu|$, $m_A$,
$m_{\tilde\chi_1^+}$, and $m_{\tilde\chi_2^+}$ versus $M_{\tilde q}$.
{}From the figure we see that the squark masses are nearly equal to
$M_{\tilde q}$, while the other masses are generally within a factor
of two or three.  The exceptions to this degeneracy include the Higgs
boson, $h$, which is always light, and the additional possibilities of
a light top squark, a light Higgs sector, and/or light gauginos.  Of
course, the gaugino masses are nearly proportional to $M_{1/2}$.  We
find that typically $m_{\tilde\chi_1^0}\simeq 0.4 M_{1/2}$,
$m_{\tilde\chi_2^0}\simeq m_{\tilde\chi_1^+}\simeq 0.8 M_{1/2}$, and
$m_{\tilde g}\simeq 2.4 M_{1/2}$, but there are substantial variations
from weak-scale threshold corrections (and from mixing for the
charginos/neutralinos).  We show the ratios $m_{\tilde\chi}/M_{1/2}$
and $m_{\tilde g}/M_{1/2}$ versus $M_{\tilde q}$ in Fig.~\ref{inos}.

In the rest of this section we discuss the one-loop corrections to the
masses of the gluino, the charginos and neutralinos, the squarks, the
sleptons, and the Higgs bosons. In the following expressions for the
mass corrections we implicitly take the real part of the
Passarino-Veltman functions.

\subsection{Gluino mass}

\begin{figure}[t]
\epsfysize=2.5in \epsffile[-140 210 0 525]{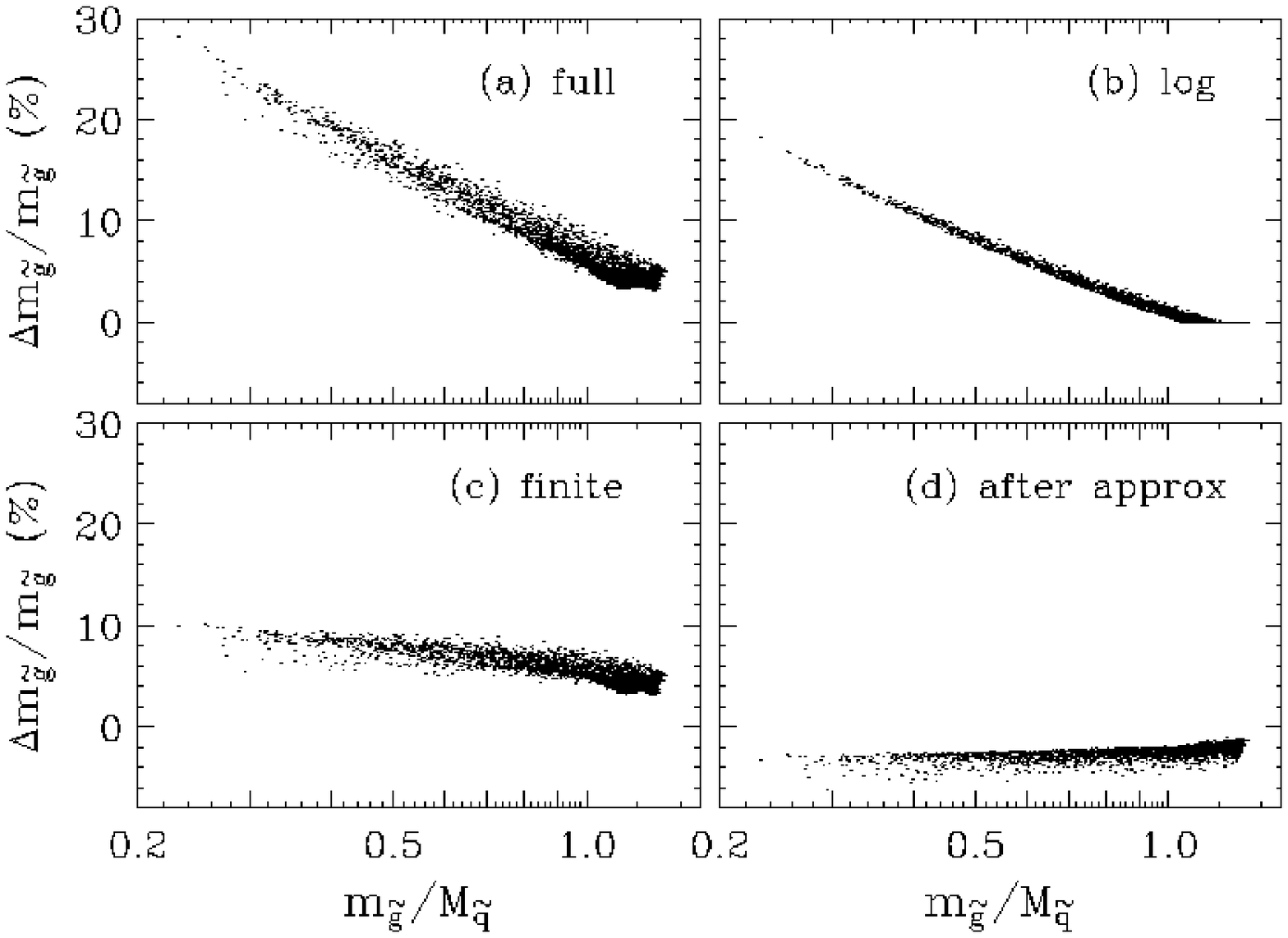}
\begin{center}
\parbox{5.5in}{
\caption[]{\small Corrections to the gluino mass versus $m_{\tilde
g}/M_{\tilde q}$.  Figure (a) shows the complete one-loop corrections;
(b) shows the leading logarithmic corrections; (c) shows the finite
corrections; and (d) shows the difference between the full one-loop
result and the approximation given in the text.
\label{gl}}}
\end{center}
\end{figure}

The gluino mass corrections are perhaps the simplest of all the mass
renormalizations.  They have previously been studied in
Refs.~\cite{MV,PP2,Kras,Donini}; for completeness we list the
corrections in Appendix D.  The gluino mass corrections arise from
gluon/gluino and quark/squark loops.  The corrections can be rather
large, so we include them in a way which automatically incorporates
the one-loop renormalization group resummation,
\begin{equation}
m_{\tilde g}\ =\ M_3(Q) \left[\, 1 - \left({\Delta M_3\over M_3}
\right)^{g\tilde g} - \left({\Delta M_3\over M_3}\right)^{q\tilde q}\,
\right]^{-1}\ .
\end{equation}
The gluon/gluino loop gives
\begin{eqnarray}
\left({\Delta M_3 \over M_3}\right)^{g\tilde g}\ &=& \ {3
g^2_3\over8\pi^2}\ \left[\,2B_0(M_3,M_3,0) -
B_1(M_3,M_3,0)\,\right] \nonumber \\ &=&\ {3 g^2_3 \over 16\pi^2}
\ \left[\,3\ln\left({Q^2\over M_3^2}\right) + 5\,\right]\ .
\end{eqnarray}
The quark/squark loop can be simplified by assuming that all quarks
have zero mass, and that all squarks have a common mass, which we take
to be $M_{Q_1}$, the soft mass of the first generation of left-handed
squarks.  We find
\begin{equation}
\left({\Delta M_3\over M_3}\right)^{q\tilde q}\ = \ -\ {3
g^2_3\over4\pi^2}\ B_{1}(M_3,0,M_{Q_1})\ .
\end{equation}
Here
\begin{equation}
B_1(p,0,m)\ =\ -{1\over2}\ln\left({M^2\over Q^2 }\right) + 1 - {1\over
2x} \left[\, 1+{(x-1)^2\over x}\ln |x-1|\, \right] +
{1\over2}\theta(x-1)\ln x\ ,
\label{b1_p0m2}
\end{equation}
where $M = \max(p^2,m^2)$ and $x=p^2/m^2$.  As usual, the full mass
renormalization contains logarithmic and finite contributions.

The gluino mass corrections are shown in Fig.~\ref{gl}.  In the figure
we define the tree-level gluino mass to be $M_3(M_3)$, and we evaluate
the one-loop mass at the scale $Q = M_{\tilde q}$.  Because we resum
the correction, varying the scale from $M_{\tilde q}/2$ to $2M_{\tilde
q}$ changes the one-loop mass by at most $\pm1\%$.

{}From the figure we see that the leading logarithmic correction can
be as large as 20\%, while the finite correction ranges from 3 to
10\%.  The finite contribution is largest in the region where the
logarithm is largest, so the leading logarithm approximation is
nowhere good.  On the other hand, the approximation we provide
typically holds to a few percent.  It is off by as much as 6\% in the
region where the full correction is 30\%.  In this region we expect
the two-loop correction to be of order 6\%.

\subsection{Neutralino and Chargino Masses}

The complete set of corrections to the neutralino and chargino masses
\cite{PP1,PP2} is given in Appendix D.  In this section we present a
set of approximations to these corrections.  These approximations are
more involved than those discussed above because there are no color
corrections that would dominate the results.

Our approximation is as follows.  We start by assuming that $|\mu|
\,>\,M_1,M_2, M_Z.$ (We find that $M_Z^2/\mu^2$ and $M_2^2/\mu^2$ are
less than 0.53; see Fig.~\ref{m vs msq}).  We work with the
undiagonalized tree-level (chargino or neutralino) mass matrix, and
correct the diagonal entries only, that is, the parameters $M_1$,
$M_2$, and $\mu$.  This approximation neglects the {\em corrections}
to the off-diagonal entries of the mass matrices, which leads to an
error of order $(\alpha/4\pi)M^2_Z/\mu^2$ in the masses.

We simplify our expressions by setting all loop masses and external
momenta to their diagonal values, i.e. we set $m_{\tilde\chi_1^0} =
M_1$, etc.  We neglect all Yukawa couplings except $\lambda_t$ and
$\lambda_b$.  We also ignore the mixings of the charginos and
neutralinos in the radiative corrections. This also leads to an error
of order $(\alpha/4\pi)M_Z^2/\mu^2$ in our final result.

We further simplify our expressions by setting all quark masses to
zero, and by assuming that all squarks are degenerate with mass
$M_{Q_1}$, and that all sleptons are degenerate as well with mass
$M_{L_1}$.  We also take the Higgs masses to be $m_h = M_Z$ and $m_H =
m_{H^+} = m_A$.  This means that we also neglect terms of order
$(\alpha/4\pi)M_Z^2/m_A^2$.

In this limit, the dominant correction to $M_1$ comes from
quark/squark, chargino/charged-Higgs and neutralino/neutral-Higgs
loops.  We find
\begin{eqnarray}
\left({\Delta M_1 \over M_1}\right)\ &=& \ -{g'^2 \over 16
\pi^2}\,\bigg\{ \,11 B_1(M_1,0,M_{Q_1}) + 9 B_1(M_1,0,M_{L_1})
\nonumber \\ &&\ +\ {\mu\over M_1} \sin(2\beta)\ \bigg( \,
B_0(M_1,\mu,m_A) - B_0(M_1,\mu,M_Z)\,\bigg) \nonumber \\ &&\ +
\ B_1(M_1,\mu,m_A) + B_1(M_1,\mu,M_Z) \, \bigg\}\ .
\label{M1a}
\end{eqnarray}
Since $M_Z,\ M_1 \ll \mu$, we can simplify this
expression by setting $M_Z = M_1 = 0$ inside the $B$ functions.  This
gives
\begin{eqnarray}
\left({\Delta M_1 \over M_1}\right)\ &=&\ {g'^2 \over 32 \pi^2}\,
\bigg\{\,11 \theta_{M_1 M_{Q_1}} + 9 \theta_{M_1 M_{L_1}} +
\theta_{M_1 \mu M_Z} - 2B_1(0,\mu,m_A) \nonumber \\ &&\ +\ {2 \mu\over
M_1} \sin(2\beta)\ \bigg(\,B_0(0,\mu,0)\,-\,B_0(0,\mu,m_A)\,\bigg)\ -
\ {23 \over 2}\, \bigg\}\ ,
\label{M1b}
\end{eqnarray}
where $B_0(0,m_1,m_2)$ and $B_1(0,m_1,m_2)$ were defined in
(\ref{b0(0)}) and (\ref{b1(0)}), and $\theta_{m_1 \ldots m_2} \equiv
\ln(M^2/Q^2)$ with $M^2 = {\rm max}(m_1^2,...,m_2^2)$.  The form of
the finite corrections depends on the assumed hierarchy in the
low-energy spectrum, but the leading logarithms are always correctly
given by the $\theta$ terms.  We set the first subscript of a $\theta$
term, $m_1$, equal to the external momentum.  Note that when the
renormalization scale equals the external momentum, $Q=m_1$, the theta
function reduces to the familiar form, $\theta_{m_1 m_2} =
\ln(m^2_2/Q^2)\,\theta(m^2_2 - Q^2)$.

The leading logarithmic corrections are easy to read from
Eq.~(\ref{M1b}).\footnote{The logarithmic part of $-2B_1(0,\mu,m_A)$
in Eq.~(\ref{M1b}) is given by $\theta_{M_1\mu m_A}$.}  Note that the
terms proportional to $\sin(2\beta)$ are enhanced by the ratio
$\mu/M_1$.  These finite terms are completely missed in the
run-and-match approach because they do not contribute to the beta
function.

In a similar way, we approximate the corrections to $M_2$ from
quark/squark and Higgs loops.  They are
\begin{eqnarray}
\left({\Delta M_2 \over M_2 }\right)\ &=&\ -\ {g^2 \over 16
\pi^2}\,\bigg\{ \,9 B_1(M_2,0,M_{Q_1}) + 3 B_1(M_2,0,M_{L_1})
\nonumber \\ &&\ +\ {\mu\over M_2} \sin(2\beta)\ \bigg(\,
B_0(M_2,\mu,m_A) - B_0(M_2,\mu,M_Z)\,\bigg) \nonumber \\ &&\ +
\ B_1(M_2,\mu,m_A) + B_1(M_2,\mu,M_Z) \, \bigg\}\ .
\label{mchi02_1}
\end{eqnarray}
Setting $M_Z = M_2 = 0$ inside the $B$ functions, we find
\begin{eqnarray}
\left({\Delta M_2 \over M_2 }\right)\ &=&\ {g^2 \over 32 \pi^2}\,
\bigg\{\,9 \theta_{M_2M_{Q_1}} + 3 \theta_{M_2M_{L_1}}\ +
\ \theta_{M_2\mu M_Z} - 2B_1(0,\mu,m_A) \nonumber \\ &&\ +\ {2 \mu\over
M_2} \sin(2\beta)\ \bigg(\,B_0(0,\mu,0) \,-\, B_0(0,\mu,m_A) \,\bigg)
\ -\ {15 \over 2}\, \bigg\}\ .
\label{mchi02_1_2}
\end{eqnarray}

There are additional corrections to $M_2$ from gauge boson loops.
Because $M_2$ enters both the chargino and the neutralino mass
matrices, the corrections differ slightly for the two cases.  However,
to the order of interest, it suffices to use the neutralino result,
\begin{equation}
\left({\Delta M_2 \over M_2 } \right)^{\rm gauge} \ =\ {g^2\over 4
\pi^2} \bigg\{\, 2 B_0(M_2,M_2,M_W) - B_1(M_2,M_2,M_W) \,\bigg\}\ .
\end{equation}
Because $M_2$ is of order $M_W$, one must use the full $B$ functions
in this expression.  Alternatively, one can use the following
empirical fit which works to better than 1\%,
\begin{eqnarray}
\left({\Delta M_2 \over M_2 } \right)^{\rm gauge} &=& -\ {g^2\over 4
\pi^2} \bigg\{\,{3\over2} \theta_{M_2M_W} + \theta(M_W-M_2) \left(
1.57 {M_2 \over M_W} - 1.85 \right) \nonumber \\ && -
\ \theta(M_2-M_W)\, \left[ \, 0.54 \ln \left({M_2 \over M_W} -
0.8\right) + 1.15 \,\right] \,\bigg\} \ .
\end{eqnarray}

The corrections to $\mu$ are obtained in a similar manner.  In the
limit $g'^2 \ll g^2$, we find
\begin{eqnarray}
\label{deltamu}
\left({\Delta \mu \over \mu }\right)\ &=& \ -\ {3\over 32 \pi^2}
\bigg[\, (\lambda_b^2 + \lambda_t^2) B_1(\mu,0,M_{Q_3}) + \lambda_t^2
B_1(\mu,0,M_{U_3}) + \lambda_b^2 B_1(\mu,0,M_{D_3})\, \bigg]\\ &&\ -
\ {3g^2 \over 64\pi^2} \bigg[ B_1(\mu,M_2,m_A) + B_1(\mu,M_2,M_Z) +2
B_1(\mu,\mu,M_Z) - 4 B_0(\mu,\mu,M_Z)\, \bigg] \ .\nonumber
\end{eqnarray}
As above, we set $M_Z = M_2 = 0$ inside the $B$ function, in which
case (\ref{deltamu}) reduces to
\begin{eqnarray}
\left({\Delta \mu \over \mu }\right)\ &=& \ -\ {3\over 32 \pi^2}
\bigg[\, (\lambda_b^2 + \lambda_t^2) B_1(\mu,0,M_{Q_3}) + \lambda_t^2
B_1(\mu,0,M_{U_3}) + \lambda_b^2 B_1(\mu,0,M_{D_3})\, \bigg] \nonumber
\\ &&\ +\ {3g^2 \over 64\pi^2} \bigg[ {1\over 2} \theta_{\mu M_2M_Z}
\, - \, 3 \theta_{\mu M_Z} \,-\,B_1(\mu,0,m_A) \, +\,4\,\bigg]\ .
\end{eqnarray}
The expression for $B_1(p,0,m)$ is given in Eq.~(\ref{b1_p0m2}).

\begin{figure}[t]
\epsfysize=2.5in \epsffile[-115 220 600 535]{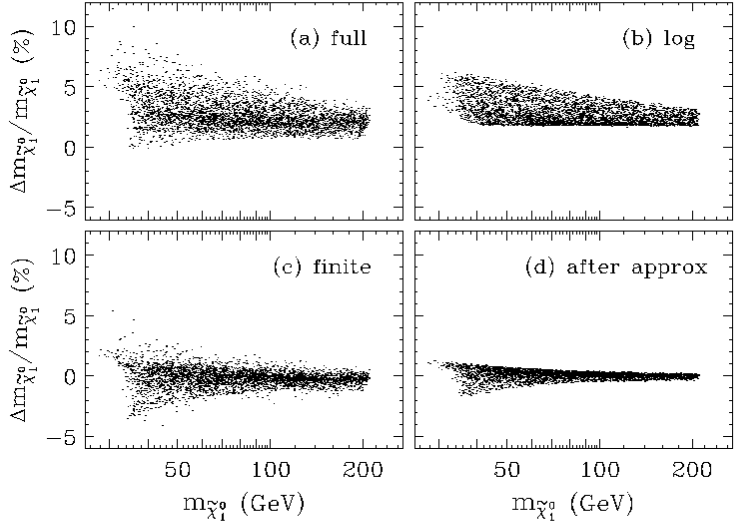}
\begin{center}
\parbox{5.5in}{
\caption[]{\small Corrections to the lightest neutralino mass, as in
Fig.~\ref{gl}.
\label{n1}}}
\end{center}
\end{figure}

In Fig.~\ref{n1} we show the corrections to the lightest neutralino
mass.  In Fig.~\ref{n1}(a) we show the full correction in percent,
with the tree-level mass defined as the eigenvalue of the mass matrix,
where the running parameters $M_1,\ M_2,$ and $\mu$ are evaluated at
their own scale. (The tree-level mass matrices also contain
$\tan\beta$ at $M_Z$ and the $W$- and $Z$-boson pole masses.)  The
one-loop masses have negligible scale dependence.

As usual, the full corrections are made up of logarithmic and finite
pieces.  The logarithmic corrections are shown in Fig.~\ref{n1}(b) and
the finite corrections are shown in Fig.~\ref{n1}(c). Note that the
finite corrections can be more than half as large as the logarithmic
corrections.  Indeed, the finite corrections can be larger than 5\% in
the small $M_1$ region, primarily because of the Higgsino-loop term
proportional to $\mu$.  In Fig.~\ref{n1}(d) we show the difference
between the full one-loop result and our approximation. Here, and in
the following two figures, the logarithmic corrections
[Fig.~\ref{n1}(b)] include an explicit sum over the soft squark and
slepton masses, while the approximations [Fig.~\ref{n1}(d)] use a
single soft squark or slepton mass.

\begin{figure}[t]
\epsfysize=2.5in \epsffile[-140 220 0 535]{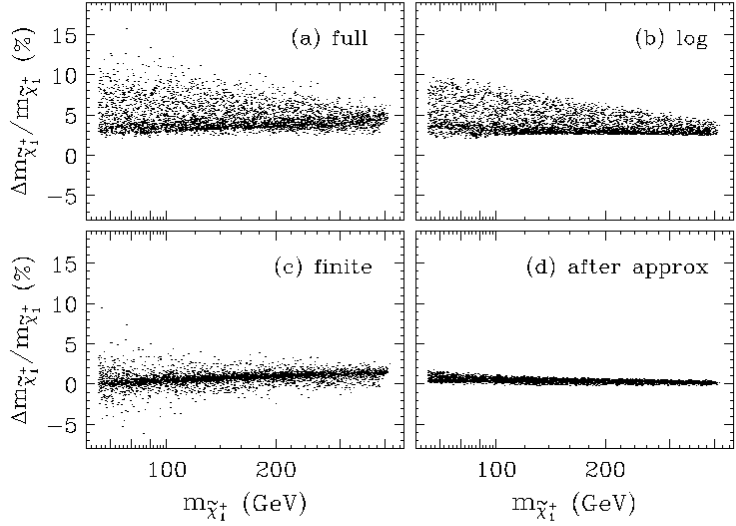}
\begin{center}
\parbox{5.5in}{
\caption[]{\small Corrections to the lightest chargino mass, as in
Fig.~\ref{gl}.
\label{c1}}}
\end{center}
\end{figure}

Figure \ref{c1} shows, similarly, the corrections to the lightest
chargino mass. Again there is a term proportional to $\mu$ which
dominates the finite corrections when $M_2$ is small.  In this region,
the finite corrections can be as large as 10\%.  In Fig.~\ref{c1}(d)
we show that the difference between our approximate correction and the
full one-loop mass is less than 2\%.  These corrections are
quantitatively similar to the corrections to the second-lightest
neutralino mass.

\begin{figure}[t]
\epsfysize=2.5in \epsffile[-123 220 17 535]{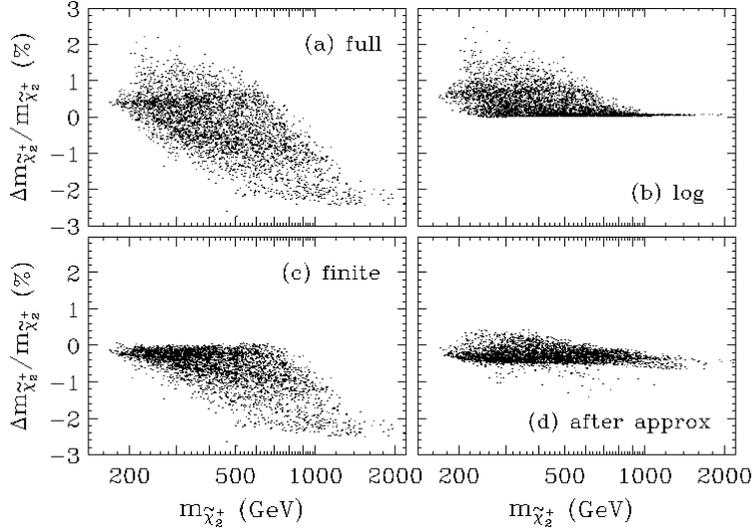}
\begin{center}
\parbox{5.5in}{
\caption[]{\small Corrections to the heaviest chargino mass, as in
Fig.~\ref{gl}.
\label{c2}}}
\end{center}
\end{figure}

The corrections to the heavy chargino mass are shown in Fig.~\ref{c2}.
These corrections are less than a few percent, as are the corrections
for the two heaviest neutralino masses.  The logarithmic corrections
are in the range 0 to 2.5\%, and the finite corrections are in the
range 0 to $-3\%$.  Figure \ref{c2}(d) shows that our approximation
for the heavy chargino mass generally holds to better than 0.5\%.  Our
approximation also works to typically better than 1\% for the two
heaviest neutralino masses, but it can be off by nearly 2\%.

\subsection{Squark masses}

\begin{figure}[t]
\epsfysize=1.5in \epsffile[-70 405 590 575]{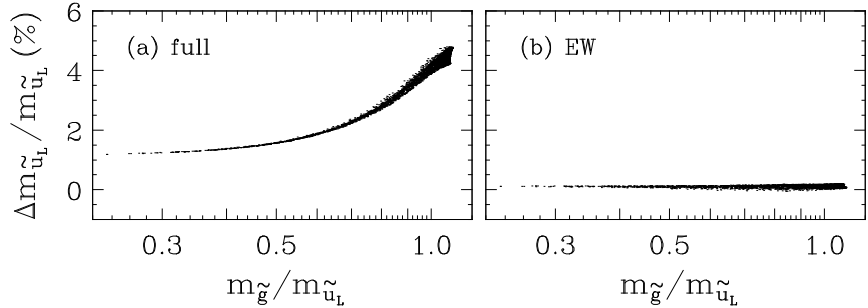}
\begin{center}
\parbox{5.5in}{
\caption[]{\small (a) Full one-loop corrections to the first generation
squark mass, $m_{\tilde u_L}$, versus the ratio $m_{\tilde
g}/m_{\tilde u_L}$.  (b) The difference between the full corrections
and the approximation in the text, versus $m_{\tilde g}/m_{\tilde
u_L}$.  (These are essentially the electroweak corrections.)\label{lsq}}}
\end{center}
\end{figure}

The first two generations of squarks receive QCD \cite{Martin} and
electroweak corrections.  However, it is a very good approximation
to ignore the electroweak graphs, since the dominant corrections
come from  gluon/squark and gluino/quark loops.  Neglecting the
quark masses, these corrections are as follows,
\begin{equation}
m_{\tilde q}^2\ =\ \hat m_{\tilde q}^2(Q) \left[\,1 + \left({\Delta
m_{\tilde q}^2 \over m_{\tilde q}^2}\right)\,\right]\ ,
\end{equation}
where
\begin{eqnarray}
\left( {\Delta m_{\tilde q}^2 \over m_{\tilde q}^2} \right) &=&
\ {g^2_3 \over 6\pi^2} \ \bigg[\,2 B_1(m_{\tilde q},m_{\tilde q},0) +
{A_0(m_{\tilde g})\over m_{\tilde q}^2} - (1 - x) B_0(m_{\tilde q},
m_{\tilde g},0)\, \bigg] \nonumber \\ &=&\ {g^2_3 \over 6\pi^2}
\bigg[\,1 + 3x + (x-1)^2\ln|x-1| - x^2\ln x\ + 2x\ln\left({Q^2\over
m_{\tilde q}^2}\right)\,\bigg]~,
\label{QCDcorr}
\end{eqnarray}
and $x=m_{\tilde g}^2/m_{\tilde q}^2$.

For the case of universal boundary conditions the gluino mass is less
than or roughly equal to the squark mass, so the correction
(\ref{QCDcorr}) is essentially finite at $Q = m_{\tilde q}$.  From
Fig.~\ref{lsq} we see that it varies from around 1\% for $x \ll 1$ to
between 4 and 5\% for $x \simeq 1$.  We also see that the electroweak
corrections are small, less than 0.5\%.

\begin{figure}[t]
\epsfysize=2.5in \epsffile[-140 220 0 535]{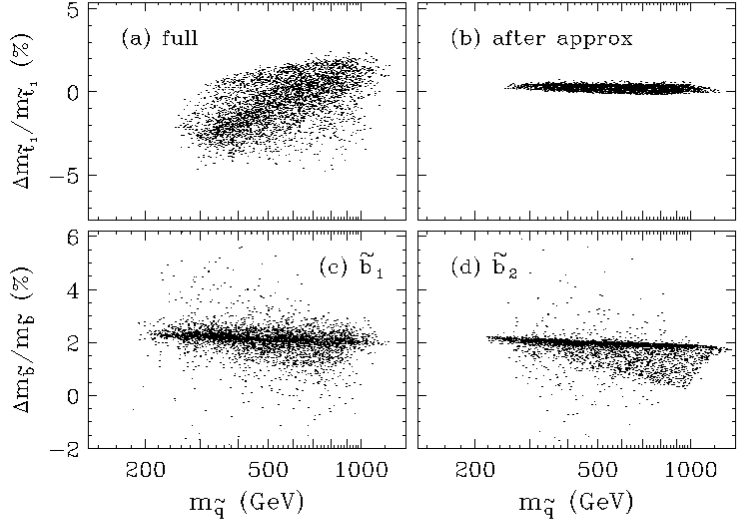}
\begin{center}
\parbox{5.5in}{
\caption[]{\small (a) The corrections to the heavy top squark mass
versus its mass.  (b) The difference between the full one-loop heavy
top squark mass and the approximation, Eq.~(\ref{top sq app}).  (c)
Same as (a), for ${\tilde b}_1$. (d) Same as (a), for ${\tilde b}_2$.
\label{tb}}}
\end{center}
\end{figure}

The third generation squark masses receive Yukawa corrections on the
order of, and opposite in sign to, the QCD corrections.  In
Fig.~\ref{tb} we show the full corrections to the third generation
heavy squark masses.  As usual, the tree-level masses are
defined in terms of the gauge-boson and quark pole masses, as well as
the soft masses $M_Q(M_Q)$, $M_U(M_U)$, and $M_D(M_D)$. The tree-level
mass matrices also contain $\tan\beta(M_Z)$, $\mu(\mu)$, and $A_i({\rm
max} (|A_i|,M_Z))$, where $A_i$ denotes the top or bottom $A$-term.
(Our convention for the third generation squarks is to associate the
subscript 1 with the mostly left-handed squark.  Since the light top
squark is predominantly right-handed, its mass is denoted $m_{\tilde
t_2}$.)

{}From Fig.~\ref{tb} we see that the heavy top squark mass receives
corrections in the range $-5$ to 2\%, while the bottom squark masses
receive corrections mostly in the 0 to 3\% range.  We note that in
none of these cases does the leading logarithm approximation work
well: as is the case for all the squarks and sleptons, these
corrections are essentially non-logarithmic. (The light top squark
mass does receive some substantial logarithmic corrections, but they
are generally not larger than the finite corrections.)

We will now present our approximation for the top squark mass matrix.
We will derive our approximation for the case of the light top squark,
but it also works quite well for the heavy top squark (see
Fig.~\ref{tb}(b)).  The mass of the light top squark receives
potentially large additive corrections proportional to the the strong
coupling and the top and bottom Yukawa couplings.  We approximate the
corrections to $m_{\tilde t_2}$ by neglecting $g, \ g'$ and the Yukawa
couplings of the first two generations.  We also neglect all quark
masses except $m_t$, which eliminates all sfermion mixing except for
that of the top squarks.

We neglect the mixing of charginos and neutralinos, so the two heavy
neutralinos and the heavy chargino all have mass $|\mu|$. We also make
the approximations $m_h = M_Z$ and $m_H = m_{H^+} = m_A$. Finally, we
set $p=0$ in the $B$-functions if any of the other arguments is much
bigger than $m_{{\tilde t}_2}$.

This gives the following expressions for the one-loop corrections to
the top squark mass matrix,
\begin{equation}
{\cal M}_{\tilde t}^2\ =\ \hat {\cal M}_{\tilde t}^2(Q)\ +\ \pmatrix{
\Delta M^2_{LL} & \Delta M^2_{LR} \cr \Delta M^2_{LR} & \Delta
M^2_{RR}}\ .\label{top sq app}
\end{equation}
The $\Delta M^2$ entries are as follows:
\begin{eqnarray}
\Delta M^2_{LL} & =& \ {g^2_3 \over 6\pi^2}\ \bigg\{ 2 m_{{\tilde
t}_2}^2 \left[ c_t^2 B_1(m_{{\tilde t}_2},m_{{\tilde t}_1},0) \ +
\ s_t^2 B_1(m_{{\tilde t}_2},m_{{\tilde t}_2},0) \right] \nonumber \\ &&
\qquad +\ A_0(m_{\tilde g})\ +\ A_0(m_t) \ -\ ( m_{{\tilde t}_2}^2 -
m_{\tilde g}^2 - m_t^2) B_0(0,m_{\tilde g},m_t) \bigg\} \nonumber \\
&-&{1\over16\pi^2} \bigg[ \lambda_t^2 s_t^2 A_0(m_{{\tilde t}_1})\ +
\ \lambda_b^2 A_0(m_{\tilde b}) \nonumber \\ && \qquad -\ 2 (\lambda_t^2 +
\lambda_b^2) A_0(\mu) \ +\ (\lambda_t^2 c_{\beta}^2 + \lambda_b^2
s_{\beta}^2) A_0(m_A) \bigg] \nonumber \\ &-& {\lambda_t^2\over32\pi^2}
\bigg[ \Lambda(\theta_t,\beta) B_0(0,m_{{\tilde t}_1},m_A) \ +
\ \Lambda(\theta_t-{\pi\over2},\beta)B_0(0,0,m_A) \nonumber \\ & &
\qquad +\ \Lambda(\theta_t,\beta-{\pi\over2}) B_0(0,m_{{\tilde
t}_1},0) \ +\ \Lambda(\theta_t-{\pi\over2},\beta-{\pi\over2})
B_0(m_{{\tilde t}_2},m_{{\tilde t}_2},m_Z) \bigg] \nonumber \\ &-&
{1\over16\pi^2} \bigg[ \left( \lambda_t^2 m_t^2 c_\beta^2 \ +
\ \lambda_b^2 (\mu c_\beta - A_b s_\beta)^2 \right) B_0(0,m_{\tilde
b},m_A) \nonumber \\ && \qquad + \ \left( \lambda_t^2 m_t^2 s_\beta^2
\ +\ \lambda_b^2 (\mu s_\beta + A_b c_\beta)^2 \right) B_0(0,m_{\tilde
b},0) \bigg] \\
\Delta M^2_{LR} &=& \ - {g^2_3 \over 6\pi^2} c_ts_t \left[
(m_{{\tilde t}_1}^2 + m_{{\tilde t}_2}^2) B_0(m_{{\tilde t}_2},
m_{{\tilde t}_1},0) \ +\ 2 m_{{\tilde t}_2}^2
B_0(m_{{\tilde t}_2},m_{{\tilde t}_2},0) \right]
\nonumber \\ &-& {g^2_3 \over 3\pi^2} m_t m_{\tilde g}
B_0(0,m_t,m_{\tilde g}) \ -\ {3\lambda_t^2\over16\pi^2} c_t s_t
A_0(m_{{\tilde t}_1}) \nonumber \\ &-& {\lambda_t^2\over32\pi^2} \bigg[
\Omega( \theta_t,\beta) B_0(0,m_{{\tilde t}_1},m_A) \ +
\ \Omega(-\theta_t,\beta) B_0(0,0,m_A) \nonumber \\ && \qquad +
\ \Omega( \theta_t,{\pi\over2}+\beta)B_0(0,m_{{\tilde t}_1},0) \ +
\ \Omega(-\theta_t,{\pi\over2}+\beta)B_0(m_{{\tilde t}_2},
m_{{\tilde t}_2},M_Z) \bigg] \nonumber \\
&-&{1\over16\pi^2} \bigg[ -\biggl(
 \lambda_t^2 m_t c_\beta (\mu s_\beta - A_t c_\beta) \ +\ \lambda_b^2
m_t s_\beta (\mu c_\beta - A_b s_\beta) \biggr)
B_0(0,m_{\tilde b},m_A)\nonumber\\
&& \qquad +\ \lambda_t^2 m_t s_\beta (\mu c_\beta
 + A_t s_\beta) B_0(0,m_{\tilde b} ,0) \bigg] \\
\Delta M^2_{RR} &=& \ {g^2_3 \over 6\pi^2}\ \bigg\{ 2 m_{{\tilde
 t}_2}^2 \left[ s_t^2 B_1(m_{{\tilde t}_2},m_{{\tilde t}_1},0) \ +
\ c_t^2 B_1(m_{{\tilde t}_2},m_{{\tilde t}_2},0) \right] \nonumber \\ &&
 \qquad +\ A_0(m_{\tilde g})\ + \ A_0(m_t) \ -\ ( m_{{\tilde t}_2}^2 -
 m_{\tilde g}^2 - m_t^2) B_0(0,m_{\tilde g},m_t) \bigg\} \nonumber \\
 &-&{\lambda_t^2\over16\pi^2} \left[ c_t^2 A_0(m_{{\tilde t}_1})\ +
\ A_0(m_{\tilde b}) \ -\ 4 A_0(\mu) \ +\ 2 c_{\beta}^2 A_0(m_A) \right]
 \nonumber \\ &-& {\lambda_t^2\over32\pi^2} \bigg[
 \Lambda({\pi\over2}-\theta_t,\beta) B_0(0,m_{{\tilde t}_1},m_A) \ +
\ \Lambda(-\theta_t,\beta)B_0(0,0,m_A) \nonumber \\ & & \qquad +
\ \Lambda({\pi\over2}-\theta_t,\beta-{\pi\over2})B_0(0,m_{{\tilde
 t}_1},0) \ +\ \Lambda(-\theta_t,\beta-{\pi\over2}) B_0(m_{{\tilde
 t}_2},m_{{\tilde t}_2},m_Z) \bigg] \nonumber \\ &-& {1\over16\pi^2}
 \bigg[ \left( \lambda_b^2 m_t^2 s_\beta^2 +\lambda_t^2 (\mu s_\beta -
 A_t c_\beta)^2 \right) B_0(0,m_{\tilde b},m_A) \nonumber \\ && \qquad
 + \ \lambda_t^2 (\mu c_\beta + A_t s_\beta)^2 B_0(0,m_{\tilde b},0)
 \bigg] \ .
\label{stop self}
\end{eqnarray}
We have defined the two functions
\begin{eqnarray}
\Lambda(\theta_t,\beta) & = & \left( 2 m_t \cos \beta \cos\theta_t\ -
\ (\mu \sin \beta - A_t \cos \beta ) \sin\theta_t\right)^2 \nonumber \\
&&\qquad +\ (\mu \sin \beta - A_t \cos \beta )^2 \sin^2\theta_t \\
\Omega(\theta_t,\beta) & = & 2 m_t^2 \cos^2 \beta \sin 2\theta_t
\ -\ 2 m_t \cos\beta (\mu \sin \beta - A_t \cos \beta ) \ .
\end{eqnarray}
Note that the running mass matrix $\hat{\cal M}^2_t(Q)$ in
Eq.~(\ref{top sq app}) contains the soft masses $M_Q$ and $M_U$
(as well as $\mu$, $A_t$, etc.) at some common scale, $Q$.  In the
limit $\lambda_b \rightarrow 0$, these expressions are equivalent
to the results of Ref.~\cite{Donini}, with certain external
momenta set to zero.

\begin{figure}[t]
\epsfysize=2.5in \epsffile[-140 220 0 535]{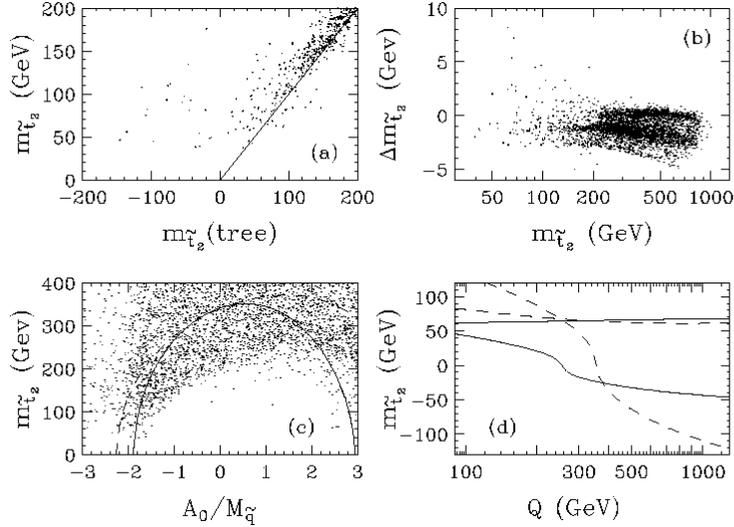}
\begin{center}
\parbox{5.5in}{
\caption[]{\small (a) The full one-loop light top
squark mass, versus the tree-level mass (in GeV). On the x-axis we
plot sign$(m_{\tilde t_2}^2)|m_{\tilde t_2}^2|^{1/2}$, so a negative
tree-level mass corresponds to $m_{\tilde t_2}^2 < 0$.  (b) The
difference between the full correction and the approximation in the
text, versus the one-loop mass.  (c) The light top squark mass
at one loop, versus $A_0/M_{\tilde q}$.  The solid line corresponds to
point (I) in the text, the dashed to point (II).  (d) The running and
one-loop light top squark mass versus the renormalization scale $Q$,
for the choice of parameters (I) (solid) and (II) (dashed) in the
text, with $A_0/M_{\tilde q}=-1.83$ and $-2.2$, respectively. The
running mass curves each have points where $m_{\tilde t_2}^2$ becomes
negative.  In these cases we plot the signed square-root of the
mass-squared, as in (a).
\label{stop}}}
\end{center}
\end{figure}

These approximations depend on the mass of the light top squark.
Normally, one would take it to be the tree-level mass.  For the case
at hand, however, the choice is more subtle because for very light top
squarks, the radiative corrections can be quite large.  In fact, the
radiative correction can change the top squark mass squared from
negative to positive.  Therefore we shall take $m_{\tilde t_2}$ to be
the one-loop pole mass, which we find by iteration.  (We find the
one-loop top squark mixing angle by iteration as well.)  We show the
light top squark one-loop pole mass versus the tree-level mass in
Fig.~\ref{stop}(a), where the tree-level mass is the eigenvalue of the
mass matrix which contains the running parameters $M_U^2$ and $M_Q^2$
evaluated at their own scale (or $M_Z$, whichever is larger; the
tree-level mass matrix also contains the top quark and $Z$-boson pole
masses, $\tan\beta(M_Z)$, $\mu(\mu)$, and $A_t({\rm max}(|A_t|,M_Z))$.
In Fig.~\ref{stop}(b) we see that our approximation for the light top
squark mass holds to within 10 GeV.

With the present unification assumptions, a top squark with mass less
than $M_Z$ requires that the RR term in the mass matrix be small and
that the LR element, proportional to the $A$-term, be large.  The
light top squark mass results from a cancellation between the diagonal
and off-diagonal terms, which requires a fine tuning.  We illustrate
this in Fig.~\ref{stop}(c), where we plot the light top squark
one-loop mass versus $A_0/M_{\tilde q}$. On the same plot we show the
curves corresponding to two choices of parameters, (I) $\tan\beta=20,
\ M_0=500$ GeV, $M_{1/2}=100$ GeV, and $\mu<0$, and (II) $\tan\beta=5,
\ M_0=100$ GeV, $M_{1/2}=200$ GeV, and $\mu>0$.  Whether at tree-level
or one-loop, the parameter $A_0$ must be tuned to one part in 75
to obtain a light top squark mass below 50 GeV.

We see from Fig.~\ref{stop}(a) that the light top squark mass-squared
can be raised from $-$(100 GeV)$^2$ at tree-level to over (100
GeV)$^2$ at one loop.  For such large corrections, it is important to
keep in mind that two-loop effects might be important.  The size of
these effects can be estimated by the scale dependence of the one-loop
mass.  In Fig.~\ref{stop}(d) we show the scale dependence at the
points (I) and (II), with $A_0=-985$ GeV and $A_0=-907$ GeV,
respectively.  We see that as the renormalization scale increases from
100 to 1000 GeV, the running masses vary over a wide range.  In
contrast, the scale dependence of the one-loop masses is quite mild.

\subsection{Slepton masses}

\begin{figure}[t]
\epsfysize=2.5in \epsffile[-140 210 0 525]{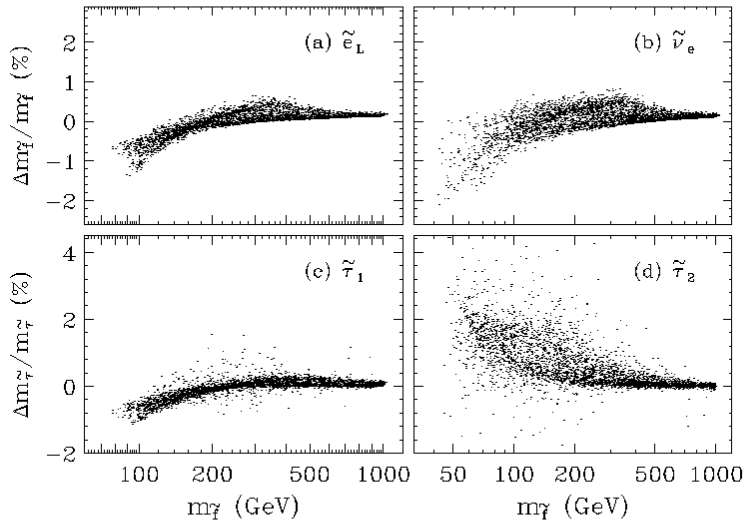}
\begin{center}
\parbox{5.5in}{
\caption[]{\small (a) The full set of corrections to the left-handed
selectron mass versus its mass.  (b) The complete corrections to the
electron sneutrino mass versus its mass.  The complete corrections to
the predominantly (c) left-handed and (d) right-handed tau slepton
masses.
\label{sl}}}
\end{center}
\end{figure}

Corrections to the SU(2) mass sum rules were studied in
Ref.~\cite{Yamada}.  They suggest that the corrections to the slepton
masses are small.  We find that the corrections to the left-handed
electron or muon slepton masses are typically in the range $\pm1\%$,
as illustrated in Fig.~\ref{sl}(a).  The right-handed electron and
muon slepton mass corrections are larger, but still less than 1.7\%.
The sneutrino mass corrections are essentially identical for all three
generations.  The full correction is typically in the range $\pm1\%$,
and reaches at most $-2.5\%$ at $m_{\tilde \nu}\simeq50$ GeV, as shown
in Fig.~\ref{sl}(b). The (predominantly) left- and right-handed tau
slepton corrections are similar to the corrections of the first two
generation charged slepton masses. However, from Figs.~\ref{sl}(c-d)
we see that the scatter plots show less uniformity because of the
additional Yukawa coupling corrections.  We emphasize that in the
corrections to the slepton masses, the leading logarithmic
approximation \cite{Lahanas} typically gives zero correction.  (The
mass $m_{\tilde\tau_2}$ receives some logarithmic corrections.
However, the finite corrections are typically of the same order as or
larger than the logarithmic corrections.)

\subsection{Higgs boson masses}

We first discuss the corrections to the heavy Higgs boson masses
($m_A,\ m_H,\ m_{H^+}$), and then we consider the one-loop light Higgs
boson mass.  The full one-loop corrections to the Higgs boson masses
appear in Ref.~\cite{CPR}.

As usual, we parametrize all the Higgs boson masses at tree-level in
terms of the CP-odd Higgs boson mass, $m_A$, and $\tan\beta$.  To
compute the one-loop mass $m_A$, we first take the soft masses
$m_{H_1}^2(Q)$ and $m_{H_2}^2(Q)$ as outputs from the renormalization
group equations, and apply corrections from the electroweak symmetry
breaking conditions to obtain the \mbox{\footnotesize$\overline{\rm DR}~$}
running parameters $\hat m_A^2(Q)$ and $\mu^2(Q)$ (see Appendix
E).  We then apply further corrections to obtain the CP-odd Higgs
boson pole mass, $m_A$, from the running mass, $\hat m_A(Q)$,
\begin{equation}
m_A^2\ =\ \hat m_A^2(Q)\ -\ {\cal R}e\, \Pi_{AA}(m_A^2) \ +
\ c^2_{\beta} {t_2 \over v_2} \ +\ s^2_{\beta} {t_1 \over v_1}\ ,
\label{dma}
\end{equation}
where $t_1$ and $t_2$ are the tadpole contributions listed in Appendix
E.  Note that we treat the Higgs mass analogously to the superpartner
masses, in that we compare the pole mass with the running mass.
However, the Higgs mass is different because the tadpole, or effective
potential, corrections must be added to the ``tree-level'' running
mass to obtain the \mbox{\footnotesize$\overline{\rm DR}~$} running
mass, $\hat m_A(Q)$.

\begin{figure}[t]
\epsfysize=1.5in \epsffile[-20 425 600 565]{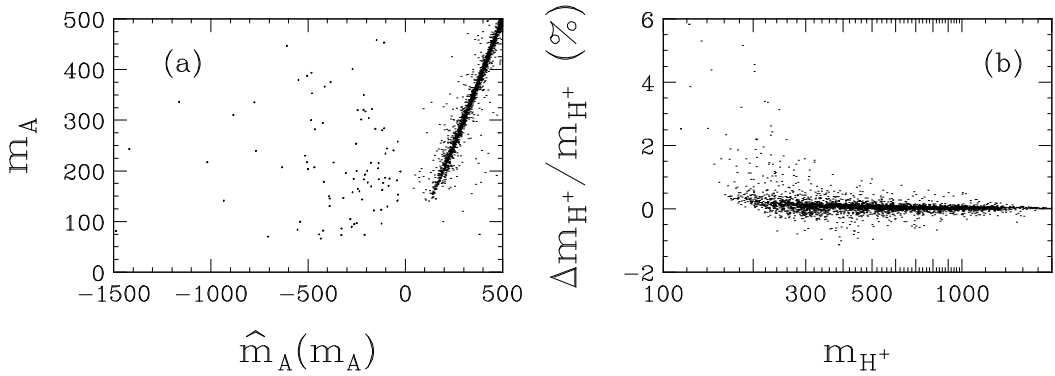}
\begin{center}
\parbox{5.5in}{
\caption[]{\small (a) The one-loop pole mass, $m_A$, versus the
running mass at the scale $m_A$ (the ordinate is the signed
square-root of the mass-squared, sign$(\hat m_A^2)|\hat m_A^2|^
{1/2}$); (b) the corrections to the charged Higgs mass, $m_{H^+}$,
versus $m_{H^+}$.
\label{ma}}}
\end{center}
\end{figure}

The difference between the running mass and the pole mass can be
quite substantial; as for the light top squark, the radiative
corrections can change a negative mass-squared running mass into a
positive mass-squared pole mass.  In Fig.~\ref{ma}(a) we show the
one-loop pole mass, $m_A$, evaluated at $Q = M_{\tilde q}$, versus the
running mass, $\hat m_A(m_A)$.

{}From Fig.~\ref{ma}(a) we see that there are points in parameter
space where the running mass-squared is $-1$ TeV$^2$, while the
one-loop mass is over 300 GeV.  (If one considers the running mass at
the scale $M_Z$, the largest corrections are even more extreme.)
These large corrections arise from terms enhanced by $\tan\beta$.  In
fact, all of the points in Fig.~\ref{ma}(a) with $\hat m_A^2(m_A)<0$
occur for $\tan\beta>25$.  The $\tan\beta$ enhanced contributions come
only from the last term in (\ref{dma}) since $1/v_1$ scales like
$\tan\beta$ for large $\tan\beta$.  Therefore the $\tan\beta$ enhanced
corrections are simply
\begin{eqnarray}
m_A^2\ -\ \hat m_A^2(Q) &=&  {3s_{\beta}^2\mu\tan\beta \over
16\pi^2} \bigg\{\, \lambda_t^2 A_t
\ B_0(0,m_{{\tilde t}_1},m_{{\tilde t}_2}) \ + \ \lambda_b^2A_b
\ B_0(0,m_{{\tilde b}_1},m_{{\tilde b}_2}) \nonumber\\ &&\ +
\ g^2M_2\left(1-{M_2^2\over\mu^2}\right) B_0(0,\mu,M_2)\bigg\}~,
\label{ma app}
\end{eqnarray}
where $B_0(0,m_1,m_2)$ is given in (\ref{b0(0)}).
These terms account for the large corrections seen in
Fig.~\ref{ma}(a).

We parametrize the other heavy Higgs boson masses in terms of the pole
mass, $m_A$.  The corrections to the heavy Higgs boson masses,
$m_{H^+}$ and $m_H$, turn out to be quite small.  For example, the
corrections to $m_{H^+}$ are typically less than 1\%, as shown in
Fig.~\ref{ma}(b).  Corrections to the charged Higgs mass are the
subject of Ref.~\cite{H+}.

\begin{figure}[t]
\epsfysize=2.5in \epsffile[-140 220 0 535]{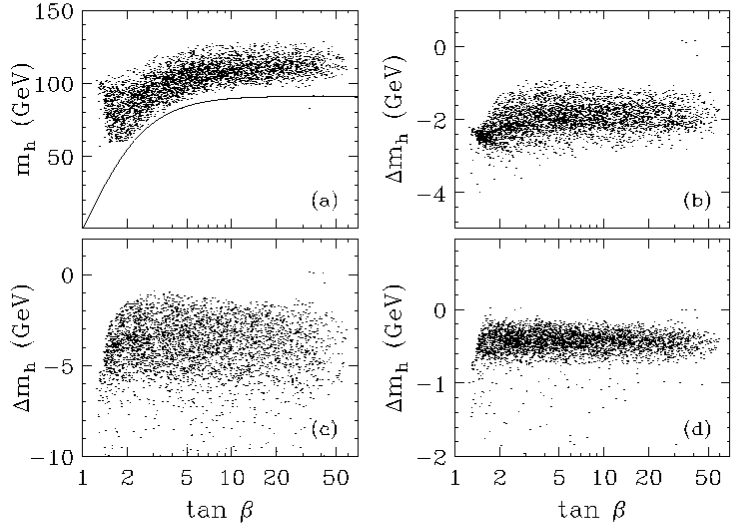}
\begin{center}
\parbox{5.5in}{
\caption[]{\small Figure (a) shows the full one-loop light Higgs mass
versus $\tan\beta$.  The line indicates the tree-level bound
$m_h<M_Z|\cos2\beta|$.  Figure (b) shows the contribution from the
gauge/Higgs/gaugino/Higgsino loops; (c) shows the difference between
the full one-loop result and Dabelstein's approximation; and (d) shows
the difference between the full one-loop mass evaluated at the scales
$M_{\tilde q}$ and $M_{\tilde q}/2$.
\label{mh}}}
\end{center}
\end{figure}

The corrections to the light Higgs boson mass, $m^2_h$, have been
studied extensively in the literature \cite{Higgs mass,CPR,dab}.
Here we show our results for the full one-loop pole mass over the
parameter space associated with radiative electroweak symmetry
breaking and universal unification-scale boundary conditions.  We
also show the size of a set of corrections which are often neglected,
and the accuracy of Dabelstein's approximation \cite{dab}.

In Fig.~\ref{mh}(a) we show the one-loop light Higgs boson mass versus
$\tan\beta$, as well as the upper limit at tree level,
$m_h=M_Z|\cos2\beta|$.  The upper limit on the one-loop Higgs mass
depends sensitively on the top quark mass, and somewhat less
sensitively on the squark masses.  In the parameter space we consider
the squark masses are less than about 1 TeV.  With $m_t=175$ GeV, we
find $m_h < 130$ GeV.

In Fig.~\ref{mh}(b) we show the gauge/Higgs/gaugino/Higgsino
contribution to the Higgs mass, which is typically $-2$ GeV.  In
Fig.~\ref{mh}(c) we show the difference between the one-loop result
and Dabelstein's approximation, Eq.~(4.9) of Ref.~\cite{dab}, which only
includes the top sector.  We see that this approximation is typically
2 to 6 GeV larger than the full one-loop mass.

In any pole mass, the scale dependence formally cancels.  However, at
any given order, there are usually higher-order corrections which do
not cancel.  For example, when we vary the scale in the Higgs mass
calculation, we change the tree-level
\mbox{\footnotesize$\overline{\rm DR}~$} mass.  To one-loop order,
this variation is canceled by the change in the self-energy.  However,
as the scale varies, the couplings and masses in the self-energies
also change.  For the case of the Higgs mass, the change in the top
quark Yukawa coupling gives rise to a two-loop ${\cal O}
(\lambda_t^4)$ scale dependence in our one-loop results.  In
Fig.~\ref{mh}(d) we show the difference between the full one-loop
result evaluated at the scale $M_{\tilde q}$ and $M_{\tilde q}/2$.  We
see that the scale dependence is usually small, in the range 0 to $-1$
GeV. The fact that it is negative follows from the dependence of
$\tan\beta$ on $Q$, and from the dependence of $\lambda_t$ on
$\tan\beta$.

\section{Conclusions}

In this paper we computed one-loop radiative corrections in the
minimal supersymmetric standard model.  We took as inputs the
electromagnetic coupling at zero momentum, $\alpha_{\rm em}$, the
Fermi constant, $G_\mu$, the $Z$-boson pole mass, $M_Z$, the strong
coupling in the \mbox{\footnotesize$\overline{\rm MS}~$} scheme at the
scale $M_Z$, $\alpha_s(M_Z)$, and the quark and lepton masses.  From
these we computed the $W$-boson mass, $M_W$, as well as the one-loop
value of the effective weak mixing angle, $\sin^2\theta^{\rm
lept}_{\rm eff}$, as a function of the supersymmetric parameters.

We studied the size of the corrections in a reduced parameter space
associated with the unification of the soft breaking parameters and
radiative electroweak symmetry breaking.  We found that supersymmetric
radiative corrections can reduce $\sin^2\theta^{\rm lept}_{\rm eff}$
by as much as $-1.6 \times 10^{-3}$ with respect to the standard-model
value.  Similarly, we found that they can increase $M_W$ by as much as
250 MeV.  Because of decoupling, the points with the largest
deviations are also the points with the lightest superpartners.  As
direct searches increase the limits on the superparticle masses, the
size of the supersymmetric radiative corrections will decrease.
Indeed, if superparticles are not discovered at LEP 2, we found that
the maximum size of the supersymmetric radiative corrections will be
reduced by a factor of two.

The apparent unification of the  SU(3), SU(2), and U(1) coupling
constants is a major piece of evidence in favor of supersymmetry.  At
next-to-leading order, the weak- and unification-scale threshold
corrections come into play.  The weak-scale thresholds {\it decrease}
the one-loop weak mixing angle.  This leads to an {\it increase} in
the predicted value of the strong coupling, $\alpha_s(M_Z)$.  As we
have seen, for squark masses less than one TeV, a unification-scale
threshold of $-1$ to $-3\%$ is necessary to bring $\alpha_s(M_Z)$ into
accord with experiment.

The size of the unification-scale thresholds places an important
constraint on unified model building.  In any unified model, the
unification-scale thresholds can be calculated as a function of the
grand unification parameters.  One can see whether the model is
consistent with a unification-scale threshold of about $-2\%$.  In
this paper we studied the minimal SU(5) model and the missing doublet
SU(5) model.  We found that the former was not compatible with gauge
coupling unification, while the latter was.

Grand unified theories also predict the unification of certain Yukawa
couplings, and in a similar fashion, the mismatch of the Yukawa
couplings at the unification scale can be used to constrain
unification-scale physics.  To this end it is necessary to extract as
precisely as possible the \mbox{\footnotesize$\overline{\rm DR}~$}
Yukawa couplings from the fermion pole masses.  In this paper we
presented full one-loop relations between the two, as well as
approximations that work at the $\cal O$(1\%) level.  We studied the
substantial (up to 50\%) $\tan\beta$-enhanced corrections to the
bottom quark mass, as well as the corrections to the top and tau
masses, which are of order 5 percent.

Supersymmetry also predicts relations between the masses and couplings
of the supersymmetric particles.  Indeed, if new particles are
discovered at future colliders, it will be necessary to check these
relations to see whether the new particles are in fact supersymmetric
partners \cite{mssm studies}.  The radiative corrections to the
supersymmetric mass spectrum presented in this paper will be an
essential element in these determinations.

The corrections to the supersymmetric mass spectrum will be used in
(at least) two ways.  First, they will be used to correct the
tree-level mass sum rules \cite{Yamada,sumrules} which test
supersymmetry at the weak scale.  Second, they will be needed to
extract the underlying soft parameters from the physical observables.
The soft parameters can then be run to higher scales, to test for
unification and possibly to shed light on the origin of the
supersymmetry breaking.

The corrections to the supersymmetric masses in the spin-1/2 sector
include potentially 30\% corrections to the gluino mass, as well
as $\cal O$(5\%) corrections to the neutralino and chargino masses.
In the spin-0 sector, the famous quadratic divergences give rise to
large corrections to the scalar masses.  These corrections can lift the
running mass-squared of the light top squark from $-(100$ GeV$)^2$ to
(100 GeV)$^2$.  Even more dramatically, large $\tan\beta$ corrections
can lift the mass-squared of the CP-odd Higgs boson from, e.g.,
$-(1$ TeV)$^2$ to (300 GeV)$^2$.

Radiative corrections also have an important effect on the mass of the
lightest Higgs boson, $h$. In the parameter space we consider, they
effectively change the sign of the tree-level bound from $m_h < M_Z
|\cos2\beta|$ to $m_h > M_Z |\cos2\beta|$.  We found the light Higgs
mass was raised to at most 130 GeV.  The corrections to the rest of
the scalar masses are smaller.  For example, we found 1 to 5\%
corrections to the first two generation squark masses, and $\cal
O$(1\%) corrections to the slepton masses.

In the paper we presented approximations to many of the formulae for
the supersymmetric mass corrections.  These approximations, often good
to better than a couple of percent, provide useful substitutes for
the full corrections.

\vspace{.7cm}
{\noindent\Large\bf Acknowledgements}
\vspace{.7cm}

D.M.P. thanks M. Peskin, T. Rizzo and J. Wells for useful discussions.
\vspace{.7cm}

{\noindent\Large\bf Appendix A: Tree-level masses}
\vspace{.7cm}
\setcounter{equation}{0}
\renewcommand{\theequation}{A.\arabic{equation}}

In this appendix we define the tree-level masses. These tree-level
relations also hold for the running \mbox{\footnotesize$\overline{\rm
DR}~$} parameters at a common scale, $Q$.  For the most part we
follow the conventions of Ref.~\cite{H&K}.

The up- and down-type quark and charged-lepton masses are related to
the Yukawa couplings and the vev's $v_1$ and $v_2$ by
\begin{equation}
m_u\ =\ {1\over\sqrt2}\lambda_uv_2~, \qquad\qquad\qquad m_d\ =
\ {1\over\sqrt2}\lambda_dv_1~.\label{mf}
\end{equation}
The ratio of vev's $v_2/v_1$ is denoted $\tan\beta$.  The tree-level
gauge boson masses are
\begin{equation}
M_W^2 \ =\ {1\over4}g^2\left(v_1^2 + v_2^2\right)\ ,
\qquad\qquad\qquad\label{mwmz} M_Z^2 \ =
\ {1\over4}\left(g'^2+g^2\right)\left(v_1^2 + v_2^2\right)\ ,
\end{equation}
where $g$ and $g'$ are the $SU(2)$ and $U(1)$ gauge couplings.

The Lagrangian contains the neutralino mass matrix as
$-{\tilde\psi^0}{}^T{\cal M}_{\tilde\psi^0}\tilde\psi^0$ + h.c., where
$\tilde\psi^0 =$ $(-i\tilde b,$ $-i\tilde w_3,$ $\tilde h_1,$ $\tilde
h_2)^T$ and
\begin{equation}
{\cal M}_{\tilde\psi^0} \ =\ \left(\begin{array}{cccc} M_1 & 0 &
-M_Zc_\beta s_W & M_Zs_\beta s_W \\ 0 & M_2 & M_Zc_\beta c_W &
-M_Zs_\beta c_W \\ -M_Zc_\beta s_W & M_Zc_\beta c_W & 0 & \mu \\
M_Zs_\beta s_W & -M_Zs_\beta c_W & \mu & 0
\end{array} \right)\ .\label{mchi0}
\end{equation}
We use $s$ and $c$ for sine and cosine, so that
$s_\beta\equiv\sin\beta,\ c_{2\beta}\equiv\cos2\beta$, etc.  $M_1$ and
$M_2$ are the soft supersymmetry-breaking bino and wino gaugino
masses, $\mu$ is the supersymmetric Higgsino mass, and $s_W\ (c_W)$ is
the sine (cosine) of the weak mixing angle.  The neutralino masses are
found by acting on the matrix ${\cal M}_{\tilde\psi^0}$ with a unitary
matrix $N$, so that $N^*{\cal M}_{\tilde\psi^0}N^\dagger$ is a
diagonal matrix which contains the physical neutralino masses,
$m_{\tilde\chi_i^0}$.  In the usual case that one of the eigenvalues
of (\ref{mchi0}) is negative, the matrix $N$ is complex even if the
elements of ${\cal M}_{\tilde\psi^0}$ are real.

The Lagrangian contains the chargino mass matrix as
$-{\tilde\psi^-}{}^T{\cal M}_{\tilde\psi^+}\tilde\psi^+$ + h.c.,
where~~$\tilde\psi^+ = (-i\tilde w^+,\ \tilde h_2^+)^T,\ \tilde\psi^-=
(-i\tilde w^-,\ \tilde h_1^-)^T$ and
\begin{equation}
{\cal M}_{\tilde\psi^+}\ =\ \left( \begin{array}{cc} M_2 &
\sqrt2\,M_Ws_\beta\\\sqrt2\,M_Wc_\beta & -\mu\end{array}\right)\ .
\label{mchi+}
\end{equation}
The chargino masses are found by acting on the matrix ${\cal
M}_{\tilde\psi^+}$ with a biunitary transformation, so that $U^*{\cal
M}_{\tilde\psi^+}V^\dagger$ is a diagonal matrix containing the two
chargino mass eigenvalues, $m_{\tilde\chi_i^+}$. The matrices $U$ and
$V$ are easily found, as they diagonalize, respectively, the matrices
${\cal M}_{\tilde\psi^+}^*{\cal M}_{\tilde\psi^+}^{\rm T}$ and ${\cal
M}_{\tilde\psi^+}^\dagger{\cal M}_{\tilde\psi^+}$.

At tree level the gluino mass, $m_{\tilde g},$ is given by the soft
mass, $M_3$.

The tree-level squark masses are found by diagonalizing the following mass
matrices,
\begin{equation}
\label{sqm u} \left(\begin{array}{cc}
M_Q^2 + m_u^2 + g_{u_L}M_Z^2c_{2\beta} &
m_u\left(A_u+\mu\cot\beta\right)\\[2mm]
m_u\left(A_u+\mu\cot\beta\right) & M_U^2 + m_u^2 +
g_{u_R}M_Z^2c_{2\beta}
\end{array}\right)\ ,
\end{equation}
\begin{equation}
\label{sqm d}  \left(\begin{array}{cc}
M_Q^2 + m_d^2 + g_{d_L}M_Z^2c_{2\beta} &
m_d\left(A_d+\mu\tan\beta\right)\\[2mm]
m_d\left(A_d+\mu\tan\beta\right) & M_D^2 + m_d^2 +
g_{d_R}M_Z^2c_{2\beta}
\end{array}\right)~.
\end{equation}
Here $M_Q,~M_U,$ and $M_D$ are the soft supersymmetry-breaking squark
masses, and the $A_f$'s are the soft supersymmetry-breaking
$A$-terms. The slepton mass matrices are analogous.  The soft slepton
masses are denoted $M_L$ and $M_E$.  We have defined the weak
neutral-current couplings
\begin{equation}
g_f \ =\ I^f_3 \ -\ e_fs_W^2~.\label{glgr}
\end{equation}
The electric charge, hypercharge, and third component of isospin of
the sfermions are\footnote{Our convention for the right-handed
sfermion fields is the charge-conjugate that of Ref.~\cite{H&K}.}
\begin{equation}
\begin{array}{cccccccc}
\qquad&\qquad\tilde u_L&\qquad\tilde u_R&\qquad\tilde d_L&\qquad\tilde
d_R&\qquad \tilde\nu&\qquad\tilde e_L&\qquad\tilde e_R \\ [1mm] \hline
\\ e_f &\qquad ~{2\over3} &\qquad -{2\over3} &\qquad -{1\over3}
&\qquad ~{1\over3} &\qquad ~0 &\qquad -1 &\qquad ~1 \\[2mm] Y_f
&\qquad ~{1\over3} &\qquad -{4\over3} &\qquad ~{1\over3} &\qquad
{}~{2\over3} &\qquad -1 &\qquad -1 &\qquad ~2 \\[2mm] I_3^f&\qquad
{}~{1\over2} &\qquad ~0 &\qquad -{1\over2}&\qquad ~0 &\qquad ~{1\over2}
&\qquad -{1\over2}&\qquad ~0 \\[2mm] \hline\\
\end{array}\label{table1}
\end{equation}
A symbol without an $L$ or $R$ subscript refers to the $L$-field
(e.g. $e_u=2/3$).

The matrix which diagonalizes a sfermion mass matrix is denoted by
\begin{equation}
\left(\begin{array}{cc} c_f & s_f\\-s_f & c_f \end{array}\right)~,
\end{equation}
where $c_f$ is the cosine of the sfermion mixing angle,
$\cos\theta_f$, and $s_f$ the sine. These angles are given by
\begin{eqnarray}
\tan(2\theta_u) &=& {2m_u\left(A_u + \mu\cot\beta\right)\over
M_Q^2-M_U^2 + ({1\over2}-2e_us_W^2)M_Z^2 c_{2\beta}}\ ,\\
\tan(2\theta_d) &=& {2m_d\left(A_d + \mu\tan\beta\right)\over
M_Q^2-M_D^2 + (-{1\over2}-2e_ds_W^2)M_Z^2c_{2\beta}}\ .
\end{eqnarray}
Since there is no right-handed sneutrino, the slepton mixing angle for
$\tilde\nu$ satisfies $c_\nu= 1$, and the sneutrino mass is
$m_{\tilde\nu}^2 = M_L^2 + M_Z^2c_{2\beta}/2$.

Given values for $\tan\beta$ and the CP-odd Higgs-boson mass, $m_A$,
the other Higgs masses are given, at tree level, by
\begin{equation}
m_{H,h}^2 \ =\ {1\over2}\Biggl(m_A^2+M_Z^2\pm \sqrt{\left(m_A^2 +
M_Z^2\right)^2 - 4m_A^2M_Z^2c_{2\beta}^2}\Biggr)\ ,
\end{equation}
and %
\begin{equation}
m_{H^+}^2 \ =\ m_A^2\ +\ M_W^2\ .
\end{equation}
The CP-even gauge eigenstates $(s_1,\, s_2)$ are rotated by the angle
$\alpha$ into the mass eigenstates $(H,\, h)$ as follows,
\begin{equation}
\left(\begin{array}{c}H\\[2mm]h\end{array}\right) \ =
\ \left(\begin{array}{cc} c_\alpha & s_\alpha\\[2mm]-s_\alpha & c_\alpha
\end{array}\right)\left(\begin{array}{c}s_1\\[2mm]s_2\end{array}\right)~.
\label{rotate h}
\end{equation}
At tree level, the angle $\alpha$ is given by
\begin{equation}
\tan2\alpha \ =\ {m_A^2 + M_Z^2\over m_A^2-M_Z^2}\tan2\beta~.
\end{equation}

\vspace{.7cm} {\noindent\Large\bf Appendix B: One-loop scalar
functions}
\vspace{.7cm} \setcounter{equation}{0}
\renewcommand{\theequation}{B.\arabic{equation}}

The following integrals appear at one loop in a self-energy
calculation \cite{PV}:\footnote{ Our $A$ and $B$ functions differ from
those of Ref.~\cite{PV} since we use the Minkowski metric.  Also,
$A_0$, $B_1$ and $B_{22}$ differ by a sign. Equations (\ref{B0 def}
--\ref{B22 def}) contain an abuse of notation.  The first argument
of $B$-functions is the square root of the scalar $p^2$, whereas
elsewhere the $p$ represents the external momentum four-vector.}
\begin{eqnarray}
A_0(m) &=& 16\pi^2Q^{4-n}\int{d^nq\over i\,(2\pi)^n}{1\over
q^2-m^2+i\varepsilon}\\
B_0(p, m_1, m_2) &=&
16\pi^2Q^{4-n}\int{d^nq\over i\,(2\pi)^n}
{1\over\biggl[q^2-m^2_1+i\varepsilon\biggr]\biggl[
(q-p)^2-m_2^2+i\varepsilon\biggr]}
\label{B0 def}  \\
p_\mu B_1(p, m_1,m_2) &=& 16\pi^2Q^{4-n}\int
{d^nq\over i\,(2\pi)^n}{q_\mu\over\biggl[q^2-m^2_1+i\varepsilon\biggr]
\biggl[(q-p)^2-m_2^2+i\varepsilon\biggr]}\label{B1 def} \\ p_\mu p_\nu
B_{21}(p,m_1,m_2) &+& g_{\mu\nu}B_{22}(p,m_1,m_2)
\label{B22 def}\\
 &=& 16\pi^2\,Q^{4-n}\,\int{d^nq\over i\,(2\pi)^n} {q_\mu
q_\nu\over\biggl[q^2-m^2_1+i\varepsilon\biggr]\biggl[
(q-p)^2-m_2^2+i\varepsilon\biggr]}\ , \nonumber
\end{eqnarray}
where $Q$ is the renormalization scale and we regularize by
integrating in $n=4-2\epsilon$ dimensions.

The expression for $A_0$ can be integrated to give
\begin{equation}
A_0(m)\ =\ m^2\left({1\over\hat\epsilon} + 1 - \ln{m^2\over
Q^2}\right)~,\label{A}
\end{equation}
where $1/\hat\epsilon =1/\epsilon-\gamma_E+\ln 4\pi$.

The function $B_0$ can be written in the form
\begin{equation}
B_0(p, m_1, m_2) \ =\ {1\over\hat\epsilon}~-~\int_0^1 dx\ \ln{
 (1-x)~m_1^2 + x~m_2^2 -x(1-x)~p^2 - i\varepsilon\over Q^2}\ .
\label{B0}
\end{equation}
It has the analytic expression
\begin{equation}
 B_0(p, m_1, m_2) \ =\ {1\over\hat\epsilon} - \ln\left(p^2\over
Q^2\right) - f_B(x_+) - f_B(x_-)~,
\end{equation}
where
\begin{equation}
 x_{\pm}\ =\ {s \pm \sqrt{s^2 - 4p^2(m_1^2-i\varepsilon)}\over2p^2}~,
\qquad f_B(x) \ =\ \ln(1-x) - x\ln(1-x^{-1})-1~,
\end{equation}
and $s=p^2-m_2^2+m_1^2$.

All the other functions can be written in terms of $A_0$ and $B_0$.  For
example,
\begin{equation}
 B_1(p, m_1,m_2) \ =\ {1\over2p^2}\biggl[ A_0(m_2) - A_0(m_1) + (p^2
+m_1^2 -m_2^2) B_0(p, m_1, m_2)\biggr]~,
\end{equation}
and
\begin{eqnarray}
B_{22}(p, m_1,m_2) &=& {1\over 6}\ \Bigg\{\,
{1\over2}\biggl(A_0(m_1)+A_0(m_2)\biggr)
+\left(m_1^2+m_2^2-{1\over2}p^2\right)B_0(p,m_1,m_2)\nonumber \\ &+&
\ {m_2^2-m_1^2\over2p^2}\ \biggl[\,A_0(m_2)-A_0(m_1)-(m_2^2-m_1^2)
B_0(p,m_1,m_2)\,\biggr] \nonumber\\ &+&\ m_1^2 + m_2^2
-{1\over3}p^2\,\Bigg\}~.
\end{eqnarray}

We also define
\begin{eqnarray}
F(p,m_1,m_2) &=& A_0(m_1)-2A_0(m_2)- (2p^2+2m^2_1-m^2_2)B_0(p,m_1,m_2)
\ ,\\[2mm] G(p,m_1,m_2) &=&
(p^2-m_1^2-m_2^2)B_0(p,m_1,m_2)-A_0(m_1)-A_0(m_2)\ ,\\[2mm] H(p,m_1,m_2)
&=& 4B_{22}(p,m_1,m_2) + G(p,m_1,m_2)\ ,\\[1mm] \tilde
B_{22}(p,m_1,m_2) &=& B_{22}(p,m_1,m_2) - {1\over4}A_0(m_1) -
{1\over4}A_0(m_2)~.\label{B22}
\end{eqnarray}
The functions $F$ and $G$ arise in scalar self-energies, with either a
vector boson and a scalar or fermions in the loop, while $H$ and
$\tilde B_{22}$ occur in vector-boson self-energies, with either
fermions or scalars in the loop.

\vspace{.7cm} {\noindent\Large\bf Appendix C:\ The gauge couplings}
\vspace{.7cm} \setcounter{equation}{0}
\renewcommand{\theequation}{C.\arabic{equation}}

In the remaining appendices we denote $\hat
s^2\equiv\sin^2\hat\theta_W$, where $\hat\theta_W$ is the
\mbox{\footnotesize$\overline{\rm DR}~$} weak mixing angle, and
$s^2\equiv\sin^2 \theta_W=1-M_W^2/M_Z^2$, where $\theta_W$ is the
``on-shell" weak mixing angle and $M_W,$ $M_Z$ are the gauge-boson
pole masses.

The \mbox{\footnotesize$\overline{\rm DR}~$} electromagnetic coupling
is given by
\begin{equation}
\hat\alpha\ =\ {\alpha_{em}\over1-\Delta\hat\alpha}
\ ,\qquad\qquad\alpha_{em} \ =\ {1\over137.036}\ ,
\end{equation}
where\footnote{The coefficients of $\ln(M_W/M_Z)$ in the expressions
for $\Delta\hat\alpha$ in Refs. \cite{CPP,BMP} are both incorrect.}
\begin{eqnarray}&&
\Delta\hat\alpha\ \ =\ \ 0.0682 \pm 0.0007 \ -
\ {\alpha_{em}\over2\pi}\Biggl\{ -7\ln\left(M_W\over M_Z\right) \ +
\ {16\over9}\ln\left(m_t\over M_Z\right) \ +
\ {1\over3}\ln\left(m_{H^+}\over M_Z\right) \nonumber \\ &&\ +
\ {4\over9}\sum_u\sum_{i=1}^2\ln\left(m_{\tilde u_i}\over M_Z\right)
\ +\ {1\over9}\sum_d\sum_{i=1}^2\ln\left(m_{\tilde d_i}\over M_Z\right)
\ +\ {1\over3}\sum_e\sum_{i=1}^2\ln\left(m_{\tilde e_i}\over
M_Z\right) \ +\ {4\over3}\sum_{i=1}^2\ln\left(m_{\tilde\chi_i^+}\over
M_Z\right)\Biggr\}\ , \nonumber\label{da}\\
\end{eqnarray}
and $\sum_u$ indicates a sum over $u,c,t$, and similarly for $\sum_d,
\ \sum_e$.  In this expression, the number 0.0682 includes the two-loop
QED and QCD corrections given in Ref.~\cite{FKS}, as well as the
five-flavor contribution $\Delta\alpha^{(5)}_{\rm
had}(M_Z^2)=0.0280\pm0.0007$ of Ref.~\cite{EJ}.

The \mbox{\footnotesize$\overline{\rm DR}~$} weak mixing angle is
given by \cite{DFS}
\begin{equation}
\hat c^2\hat s^2 \ =
\ {\pi\hat\alpha\over\sqrt2\,M_Z^2\,G_\mu\,(1-\Delta\hat r)}\ ,\qquad
\Delta\hat r \ =\ \hat\rho\ {\Pi_{WW}^T(0) \over M_W^2} \ -\ {\cal
R}\!e\ {\Pi_{ZZ}^T(M_Z^2)\over M_Z^2}\ +\ \delta_{\rm VB}\ ,
\end{equation}
where $\hat\rho$ is defined to be $c^2/\hat c^2$, and $\delta_{\rm VB}$
denotes the nonuniversal vertex and box diagram corrections given
below.  The $W$ and $Z$ gauge-boson self-energies are given in
Eqs.~(\ref{piz}) and (\ref{piw}).  We compute $\hat\rho$ via
\cite{DFS}
\begin{equation}
\hat\rho\ =\ {1\over1-\Delta\hat\rho}\ ,\qquad\qquad\Delta\hat\rho
\ =\ {\cal R}\!e\,\Biggl[ {\Pi_{ZZ}^T(M_Z^2)\over\hat\rho\,M_Z^2} \ -
\ {\Pi_{WW}^T(M_W^2)\over M_W^2}\Biggr]\ .
\end{equation}
We deduce the leading two-loop standard model corrections to
$\Delta\hat r$ and $\Delta\hat\rho$ from Ref.~\cite{FKS},
\begin{eqnarray}
\Delta\hat r\biggr|_{\rm 2-loop} &=& {\hat\alpha\over4\pi\hat s^2\hat
c^2}{\alpha_s\over\pi} \ \Biggl[\, 2.145{m_t^2\over M_Z^2} +
0.575\ln\left(m_t\over M_Z\right) -0.224 \nonumber\\&&\qquad\qquad
-\ 0.144 {M_Z^2\over m_t^2}\,\Biggr]\ -
\ {1\over3}x_t^2\,\rho^{(2)}\!\!\left(m_\varphi\over
m_t\right)\,(1-\Delta\hat r)\hat\rho\ ,\\ \Delta\hat\rho\biggr|_{\rm
2-loop} &=& {\hat\alpha\over4\pi\hat s^2}{\alpha_s\over\pi} \ \Biggl[
\,-2.145{m_t^2\over M_W^2} + 1.262\ln\left(m_t\over M_Z\right) -2.24
\nonumber\\&&\qquad\qquad-\ 0.85 {M_Z^2\over m_t^2}\,\Biggr]\ +
\ {1\over3}x_t^2\,\rho^{(2)}\!\!\left(m_\varphi\over m_t\right)\ ,
\end{eqnarray}
where $x_t = 3G_\mu m_t^2/8\pi^2\sqrt2$ and $m_\varphi$ is the
standard model Higgs-boson mass.  For $r\le1.9$, $\rho^{(2)}(r)$ is
well approximated by \cite{rho2}
\begin{eqnarray}
\rho^{(2)}(r) &=& 19 - {33\over2}r + {43\over12}r^2 + {7\over120}r^3
-\pi\sqrt r\left(4-{3\over2}r+{3\over32}r^2 +
{1\over256}r^3\right)\nonumber\\ &-&\pi^2(2-2r+{1\over2}r^2)-\ln
r\left(3r-{1\over2}r^2\right)\ ,
\end{eqnarray}
while, for $r\ge1.9$, we use
\begin{eqnarray}
\rho^{(2)}(r) &=& \ln^2r\left({3\over2}-9r^{-1}-15r^{-2}
-48r^{-3}-168r^{-4}-612r^{-5}\right)\nonumber\\ &-&\ln
r\left({27\over2} + 4r^{-1} - {125\over4}r^{-2} - {558\over5}r^{-3}
-{8307\over20}r^{-4} - {109321\over70}r^{-5}\right)\\
&+&\pi^2\left(1-4r^{-1} -5r^{-2}-16r^{-3}-56r^{-4}-204r^{-5}
\right)\nonumber\\ &+& {49\over4} + {2\over3}r^{-1} +
{1613\over48}r^{-2} + {8757\over100}r^{-3}
+{341959\over1200}r^{-4}+{9737663\over9800}r^{-5}\nonumber\ .
\end{eqnarray}
For the case of the MSSM, we replace the function
$\rho^{(2)}(m_\varphi/m_t)$ with
\begin{equation}
\left(\cos\alpha\over\sin\beta\right)^2\rho^{(2)}\!\!\left(m_h\over
m_t\right)\ .
\end{equation}
We have not computed the corresponding $G_\mu^2m_t^4$ higher-order
contributions from the heavy Higgs bosons, but we know that they must
decouple.  Using the ansatz
\begin{equation}
\Delta\hat\rho\Big|_{\rm Heavy\ Higgs} = {1\over3}x_t^2\biggl\{\,
\left(\sin\alpha\over\sin\beta\right)^2\rho^{(2)}\!\!\left(m_H\over
m_t\right)\ -
\left(1\over\tan\beta\right)^2\rho^{(2)}\!\!\left(m_A\over
m_t\right)\,\biggr\}\ ,
\end{equation}
we find these contributions are negligible. We do not include them in
our results.

The nonuniversal contribution to $\Delta\hat r$ is made up of two
parts, one from the standard model and the other from supersymmetry,
\begin{equation}
\delta_{\rm VB}\ =\ \delta_{\rm VB}^{\rm SM}\ +
\ \delta_{\rm VB}^{\rm SUSY}\ .
\end{equation}
The standard-model part is given by the well known formula \cite{DFS}
\begin{equation}
\delta_{\rm VB}^{\rm SM}\ =\ \hat\rho\,{\hat\alpha\over4\pi\hat
s^2}\Biggl\{ 6 + {\ln c^2\over s^2}\biggl[ {7\over2}-{5\over2}s^2 -
\hat s^2\left(5-{3\over2}{c^2\over\hat c^2}\right) \biggr]\Biggr\}\ .
\end{equation}
The supersymmetric part appears in Ref.~\cite{Grifols}, and more
recently in Ref.~\cite{deltar}.  We include it here for
completeness.  It includes box diagram contributions, vertex
corrections, and external wave-function renormalizations.  We neglect
the mixing between different generations of sleptons, and we ignore
the left-right slepton mixing in the first two generations, in which
case the right-handed sleptons $\tilde e_R, \ \tilde\mu_R$ do not
contribute.  We find
\begin{equation}
\delta_{\rm VB}^{\rm SUSY}\ =\ -\ {\hat s^2\hat
c^2\over2\pi\hat\alpha}M_Z^2\,{\cal R}e\,a_1 + \delta v_e + \delta
v_\mu + {1\over2}\biggl( \delta Z_e + \delta Z_{\nu_e} + \delta Z_\mu+
\delta Z_{\nu_\mu}\biggr)\ .
\end{equation}

The wave-function and vertex corrections are
\begin{equation}
16\pi^2~\delta Z_{\nu_e} \ =\ -\ \sum_{i=1}^2
\left|b_{\tilde\chi_i^+\nu_e\tilde e_L}\right|^2
B_1(0,m_{\tilde\chi^+_i},m_{\tilde e_L}) - \sum_{j=1}^4
\left|b_{\tilde\chi_j^0\nu_e\tilde\nu_e}\right|^2
B_1(0,m_{\tilde\chi^0_j},m_{\tilde\nu_e})~,
\end{equation}
\begin{equation}
16\pi^2~\delta Z_e\ =\ -\ \sum_{i=1}^2
\left|a_{\tilde\chi_i^+e\tilde\nu_e}\right|^2
B_1(0,m_{\tilde\chi^+_i},m_{\tilde\nu_e}) \ -\ \sum_{j=1}^4
\left|b_{\tilde\chi_j^0e\tilde e_L}\right|^2
B_1(0,m_{\tilde\chi^0_j},m_{\tilde e_L})~,
\end{equation}
\begin{eqnarray}
16\pi^2~\delta v_e &=& \sum_{i=1}^2\sum_{j=1}^4
b_{\tilde\chi_i^+\nu_e\tilde e_L}b^*_{\tilde\chi_j^0e\tilde e_L}
\ \Biggl\{ - \ {\sqrt{2}\over g}a_{\tilde\chi_j^0\tilde\chi_i^+W}
m_{\tilde\chi_i^+} m_{\tilde\chi_j^0} \,C_0(m_{\tilde
e_L},m_{\tilde\chi_i^+},m_{\tilde\chi_j^0}) \nonumber\\
&&\qquad\qquad+ \ {1\over\sqrt2g}b_{\tilde\chi_j^0\tilde\chi_i^+W}
\ \biggl[\,B_0(0,m_{\tilde\chi_i^+},m_{\tilde\chi_j^0}) + m_{\tilde
e_L}^2\,C_0(m_{\tilde e_L},m_{\tilde\chi_i^+},m_{\tilde\chi_j^0}) -
{1\over2}\,\biggr] \Biggr\}\nonumber\\ &-& \sum_{i=1}^2\sum_{j=1}^4
a_{\tilde\chi_i^+e\tilde\nu_e}b_{\tilde\chi_j^0\nu_e\tilde\nu_e}
\ \Biggl\{ - \ {\sqrt2\over g}b_{\tilde\chi_j^0\tilde\chi_i^+W}
m_{\tilde\chi_i^+} m_{\tilde\chi_j^0} \,C_0(m_{\tilde\nu_e},
m_{\tilde\chi_i^+}, m_{\tilde\chi_j^0}) \nonumber\\ &&\qquad\qquad+
\ {1\over\sqrt2g} a_{\tilde\chi_j^0\tilde\chi_i^+W}
\ \biggl[\,B_0(0,m_{\tilde\chi_i^+},m_{\tilde\chi_j^0}) +
m_{\tilde\nu_e}^2 \, C_0(m_{\tilde\nu_e}, m_{\tilde\chi_i^+},
m_{\tilde\chi_j^0}) - {1\over2} \,\biggr]\Biggr\} \nonumber\\ &+&
{1\over2}\sum_{j=1}^4 b_{\tilde\chi_j^0e\tilde e_L}^*
b_{\tilde\chi_j^0\nu_e\tilde\nu_e} \ \biggl[\,B_0(0,m_{\tilde
e_L},m_{\tilde\nu_e}) +
m_{\tilde\chi_j^0}^2\,C_0(m_{\tilde\chi_j^0},m_{\tilde
e_L},m_{\tilde\nu_e}) + {1\over2} \,\biggr]~.
\end{eqnarray}
The corrections $\delta Z_{\nu_\mu},\ \delta Z_\mu$, and $\delta
v_\mu$ are obtained from these expressions by replacing $e\rightarrow
\mu$.  The $\tilde\chi$-fermion-sfermion couplings
$a_{\tilde\chi_if\tilde f_j}$ and $b_{\tilde\chi_if\tilde f_j}$ are
listed in Eqs.~(\ref{NQS}--\ref{CQS}), while the
chargino-neutralino-$W$ couplings $a_{\tilde\chi_i^0\tilde\chi_j^+W}$
and $b_{\tilde\chi_i^0\tilde\chi_j^+W}$ are defined in
Eqs.~(\ref{NCW}--\ref{NCW N}).  In these expressions, the $B_0$,
$B_1$, and $C_0$ functions are evaluated at zero momentum,
\begin{equation}
B_0(0,m_1,m_2) \ =\ {1\over\hat\epsilon} + 1 + \ln\left(Q^2\over
m_2^2\right) + {m_1^2\over m_1^2-m_2^2} \ln\left(m_2^2\over
m_1^2\right)~,
\end{equation}
\begin{equation}
B_1(0,m_1,m_2) \ =\ {1\over2}\biggl[{1\over\hat\epsilon} + 1 +
\ln\left(Q^2\over m_2^2\right) + \left({m_1^2\over m_1^2 -
m_2^2}\right)^2\ln\left({m_2^2\over m_1^2}\right) +
{1\over2}\left({m_1^2+m_2^2\over m_1^2-m_2^2}\right)\biggr]~,
\end{equation}
\begin{equation}
C_0(m_1,m_2,m_3) \ =\ {1\over m_2^2-m_3^2}\biggl[ {m_2^2\over
 m_1^2-m_2^2}\ln\left(m_2^2\over m_1^2\right) -{m_3^2\over
 m_1^2-m_3^2}\ln\left(m_3^2\over m_1^2\right)\biggr]\ ,
\end{equation}
where $Q$ is the renormalization scale.

The box diagram contributions are
\begin{eqnarray}
16\pi^2~a_1 &=& {1\over2}\sum_{i=1}^2\sum_{j=1}^4
a_{\tilde\chi_i^+\mu\tilde\nu_\mu}
b^*_{\tilde\chi_i^+\nu_e\tilde e_L}
b_{\tilde\chi_j^0\nu_\mu\tilde\nu_\mu} b_{\tilde\chi_j^0e\tilde e_L}
\,m_{\tilde\chi_i^+}\,m_{\tilde\chi_j^0}\,D_0(m_{\tilde
e_L},m_{\tilde\nu_\mu}, m_{\tilde\chi_i^+},m_{\tilde\chi_j^0})
\nonumber\\ &+&{1\over2}\sum_{i=1}^2\sum_{j=1}^4
a^*_{\tilde\chi_i^+e\tilde\nu_e} b_{\tilde\chi_i^+\nu_\mu\tilde\mu_L}
b^*_{\tilde\chi_j^0\nu_e\tilde\nu_e}
b^*_{\tilde\chi_j^0\mu\tilde\mu_L}
\,m_{\tilde\chi_i^+}\,m_{\tilde\chi_j^0}\,
D_0(m_{\tilde\mu_L},m_{\tilde\nu_e},
m_{\tilde\chi_i^+},m_{\tilde\chi_j^0}) \nonumber\\
&+&\sum_{i=1}^2\sum_{j=1}^4
b_{\tilde\chi_i^+\nu_\mu\tilde\mu_L}b^*_{\tilde\chi_i^+\nu_e\tilde
e_L} b^*_{\tilde\chi_j^0\mu\tilde\mu_L}b_{\tilde\chi_j^0e\tilde e_L}
\,D_{27}(m_{\tilde\mu_L},m_{\tilde
e_L},m_{\tilde\chi_i^+},m_{\tilde\chi_j^0})\\
&+&\sum_{i=1}^2\sum_{j=1}^4 a^*_{\tilde\chi_i^+\mu\tilde\nu_\mu}
a_{\tilde\chi_i^+e\tilde\nu_e} b_{\tilde\chi_j^0\nu_\mu\tilde\nu_\mu}
b^*_{\tilde\chi_j^0\nu_e\tilde\nu_e}
\,D_{27}(m_{\tilde\nu_\mu},m_{\tilde\nu_e}, m_{\tilde\chi_i^+},
m_{\tilde\chi_j^0})~,\nonumber
\end{eqnarray}
where the functions $D_0$ and $D_{27}$ are
\begin{equation}
D_0(m_1,m_2,m_3,m_4) \ =\ {1\over m_1^2-m_2^2}
\ \biggl[\,C_0(m_1,m_3,m_4)-C_0(m_2,m_3,m_4)\,\biggr]\ ,
\end{equation}
\begin{equation}
D_{27}(m_1,m_2,m_3,m_4) \ =\ {1\over4(m_1^2-m_2^2)}
\ \biggl[\,m_1^2\,C_0(m_1,m_3,m_4) - m_2^2\,C_0(m_2,m_3,m_4)\,\biggr]\ .
\end{equation}
We checked our box diagram calculation against
Refs.~\cite{Grifols,deltar}, and we checked our formulas for
$\delta Z$ and $\delta v$ with those of Ref.~\cite{deltar}.  Here
(there) the formulas are written in terms of the couplings
corresponding to vertices with incoming (outgoing) charginos and
neutralinos.  To compare we must make the transformation
$a_{\tilde\chi f\tilde f} \leftrightarrow b^*_{\tilde\chi f\tilde f}$,
except for couplings involving the chargino and down-type fermions,
which remain unchanged.  Also, their $\tilde\chi^+\tilde\chi^0W$
coupling differs from ours by a sign.

The effective weak mixing angle is given in terms of the
\mbox{\footnotesize$\overline{\rm DR}~$} weak mixing angle, $\hat
s^2$, via
\begin{equation}
\sin^2\theta_{\rm eff}^{\rm lept}\ =\ \hat s^2\,{\cal R}e\,\hat k_\ell
\end{equation}
where \cite{DS}
\begin{equation}
\hat k_\ell\ =\ 1\ +\ {\hat c\over\hat s}{\Pi_{Z\gamma}(M_Z^2) -
\Pi_{Z\gamma}(0)\over M_Z^2}\ +\ {\hat\alpha\hat c^2\over\pi\hat
s^2}\ln c^2 \ -\ {\hat\alpha\over4\pi\hat s^2}V_{\ell}(M_Z^2)\ ,
\end{equation}
with
\begin{equation}
V_{\ell}(M_Z^2) \ =\ {1\over2}f\left({1\over c^2}\right) \ +\ 4\hat
		c^2g\left({1\over c^2}\right) \ -\ {1-6\hat s^2+8\hat
		s^4\over4\hat c^2}\, f(1)
\end{equation}
and
\begin{eqnarray}
{\cal R}e\,f(x)&=&{2\over x}+{7\over2}-\left(3+{2\over x}\right)\ln x
+ \left(1+{1\over x}\right)^2\biggl[\,2{\rm
Li}_2\left({1\over1+x}\right)
-{\pi^2\over3}+\ln^2(1+x)\,\biggr]~,\nonumber\\ g(x)&=&\left({1\over
x}+{1\over2}\right)\left( {\tan^{-1}y\over
y}-1\right)+{9\over8}+{1\over2x} -\left(1+{1\over2x}\right){4\over
x}\left(\tan^{-1}y\right)^2~.
\end{eqnarray}
Here $y\equiv\sqrt{x/(4-x)}$, Li$_2$ is the Spence function, and
$\Pi_{Z\gamma}$ is listed in Eq.~(\ref{pizg}).  We do not include here
the nonuniversal $Z$-vertex supersymmetric contribution to
$\sin^2\theta_{\rm eff}^{\rm lept}$. The largest contributions
can be obtained from Ref.~\cite{Rb}.

\vspace{.7cm} {\noindent\Large\bf Appendix D: One-loop self-energies}
\vspace{.7cm} \setcounter{equation}{0}
\renewcommand{\theequation}{D.\arabic{equation}}

In this appendix we list all the relevant self-energy functions which
allow us to determine the one-loop fermion, gauge-boson, and
superpartner masses.  We explicitly include all of the necessary
couplings.  We perform our calculations in the 't~Hooft-Feynman gauge,
in which the Goldstone bosons and the ghosts have the same masses as
the corresponding gauge-bosons. The gauge couplings $g',\ g,$ and
$g_3$, and the Yukawa couplings $\lambda_f$ are all
\mbox{\footnotesize$\overline{\rm DR}~$} couplings.  The neutralino
mixing matrix $N$, the chargino mixing matrices $U$ and $V$, the Higgs
mixing angles $\alpha$ and $\beta$, and the sfermion mixing angles
$\theta_f$ are described in Appendix A, as are the normalizations of
the Yukawa couplings $\lambda_f$.  The self-energies are given in
terms of the Passarino-Veltman functions $A_0,\ B_0,\ B_1,\ F,\ G,\ H$,
and $\tilde B_{22}$ listed in Appendix B, Eqs.~(\ref{A}--\ref{B22}).

To streamline notation we do not write explicitly the external
momentum dependence of these functions, e.g., we write
$B_0(p,m_1,m_2)$ as $B_0(m_1,m_2)$. Throughout this appendix we write
$s$ for $\sin$, $c$ for $\cos$ and $t$ for $\tan$, so that
$s_\beta\equiv\sin\beta$, $c_{2\theta_t}\equiv\cos2\theta_t$, etc.,
and for the sfermion mixing angles, $c_u\equiv\cos\theta_u$, etc. Sub-
or superscripts $f$ denote a quark or lepton, and $q$ denotes a quark.
Inside a summation $\sum_{f_u}$, the subscript or superscript $u$
denotes all up-type (s)fermions, $u,c,t,\nu_e,\nu_\mu,\nu_\tau$, and
similarly inside a summation $\sum_{f_d}$, the script $d$ denotes all
down-type (s)fermions, $d,s,b,e,\mu,$ and $\tau$. The sum
$\sum_{f_u/f_d}$ denotes a summation over (s)quark and (s)lepton
doublets, and the sum $\Sigma_q$ denotes a sum over (s)quarks.  Some
terms are zero, for example $\lambda_\nu=0$, and terms involving the
right-handed sneutrino are absent.

In the self-energies listed below the $1/\hat\epsilon$ poles are
canceled by counterterms which relate the bare mass to the running
mass. So, in the following \mbox{\footnotesize$\overline{\rm DR}~$}
self-energies we implicitly subtract the $1/\hat\epsilon$ poles.

\vspace{.5cm} {\noindent\large\bf $Z$ and $W$ bosons}
\vspace{.5cm}

The full one-loop MSSM gauge-boson self-energies appear in
Ref.~\cite{Grifols} and subsequently in Ref.~\cite{CPR}.  The
supersymmetric contributions are listed in Refs.~\cite{DHY,CPP,dab}.

The self-energies of the gauge-bosons can be separated into transverse
and longitudinal pieces, e.g.
\begin{equation}
\Pi^{\mu\nu}_{ZZ}(p^2) \ =\ \Pi_{ZZ}^T(p^2)\biggl[g^{\mu\nu}
-{p^{\mu}p^{\nu}\over p^2}\biggr] \ +\ \Pi_{ZZ}^L(p^2)\,
{p^{\mu}p^{\nu}\over p^2}\ .
\end{equation}
The physical gauge-boson masses are the poles of the corresponding
propagators, which involve only the transverse part of the gauge-boson
self-energy,
\begin{eqnarray}
&& M_Z^2 \ =\ \hat M_Z^2(Q) \ -\ {\cal
R}e\,\Pi^T_{ZZ}(M_Z^2)~,\label{mz}\\ && M_W^2 \ =\ \hat M_W^2(Q) \ -
\ {\cal R}e\,\Pi^T_{WW}(M_W^2)~.
\end{eqnarray}
Here $\hat M_Z(Q)$ and $\hat M_W(Q)$ denote the
\mbox{\footnotesize$\overline{\rm DR}~$} running masses which are
related to the \mbox{\footnotesize$\overline{\rm DR}~$} gauge
couplings and vev's, as in Eq.~(\ref{mwmz}). The gauge-boson
self-energies are evaluated at the renormalization scale $Q$.

The transverse part of the $Z$-boson self-energy is
\begin{eqnarray}
\qquad 16\pi^2\,{\hat c^2\over g^2}\,\Pi^T_{ZZ}(p^2) &=&-
\ s^2_{\alpha\beta}\biggl[\tilde B_{22}(m_A,m_H) +\tilde
B_{22}(M_Z,m_h)-M^2_Z B_0(M_Z,m_h)\,\biggr]\nonumber\\ &-&
c^2_{\alpha\beta}\biggl[\tilde B_{22}(M_Z,m_H) +\tilde B_{22}(m_A,m_h)
-M^2_ZB_0(M_Z,m_H)\,\biggr]\nonumber\\ &-& 2\hat
c^4\biggl(2p^2+M_W^2-M^2_Z{\hat s^4\over\hat c^2}\biggr)
B_0(M_W,M_W)\nonumber\\ &-&(8\hat c^4+c^2_{2\hat\theta_W})\tilde
B_{22}(M_W,M_W) \ -\ c^2_{2\hat\theta_W}\tilde
B_{22}(m_{H^+},m_{H^+})\nonumber\\ &-& \sum_f\sum_{i,j=1}^2
4N_c^fv_{fij}^2 \tilde B_{22}(m_{\tilde f_i},m_{\tilde
f_j})\nonumber\\ &+& \sum_f
N_c^f\Biggl\{\biggl(g_{f_L}^2+g_{f_R}^2\biggr) H(m_f,m_f) \ -
\ 4g_{f_L}g_{f_R}m_f^2 B_0(m_f,m_f)\Biggr\}\nonumber\\ &+& {\hat
c^2\over2g^2}\sum_{i,j=1}^4
\Biggl\{f_{ijZ}^0H(m_{\tilde\chi^0_i},m_{\tilde\chi^0_j}) \ +
\ 2\,g_{ijZ}^0\,m_{\tilde\chi^0_i} m_{\tilde\chi^0_j}
B_0(m_{\tilde\chi^0_i},m_{\tilde\chi_j^0})\Biggr\}\nonumber\\ &+&
{\hat c^2\over g^2}\sum_{i,j=1}^2
\Biggl\{f^+_{ijZ}H(m_{\tilde\chi_i^+},m_{\tilde\chi^+_j}) \ +
\ 2\,g^+_{ijZ}\,m_{\tilde\chi^+_i}m_{\tilde\chi^+_j}
B_0(m_{\tilde\chi_i^+},m_{\tilde\chi^+_j})\Biggr\}~,\nonumber\\
\label{piz}
\end{eqnarray}
where the summation $\sum_f$ is over all quarks and leptons, and the
color factor $N_c^f$ is 3 for (s)quarks and 1 for (s)leptons.  The
notation $s_{\alpha\beta}$ denotes $\sin(\alpha-\beta)$, and
$c_{\alpha\beta}$ refers to $\cos(\alpha-\beta)$.

The sfermion-sfermion-$Z$ couplings can be written in terms of the
weak neutral-current couplings defined in Eq.~(\ref{glgr}):
\begin{equation}
v_{f11} \ =\ g_{f_L}c^2_f-g_{f_R}s^2_f~,\ \ \ v_{f22}\ =
\ g_{f_R}c^2_f-g_{f_L}s^2_f~, \ \ \ v_{f12}\ =\ v_{f21} \ =\ (g_{f_L} +
g_{f_R})c_fs_f~.
\end{equation}

The neutralino-neutralino-$Z$-boson couplings are defined by
\begin{equation}
f^0_{ijZ}\ =\ |a_{\tilde\chi_i^0\tilde\chi_j^0Z}|^2 +
|b_{\tilde\chi_i^0\tilde\chi_j^0Z}|^2, \qquad g^0_{ijZ}\ =\ 2\,{\cal
R}e\left(b^*_{\tilde\chi_i^0\tilde\chi_j^0Z}~
a_{\tilde\chi_i^0\tilde\chi_j^0Z}\right)\ ,
\label{XZ1}
\end{equation}
and analogous definitions hold for $f^+_{ijZ}$ and $g^+_{ijZ}$.  We
write the Feynman rule for the $\tilde\chi\tilde\chi Z_\mu$ vertex,
where $\tilde\chi$ is a chargino or neutralino, as
$-i\gamma_\mu(a{\cal P}_L + b{\cal P}_R)$, where ${\cal P}_{L,R}$ are
the usual chiral projectors $(1\mp\gamma_5)/2$. The couplings
involving the unrotated $\tilde\psi^0$ and $\tilde\psi^+$ fields
satisfy $b_{\tilde\psi_i^0\tilde\psi_j^0Z}
=-a_{\tilde\psi_i^0\tilde\psi_j^0Z}$ and
$b_{\tilde\psi_i^+\tilde\psi_j^+Z} =
a_{\tilde\psi_i^+\tilde\psi_j^+Z}$.  The nonzero $a$-type couplings
are
\begin{equation}
a_{\tilde\psi_3^0\tilde\psi_3^0Z} \ = \ -
\ a_{\tilde\psi_4^0\tilde\psi_4^0Z} \ =\ {g\over2\hat c}~,~~~~
a_{\tilde\psi_1^+\tilde\psi_1^+Z} \ =\ g\hat c~,~~~~
a_{\tilde\psi_2^+\tilde\psi_2^+Z} \ =\ {gc_{2\hat\theta_W}\over2\hat
c}\ .
\label{NNZ}
\end{equation}
For an {\em incoming} $\tilde\chi_i^0$ and {\em incoming}
$\tilde\chi_i^+$ we have
\begin{eqnarray}
     \label{NNZ N} && a_{\tilde\chi_i^0\tilde\chi_j^0Z} \ =
\ N_{ik}^*\,N_{jl}\, a_{\tilde\psi_k^0\tilde\psi_l^0Z}~,~~~~
b_{\tilde\chi_i^0\tilde\chi_j^0Z} \ =\ N_{ik}\,
N_{jl}^*\,b_{\tilde\psi_k^0\tilde\psi_l^0Z}~,\nonumber\\[2mm] &&
a_{\tilde\chi_i^+\tilde\chi_j^+Z} \ =\ V_{ik}^*\,V_{jl}\,
a_{\tilde\psi_k^+\tilde\psi_l^+Z}~,~~~~
b_{\tilde\chi_i^+\tilde\chi_j^+Z} \ =\ U_{ik}
\,U_{jl}^*\,b_{\tilde\psi_k^+\tilde\psi_l^+Z}~.
\end{eqnarray}
(Here and in the following formulae which specify rotations, we adopt
the summation convention for repeated indices.)

For the transverse part of the $W$-boson self-energy, we find
\begin{eqnarray}
\qquad{16\pi^2\over g^2}~\Pi^T_{WW}(p^2) &=& -\ s^2_{\alpha\beta}
\biggl[\tilde B_{22}(m_H,m_{H^+})+\tilde B_{22}(m_h,M_W)-M_W^2
B_0(m_h,M_W)\,\biggr]\nonumber\\ &-& c^2_{\alpha\beta}\biggl[\tilde
B_{22}(m_h,m_{H^+}) +\tilde
B_{22}(m_H,M_W)-M_W^2B_0(m_H,M_W)\,\biggr]\nonumber\\ &-& \tilde
B_{22}(m_A,m_{H^+})\ - \ (1+8\hat c^2)\tilde
B_{22}(M_Z,M_W)\nonumber\\ &-& \hat s^2\ \biggl[\,8\tilde
B_{22}(M_W,0)+4p^2B_0(M_W,0)\,\biggr]\nonumber\\ &-&
\biggl[\,(4p^2+M_Z^2+M_W^2)\hat c^2-M^2_Z\hat s^4\,\biggr]\,
B_0(M_Z,M_W)\nonumber\\ &+& \sum_{f_u/f_d}
\Biggl\{{1\over2}N_c^fH(m_u,m_d) \ -\ \sum_{i,j=1}^2 2N_c^fw_{fij}^2
\tilde B_{22}(m_{\tilde u_i}, m_{\tilde d_j})\Biggr\}\nonumber\\ &+&
{1\over g^2}\sum_{i=1}^4\sum_{j=1}^2 \Biggl\{f_{ijW}
H(m_{\tilde\chi^0_i},m_{\tilde\chi^+_j}) \ +
\ 2\,g_{ijW}\,m_{\tilde\chi^0_i}m_{\tilde\chi^+_j}
B_0(m_{\tilde\chi^0_i},m_{\tilde\chi^+_j})\Biggr\} \ ,\nonumber\\
\label{piw}
\end{eqnarray}
where the summation $\sum_{f_u/f_d}$ is over quark and lepton
doublets, and
\begin{eqnarray}
&&w_{f11}\ =\ c_uc_d~,\qquad w_{f12}\ =\ c_us_d~,\nonumber\\
&&w_{f21}\ =\ s_uc_d~,\qquad w_{f22}\ =\ s_us_d~.
\end{eqnarray}

The neutralino-chargino-$W$-boson couplings are
\begin{equation}
f_{ijW}\ =\ |a_{\tilde\chi_i^0\tilde\chi_j^+W}|^2 +
|b_{\tilde\chi_i^0\tilde\chi_j^+W}|^2~,~~~~g_{ijW}\ =\ 2\,{\cal
R}e\left(b^*_{\tilde\chi_i^0\tilde\chi_j^+W}~
a_{\tilde\chi_i^0\tilde\chi_j^+W}\right)~.
\label{XW1}
\end{equation}
We write the Feynman rule for the neutralino-chargino-$W_\mu$ vertex
as $-i\gamma_\mu(a{\cal P}_L+b{\cal P}_R)$, and the nonzero couplings
are
\begin{equation}
\label{NCW}
a_{\tilde\psi_2^0\tilde\psi_1^+W} \ =
\ b_{\tilde\psi_2^0\tilde\psi_1^+W} \ =\ -\ g~,~~~~
a_{\tilde\psi_4^0\tilde\psi_2^+W} \ =\ -
\ b_{\tilde\psi_3^0\tilde\psi_2^+W} \ =\ {g\over\sqrt2}~.
\end{equation}
For an {\em incoming} $\tilde\chi_i^0$ we have the couplings to mass
eigenstates,
\begin{equation}
  \label{NCW N} a_{\tilde\chi_i^0\tilde\chi_j^+W} \ =\ N_{ik}^*\,V_{jl}\,
a_{\tilde\psi_{k}^0\tilde\psi_{l}^+W}~,~~~~
b_{\tilde\chi_i^0\tilde\chi_j^+W} \ =\ N_{ik}\, U_{jl}^*\,
b_{\tilde\psi_{k}^0\tilde\psi_{l}^+W}~,
\end{equation}
while for an {\em incoming} $\tilde\chi_j^+$ we have the couplings
\begin{equation}
  \label{NCW C} a_{\tilde\chi_i^0\tilde\chi_j^+W} \ =\ N_{ik}\, V_{jl}^*
\,a_{\tilde\psi_{k}^0\tilde\psi_{l}^+W}~,~~~~
b_{\tilde\chi_i^0\tilde\chi_j^+W} \ =\ N_{ik}^*\, U_{jl} \,
b_{\tilde\psi_{k}^0\tilde\psi_{l}^+W}~.
\end{equation}

Finally, we write the mixed $Z-\gamma$ self-energy as
\begin{eqnarray}
16\pi^2~{\hat c\over eg}~\Pi_{Z\gamma}(p^2) &=& (12\hat s^2-10) \tilde
B_{22}(M_W,M_W)\ -\ 2(M_W^2+2\hat c^2p^2)B_0(M_W,M_W)\nonumber\\ &+&
\sum_f N_c^fe_f(g_{f_L}-g_{f_R})\biggl[4\tilde B_{22}(m_f,m_f)+p^2
B_{0}(m_f,m_f)\biggr]\nonumber\\ &-& 2c_{2\hat\theta_W}\tilde
B_{22}(m_{H^+},m_{H^+})\nonumber\\ &+&
{1\over2}\sum_{i=1}^2(|V_{i1}|^2+|U_{i1}|^2+2c_{2\hat\theta_W})
\biggl(4\tilde B_{22}(m_{\tilde\chi_i^+},m_{\tilde\chi_i^+})+
p^2B_0(m_{\tilde\chi_i^+},m_{\tilde\chi_i^+})\biggr) \nonumber\\ &-&
4\sum_f N_c^f e_f \biggl[(g_{f_L}c_f^2-g_{f_R}s_f^2) \tilde
B_{22}(m_{\tilde f_1},m_{\tilde f_1}) \nonumber\\ &&\qquad\qquad +
\ (g_{f_L}s_f^2-g_{f_R}c_f^2)\tilde B_{22}(m_{\tilde f_2}, m_{\tilde
f_2})\biggr] \ .
\label{pizg}
\end{eqnarray}

\vspace{.5cm} {\noindent\large\bf Quarks and leptons}
\vspace{.5cm}

The fermion masses are defined as the poles of the corresponding
fermion propagators. They are related to the
\mbox{\footnotesize$\overline{\rm DR}~$} masses, $\hat m_f,$ by the
self-energies, $\Sigma_f(p^2),$ as follows
\begin{equation}
m_f \ =\ \hat m_f(Q)\ -\ {\cal R}e\,\Sigma_f(m_f^2)~.\label{mfr}
\end{equation}
The \mbox{\footnotesize$\overline{\rm DR}~$} fermion mass $\hat m_f$
is related to the \mbox{\footnotesize$\overline{\rm DR}~$} Yukawa
coupling and vev as shown in Eq.~(\ref{mf}). Care must be taken in
evaluating the \mbox{\footnotesize$\overline{\rm DR}~$} vev. After
evaluating the \mbox{\footnotesize$\overline{\rm DR}~$} gauge
couplings $g'$ and $g$ as outlined in Appendix C, we determine the
\mbox{\footnotesize$\overline{\rm DR}~$} vev via
\begin{equation}
v^2(Q) \ =\ 4{M_Z^2 + {\cal R}e\,\Pi_{ZZ}^T(M_Z^2)\over g'^2(Q) +
g^2(Q)}~,
\end{equation}
where $Q$ is the renormalization scale (the argument of the $Z$
self-energy is the external momentum; it implicitly depends on the
scale $Q$ as well).

For the top quark, $\Sigma_t(p^2)$ is
\begin{eqnarray}
16\pi^2~{\Sigma_t(p^2)\over m_t}&=& {4g_3^2\over3}
\Biggl\{B_1(m_{\tilde g},m_{\tilde t_1})\ +\ B_1(m_{\tilde
g},m_{\tilde t_2}) \ -\ \biggl(5+3\ln{Q^2\over m^2_t}\biggr)
\nonumber\\ &&\qquad -\ s_{2\theta_t}{m_{\tilde g}\over m_t}\left(
B_0(m_{\tilde g},m_{\tilde t_1})- B_0(m_{\tilde g},m_{\tilde
t_2})\right)\Biggr\}\nonumber\\ &+&{1\over2}\lambda_t^2
\Biggl\{s_\alpha^2 \biggl[\, B_1(m_t,m_H) + B_0(m_t,m_H)\,\biggr]\ +
\ c^2_\alpha \biggl[\, B_1(m_t,m_h) + B_0(m_t,m_h)\,\biggr] \nonumber\\
&& \qquad +\ c^2_\beta \biggl[\, B_1(m_t,m_A) - B_0(m_t,m_A) \,\biggr]
\ +\ s^2_\beta \biggl[\, B_1(m_t,M_Z) -
B_0(m_t,M_Z)\,\biggr]\Biggr\}\nonumber\\ & + &{1\over2}
\biggl[\,(\lambda^2_b s^2_\beta + \lambda^2_t c^2_\beta )
B_1(m_b,m_{H^+}) \ +\ ( g^2 +\lambda_b^2c_\beta^2+\lambda^2_t
s^2_\beta) B_1(m_b,M_W) \, \biggr]\nonumber\\ &+&\lambda_b^2 c_\beta^2
\biggl[\,B_0(m_b,m_{H^+})-B_0(m_b,m_{W})\,\biggr] \ -\ (ee_t)^2
\biggl(5+3\ln{Q^2\over m^2_t}\biggr) \nonumber\\ & + & {g^2\over \hat
c^2}\biggl[\,\biggl(g_{t_L}^2+ g_{t_R}^2\biggr) B_1( m_t,M_Z) +
4g_{t_L}g_{t_R} B_0(m_t,M_Z)\,\biggr] \nonumber\\ & +
&{1\over2}\sum_{i=1}^4\sum_{j=1}^2\biggl[\,f_{it\tilde t_j}
B_1(m_{\tilde\chi^0_i},m_{\tilde t_j}) + \,g_{it\tilde
t_j}\,{m_{\tilde\chi^0_i}\over m_t} B_0(m_{\tilde\chi^0_i},m_{\tilde
t_j})\,\biggr]\nonumber\\ & +
&{1\over2}\sum_{i,j=1}^2\biggl[\,f_{it\tilde b_j}
B_1(m_{\tilde\chi^+_i},m_{\tilde b_j}) + \,g_{it\tilde
b_j}\,{m_{\tilde\chi^+_i}\over m_t} B_0(m_{\tilde\chi^+_i},m_{\tilde
b_j})\,\biggr]~.
\end{eqnarray}
The neutral current couplings $g_f$ are defined in Eq.~(\ref{glgr}).

We write the Feynman rules for the $\tilde\chi_if\tilde f_j$ couplings
as $-i(a{\cal P}_L+b{\cal P}_R)$ (for vertices involving the chargino
and down-type fermions the Feynman rule is $iC^{-1}(a{\cal P}_L+b{\cal
P}_R)$, where $C$ is the charge-conjugation matrix). We define
\begin{equation}
f_{if\tilde f_j} \ = \ |a_{\tilde\chi_if\tilde f_j}|^2 +
|b_{\tilde\chi_if\tilde f_j}|^2~,\qquad\qquad g_{if\tilde f_j} \ =
\ 2\,{\cal R}e\,(b^*_{\tilde\chi_if\tilde f_j}\,a_{\tilde\chi_if\tilde
f_j})~.\label{ffX1}
\end{equation}
In the unrotated $\tilde\psi^0,\ \tilde\psi^+$ basis, we have
\begin{eqnarray}
&& \hspace{-3.5pt} a_{\tilde\psi_1^0f\tilde f_R} \ =
\ {g'\over\sqrt2}\,Y_{f_R}~,~~~~~~~~ b_{\tilde\psi_1^0f\tilde f_L}
\ =\ {g'\over\sqrt2}\,Y_{f_L}~,\nonumber\\[2mm] &&
b_{\tilde\psi_2^0f\tilde f_L} \ =\ \sqrt2\,g\,I_3^{f_L}~,~~~~
a_{\tilde\psi_1^+d\tilde u_L} \ =\ b_{\tilde\psi_1^+u\tilde d_L}
\ =\ g~,\nonumber\\[2mm] && a_{\tilde\psi_3^0d\tilde d_L} \ =
\ b_{\tilde\psi_3^0d\tilde d_R} \ =\ -\ b_{\tilde\psi_2^+d\tilde u_L}
\ =\ -\ b_{\tilde\psi_2^+u\tilde d_R} \ =\ \lambda_d~,\label{NQS}
\nonumber\\[2mm] && a_{\tilde\psi_4^0u\tilde u_L} \ =
\ b_{\tilde\psi_4^0u\tilde u_R} \ =\ -\ a_{\tilde\psi_2^+u\tilde d_L}
\ =\ -\ a_{\tilde\psi_2^+d\tilde u_R} \ =\ \lambda_u~,
\end{eqnarray}
where the quantum numbers $Y_f$ and $I_3^f$ are listed in the table of
Eq.~(\ref{table1}). These couplings correspond to vertices with {\em
incoming} neutralinos and {\em incoming} charginos. To obtain the
couplings to the mass eigenstates $\tilde\chi_i^0$ and
$\tilde\chi_i^+$, we specify the rotations
\begin{eqnarray}
&& a_{\tilde\chi_i^0f\tilde f } \ =
\ N_{ij}^*\,a_{\tilde\psi_j^0f\tilde f }~,~~~~
b_{\tilde\chi_i^0f\tilde f } \ =\ N_{ij} \,b_{\tilde\psi_j^0f\tilde
f }~,\\[2mm] && a_{\tilde\chi_i^+f\tilde f'} \ =
\ V_{ij}^*\,a_{\tilde\psi_j^+f\tilde f'}~,~~~
b_{\tilde\chi_i^+f\tilde f'} \ =\ U_{ij} \,b_{\tilde\psi_j^+f\tilde
f'}~.
\label{CQS}
\end{eqnarray}
The couplings to the sfermion mass eigenstates are found by rotating
these couplings (both $a$- and $b$-type) by the sfermion mixing
matrix,
\begin{equation}
\left(\begin{array}{c} a_{\tilde\chi f\tilde f'_1} \\ a_{\tilde\chi
f\tilde f'_2}
\end{array}\right) \ =\ \left(\begin{array}{cc}
{}~c_{f'} & s_{f'} \\ -s_{f'} & c_{f'}
\end{array}\right) \left(\begin{array}{c}
a_{\tilde\chi f\tilde f'_L} \\ a_{\tilde\chi f\tilde f'_R}
\end{array}\right)~.
\label{NQS s}
\end{equation}

The self-energies $\Sigma_f(p^2)$ for the other up-type quarks and
leptons can be obtained from the previous formulae by obvious
substitutions.  For the bottom quark (and similarly for all down-type
fermions), one interchanges $t\leftrightarrow b$,
$c_\alpha\leftrightarrow s_\alpha,$ and $c_\beta\leftrightarrow
s_\beta$.

\vspace{.5cm}
\noindent{\large\bf Charginos and neutralinos}
\vspace{.5cm}

The complete one-loop self-energies for charginos and neutralinos are
given in \cite{PP1,PP2}; we present them here in a matrix
formulation. For the Higgs-boson contributions, $H_n^0$ refers to $H,
\ h,\ G^0$, and $A$, while $H_n^+$ represents $G^+$ and $H^+$. The
$G^0$ and $G^+$ are the Goldstone bosons; in the 't~Hooft-Feynman
gauge their masses are equal to $M_Z$ and $M_W$, respectively.

We now describe the full one-loop neutralino and chargino mass
matrices, from which we determine the one-loop masses.  The one-loop
neutralino mass matrix has the form
\begin{equation}
 {\cal M}_{\tilde\psi^0}\ +\ {1\over2}\left(\delta {\cal
M}_{\tilde\psi^0}(p^2) + \delta{\cal M}_{\tilde\psi^0}^T(p^2)\right)~,
\label{chi0-1}
\end{equation}
where
\begin{equation}
 \delta{\cal M}_{\tilde\psi^0}(p^2) \ =
\ -\ \Sigma_R^0(p^2){\cal M}_{\tilde\psi^0}
\ -\ {\cal M}_{\tilde\psi^0}\Sigma_L^0(p^2)
\ -\ \Sigma_S^0(p^2)\ .
\end{equation}
Here ${\cal M}_{\tilde\psi^{0}}$ is the tree-level neutralino mass
matrix of Eq.~(\ref{mchi0}), and the $\Sigma_{L,R,S}^{+,0}(p^2)$ are
{\em matrix} corrections.  They allow us to determine the one-loop
masses and mixing angles for arbitrary tree-level parameters.

The one-loop chargino mass matrix is as follows,
\begin{equation}
{\cal M}_{\tilde\psi^+}\ -\ \Sigma_R^+(p^2)\,{\cal M}_{\tilde\psi^+}
\ -\ {\cal M}_{\tilde\psi^+}\,\Sigma_L^+(p^2)\ -\ \Sigma_S^+(p^2)~,
\end{equation}
where ${\cal M}_{\tilde\psi^{+}}$ is the tree-level chargino mass
matrix of Eq.~(\ref{mchi+}).  The elements of ${\cal
M}_{\tilde\chi^0}$ and ${\cal M}_{\tilde\psi^+}$ contain
\mbox{\footnotesize$\overline{\rm DR}~$} parameters at the scale $Q$.
In particular, they include corrections corresponding to replacing
$M_Z$ with $\hat M_Z$, obtained from Eq.~(\ref{mz}). Similarly,
$\tan\beta$ in the tree-level matrices is $\tan\beta(Q)$. The
self-energies $\Sigma_{L,R,S}$ are also evaluated at the scale $Q$.

To obtain the mass for a given neutralino or chargino, for example
$\tilde\chi_1^0$, we first evaluate the matrix of Eq.~(\ref{chi0-1})
with the momenta $p^2=m_{\tilde\chi_1^0}^2$. We then solve for the
eigenvalues of that matrix. So, in determining four neutralino and two
chargino masses, we construct a total of six different matrices.

We compute the mass matrix corrections by evaluating two-point
diagrams with unrotated neutralinos or charginos on external legs, and
mass eigenstates inside the loop.  We obtain the couplings associated
with these diagrams by the following method. The neutralino mass
corrections involve the couplings $a_{\tilde\psi^0_k\cdots}$ which we
obtain from the various couplings $a_{\tilde\chi^0_i\cdots}$ (for {\em
incoming} $\tilde\chi_i^0$) by leaving off one factor of
$N_{ik}^*$. The neutralino mass corrections also involve the couplings
$b_{\tilde\psi_k^0\cdots}$ which we obtain from the couplings
$b_{\tilde\chi_i^0\cdots}$ (for {\em incoming} $\tilde\chi_i^0$) by
leaving off one rotation $N_{ik}$.  We obtain the couplings
$a_{\tilde\psi^+_k\cdots}$ which appear in the chargino mass
corrections from the couplings $a_{\tilde\chi_i^+\cdots}$ (for {\em
incoming} $\tilde\chi_i^+$) by leaving off one factor of $V_{ik}^*$,
and we determine the couplings $b_{\tilde\psi^+_k\cdots}$ from the
couplings $b_{\tilde\chi^+_i\cdots}$ (for {\em incoming}
$\tilde\chi_i^+$) by leaving off one factor of $U_{ik}$.

For the neutralinos, we have the one-loop correction
\begin{eqnarray}
16\pi^2\,{\Sigma_L^0}_{ij}(p^2) &=& \sum_f\sum_{k=1}^2N_c^f\,
a^*_{\tilde\psi_i^0f\tilde f_k}\, a_{\tilde\psi_j^0f\tilde f_k}
\ {\cal R}e\,B_1(m_f,m_{\tilde f_k})\nonumber\\
&+& 2\sum_{k=1}^2 a^*_{\tilde\psi_i^0\tilde\chi_k^+W} \,
a_{\tilde\psi_j^0\tilde\chi_k^+W}
\ {\cal R}e\,B_1(m_{\tilde\chi_k^+},M_W)\nonumber\\
&+& \sum_{k=1}^4 a^*_{\tilde\psi_i^0\tilde\chi_k^0Z} \,
a_{\tilde\psi_j^0\tilde\chi_k^0Z}\ {\cal R}e\,
B_1(m_{\tilde\chi_k^0},M_Z)\nonumber\\
&+& \sum_{k,n=1}^2a^*_{\tilde\psi_i^0\tilde\chi_k^+H_n^+}\,
a_{\tilde\psi_j^0\tilde\chi_k^+H_n^+} \ {\cal
R}e\,B_1(m_{\tilde\chi_k^+},M_{H_n^+})\nonumber\\
&+& {1\over2}\sum_{k,n=1}^4 a^*_{\tilde\psi_i^0\tilde\chi_k^0H_n^0} \,
a_{\tilde\psi_j^0\tilde\chi_k^0H_n^0} \ {\cal R}e\,
B_1(m_{\tilde\chi_k^0},M_{H_n^0})\ ;
\end{eqnarray}
$\Sigma_R^0$ is obtained from $\Sigma_L^0$ by replacing the couplings
$a_{\tilde\psi^0\cdots}$ with $b_{\tilde\psi^0\cdots}$.  The
$\Sigma_S^0(p^2)$ correction is given by
\begin{eqnarray}
16\pi^2\,{\Sigma_S^0}_{ij}(p^2) &=& 2\sum_f\sum_{k=1}^2N_c^f\,
b^*_{\tilde\psi_i^0f\tilde f_k}\,a_{\tilde\psi_j^0f\tilde f_k}\, m_f
\ {\cal R}e\,B_0(m_f,m_{\tilde f_k})\nonumber\\
&-& 8\sum_{k=1}^2
b^*_{\tilde\psi_i^0\tilde\chi_k^+W}\,a_{\tilde\psi_j^0\tilde\chi_k^+W}\,
m_{\tilde\chi_k^+}\ {\cal R}e\,B_0(m_{\tilde\chi_k^+},M_W)\nonumber\\
&-& 4\sum_{k=1}^4
b^*_{\tilde\psi_i^0\tilde\chi_k^0Z}\,a_{\tilde\psi_j^0\tilde\chi_k^0Z}\,
m_{\tilde\chi_k^0}\ {\cal R}e\,B_0(m_{\tilde\chi_k^0},M_Z)\nonumber\\
&+& 2\sum_{k,n=1}^2
b^*_{\tilde\psi_i^0\tilde\chi_k^+H_n^+}\,
a_{\tilde\psi_j^0\tilde\chi_k^+H_n^+}\, m_{\tilde\chi_k^+}\ {\cal
R}e\,B_0(m_{\tilde\chi_k^+},M_{H_n^+})\nonumber\\
&+& \sum_{k,n=1}^4 b^*_{\tilde\psi_i^0\tilde\chi_k^0H_n^0} \,
a_{\tilde\psi_j^0\tilde\chi_k^0H_n^0}\, m_{\tilde\chi_k^0}\ {\cal
R}e\,B_0(m_{\tilde\chi_k^0},M_{H_n^0})\ .
\end{eqnarray}

The chargino mass corrections are given by similar formulae,
\begin{eqnarray}
16\pi^2~{\Sigma_L^+}_{ij}(p^2) &=& {1\over2}\sum_f\sum_{k=1}^2N_c^f\,
a^*_{\tilde\psi_i^+f\tilde f'_k}\,a_{\tilde\psi_j^+f\tilde f'_k}
\ {\cal R}e\,B_1(m_f,m_{\tilde f'_k})\nonumber\\
&+& \sum_{k=1}^4
a^*_{\tilde\chi_k^0\tilde\psi_i^+W}\,a_{\tilde\chi_k^0\tilde\psi_j^+W}
\ {\cal R}e\,B_1(m_{\tilde\chi_k^0},M_W)\nonumber\\
&+& \sum_{k=1}^2
a^*_{\tilde\psi_i^+\tilde\chi_k^+Z}\,a_{\tilde\psi_j^+\tilde\chi_k^+Z}
\ {\cal R}e\,B_1(m_{\tilde\chi_k^+},M_Z)\nonumber\\
&+& \sum_{k=1}^2
a^*_{\tilde\psi_i^+\tilde\chi_k^+\gamma}\,a_{\tilde\psi_j^+
\tilde\chi_k^+\gamma}\ {\cal R}e\,B_1(m_{\tilde\chi_k^+},0)\nonumber\\
&+& {1\over2}\sum_{k=1}^4\sum_{n=1}^2
a^*_{\tilde\chi_k^0\tilde\psi_i^+H_n^+}
\,a_{\tilde\chi_k^0\tilde\psi_j^+H_n^+} \ {\cal
R}e\,B_1(m_{\tilde\chi_k^0},M_{H_n^+})\nonumber\\
&+& {1\over2}\sum_{k=1}^2\sum_{n=1}^4
a^*_{\tilde\psi_i^+\tilde\chi_k^+H_n^0}\,
a_{\tilde\psi_j^+\tilde\chi_k^+H_n^0} \ {\cal
R}e\,B_1(m_{\tilde\chi_k^+},M_{H_n^0})\ ;
\end{eqnarray}
$\Sigma_R^+(p^2)$ is obtained from $\Sigma_L^+(p^2)$ by substituting
$a_{\tilde\psi_i^+\cdots}$ with $b_{\tilde\psi_i^+\cdots}$.
$\Sigma_S^+(p^2)$ is given by the following formula,
\begin{eqnarray}
16\pi^2~{\Sigma_S^+}_{ij}(p^2) &=& \sum_f\sum_{k=1}^2N_c^f\,
b^*_{\tilde\psi_i^+f\tilde f'_k}\, a_{\tilde\psi_j^+f\tilde f'_k}\,m_f
\ {\cal R}e\,B_0(m_f,m_{\tilde f'_k})\nonumber\\
&-& 4\sum_{k=1}^4
b^*_{\tilde\chi_k^0\tilde\psi_i^+W}\,a_{\tilde\chi_k^0\tilde\psi_j^+W}\,
m_{\tilde\chi_k^0}\ {\cal R}e\,B_0(m_{\tilde\chi_k^0},M_W)\nonumber\\
&-& 4\sum_{k=1}^2
b^*_{\tilde\psi_i^+\tilde\chi_k^+Z}\,a_{\tilde\psi_j^+\tilde\chi_k^+Z}\,
m_{\tilde\chi_k^+}\ {\cal R}e\,B_0(m_{\tilde\chi_k^+},M_Z)\nonumber\\
&-& 4\sum_{k=1}^2
b^*_{\tilde\psi_i^+\tilde\chi_k^+\gamma}\,
a_{\tilde\psi_j^+\tilde\chi_k^+\gamma}\, m_{\tilde\chi_k^+}\ {\cal
R}e\,B_0(m_{\tilde\chi_k^+},0)\nonumber\\
&+& \sum_{k=1}^4\sum_{n=1}^2 b^*_{\tilde\chi_k^0\tilde\psi_i^+H_n^+}\,
a_{\tilde\chi_k^0\tilde\psi_j^+H_n^+}\,m_{\tilde\chi_k^0} \ {\cal
R}e\,B_0(m_{\tilde\chi_k^0},M_{H_n^+})\nonumber\\
&+& \sum_{k=1}^2\sum_{n=1}^4 b^*_{\tilde\psi_i^+\tilde\chi_k^+H_n^0} \,
a_{\tilde\psi_j^+\tilde\chi_k^+H_n^0}\, m_{\tilde\chi_k^+}\ {\cal
R}e\,B_0(m_{\tilde\chi_k^+},M_{H_n^0})\ .
\end{eqnarray}

In these expressions, the color factor $N_c^f$ is 3 for (s)quarks, and
1 for (s)leptons.  The $\tilde\psi f\tilde f$ couplings are listed in
Eqs.~(\ref{NQS}--\ref{NQS s}), and the $\tilde\psi\tilde\chi Z$ and
$\tilde\psi\tilde\chi W$ couplings are given in
Eqs.~(\ref{NNZ}--\ref{NNZ N}, \ \ref{NCW}--\ref{NCW C}).  We determine
the $\tilde\psi^+\tilde\chi^+\gamma$ couplings from the following
equations, which apply for {\em incoming} $\tilde\chi_i^+$,
\begin{equation}
a_{\tilde\chi_i^+\tilde\chi_j^+\gamma} \ =\ eV_{ik}^*V_{jk}\ =
\ e\,\delta_{ij} ~,~~~~ b_{\tilde\chi_i^+\tilde\chi_j^+\gamma} \ =
\ eU_{ik} U_{jk}^*\ =\ e\,\delta_{ij}~,
\end{equation}
where we write the chargino-chargino-photon Feynman rule as
$-i\gamma_\mu(a{\cal P}_L+b{\cal P}_R).$

We next list the $\tilde\chi\tilde\chi$-Higgs-boson couplings.  We
write these couplings in the unrotated Higgs basis $(s_1,\,s_2)$,
$(p_1,\,p_2)$, and $(h_1^+,\,h_2^+)$.  These fields are rotated to
obtain the mass eigenstate fields.  The $(H,\,h)$ rotation is given in
Eq.~(\ref{rotate h}), while for $(G^0,\,A)$ and $(G^+,\,H^+)$ we have
\begin{equation}
\left(\begin{array}{c}G^0\\[2mm]A\end{array}\right) \ =
\ \left(\begin{array}{cc} c_\beta & s_\beta\\[2mm]-s_\beta & c_\beta
\end{array}\right)\left(\begin{array}{c}p_1\\[2mm]p_2\end{array}\right)~,
\qquad\qquad\left(\begin{array}{c}G^+\\[2mm]H^+\end{array}\right) \ =
\ \left(\begin{array}{cc} c_\beta & s_\beta\\[2mm]-s_\beta & c_\beta
\end{array}\right) \left( \begin{array}{c}h_1^+
\\[2mm]h_2^+\end{array}\right)~.
\end{equation}
We write the Feynman rules for the $\tilde\chi^0\tilde\chi^0s_k$
couplings as $-i(a{\cal P}_L+b{\cal P}_R)$ and for
$\tilde\chi^0\tilde\chi^0p_k$ as $(a{\cal P}_L+b{\cal P}_R)$.  These
couplings are symmetric under $i\leftrightarrow j$ and satisfy
$b_{\tilde\psi_i^0\tilde\psi_j^0s_k} =
a_{\tilde\psi_i^0\tilde\psi_j^0s_k}$ and
$b_{\tilde\psi_i^0\tilde\psi_j^0p_k} =
-a_{\tilde\psi_i^0\tilde\psi_j^0p_k}$.  The nonvanishing $a$-couplings
are
\begin{eqnarray}
\label{NNS}
&& -\ a_{\tilde\psi_1^0\tilde\psi_3^0s_1} \ =
\ a_{\tilde\psi_1^0\tilde\psi_4^0s_2} \ = \ {g'\over2}~, {}~~~~
a_{\tilde\psi_2^0\tilde\psi_3^0s_1} \ =\ -
\ a_{\tilde\psi_2^0\tilde\psi_4^0s_2} \ = {g\over2}~,\\[2mm] &&
a_{\tilde\psi_1^0\tilde\psi_3^0p_1} \ =
\ a_{\tilde\psi_1^0\tilde\psi_4^0p_2} \ =\ - \ {g'\over2}~, {}~~~~
a_{\tilde\psi_2^0\tilde\psi_3^0p_1} \ =
\ a_{\tilde\psi_2^0\tilde\psi_4^0p_2} \ =\ {g\over2}~.
\label{NNP}
\end{eqnarray}
The couplings to {\em incoming} neutralino mass eigenstates
$\tilde\chi_i^0$ are
\begin{equation}
\label{NNS N}
a_{\tilde\chi_i^0\tilde\chi_j^0s_n} \ =\ N_{ik}^*\,N_{jl}^*\,
a_{\tilde\psi_k^0\tilde\psi_l^0s_n}~,~~~~
b_{\tilde\chi_i^0\tilde\chi_j^0s_n} \ =\ N_{ik} \,N_{jl} \,
b_{\tilde\psi_k^0\tilde\psi_l^0s_n}~,
\end{equation}
and likewise for $p_n$ couplings.  The couplings to Higgs-boson mass
eigenstates are found by rotating these couplings,
\begin{equation}
\left(\begin{array}{c} a_{\tilde\chi^0\tilde\chi^0H} \\[2mm]
a_{\tilde\chi^0\tilde\chi^0h} \end{array}\right) \ =
\ \left(\begin{array}{cc}c_\alpha & s_\alpha\\[2mm] -s_\alpha &
c_\alpha\end{array}\right)\left(\begin{array}{c}
a_{\tilde\chi^0\tilde\chi^0s_1}\\[2mm]
a_{\tilde\chi^0\tilde\chi^0s_2}\end{array}\right)~,~~~~
\label{NNS A}
\left(\begin{array}{c} a_{\tilde\chi^0\tilde\chi^0G^0}\\[2mm]
a_{\tilde\chi^0\tilde\chi^0A} \end{array}\right) \ =
\ \left(\begin{array}{cc}c_\beta & s_\beta\\[2mm] -s_\beta &
c_\beta\end{array}\right)\left(\begin{array}{c}
a_{\tilde\chi^0\tilde\chi^0p_1} \\[2mm]
a_{\tilde\chi^0\tilde\chi^0p_2}\end{array}\right)~,
\end{equation}
and likewise for the $b$-couplings.

We write the Feynman rules for the
$\tilde\chi^+\tilde\chi^+$-neutral-Higgs couplings as $-i(a{\cal
P}_L+b{\cal P}_R)$ for couplings with CP-even $s$-fields, and $(a{\cal
P}_L+b{\cal P}_R)$ for couplings with CP-odd $p$-fields.  These
couplings satisfy $b_{\tilde\psi_i^+\tilde\psi_j^+s_n} =
a_{\tilde\psi_j^+\tilde\psi_i^+s_n}$ and
$b_{\tilde\psi_i^+\tilde\psi_j^+p_n} =
-a_{\tilde\psi_j^+\tilde\psi_i^+p_n}$. The nonzero $a$-couplings are
\begin{equation}
\label{CCH}
  a_{\tilde\psi_1^+\tilde\psi_2^+s_1} \ =
\ a_{\tilde\psi_2^+\tilde\psi_1^+s_2} \ =
\ a_{\tilde\psi_1^+\tilde\psi_2^+p_1} \ = \ -
\ a_{\tilde\psi_2^+\tilde\psi_1^+p_2} \ =\ {g\over\sqrt2}~.
\end{equation}
The couplings to {\em incoming} $\tilde\chi_i^+$ are obtained from
these as follows,
\begin{equation}
\label{CCH U}
a_{\tilde\chi_i^+\tilde\chi_j^+s_n} \ =\ V_{ik}^*\,U_{jl}^*\,
a_{\tilde\psi_k^+\tilde\psi_l^+s_n}~,~~~~
b_{\tilde\chi_i^+\tilde\chi_j^+s_n} \ =\ U_{ik} \,V_{jl} \,
b_{\tilde\psi_k^+\tilde\psi_l^+s_n}~,
\end{equation}
and the same rotations apply for the $p_n$-couplings.  To find the
couplings to Higgs-boson mass eigenstates, we rotate these couplings
by the angle $\alpha$ or $\beta$, just as for the
$\tilde\chi^0\tilde\chi^0s$ and $\tilde\chi^0\tilde\chi^0p$ couplings
in Eq.~(\ref{NNS A}).

The $\tilde\chi^0\tilde\chi^+$--charged-Higgs-boson vertex Feynman
rules are written $-i(a{\cal P}_L+b{\cal P}_R)$, where, for {\em
incoming} $\tilde\psi_i^0$, we have
\begin{equation}
\label{NCH}
     a_{\tilde\psi_1^0\tilde\psi_2^+h_1^+} \ =
\ b_{\tilde\psi_1^0\tilde\psi_2^+h_2^+} \ =\ {g'\over\sqrt2}~, {}~~~~
a_{\tilde\psi_2^0\tilde\psi_2^+h_1^+} \ =
\ b_{\tilde\psi_2^0\tilde\psi_2^+h_2^+} \ =\ {g\over\sqrt2}~, {}~~~~
a_{\tilde\psi_3^0\tilde\psi_1^+h_1^+} \ =
\ -b_{\tilde\psi_4^0\tilde\psi_1^+h_2^+} \ =\ -\ g~.
\end{equation}
To obtain the couplings to chargino and neutralino mass eigenstates
with an {\em incoming} neutralino $\tilde\chi_i^0$, we rotate these
couplings as
\begin{equation}
\label{NCH N}
a_{\tilde\chi_i^0\tilde\chi_j^+h_n^+} \ =
\ N_{ik}^* \, U_{jl}^* \, a_{\tilde\psi_k^0\tilde\psi_l^+h_n^+}~,~~~~
b_{\tilde\chi_i^0\tilde\chi_j^+h_n^+} \ =\ N_{ik} \, V_{jl} \,
b_{\tilde\psi_k^0\tilde\psi_l^+h_n^+}~,
\end{equation}
while for an {\em incoming} chargino $\tilde\chi_j^+,$ we rotate them
as
\begin{equation}
a_{\tilde\chi_i^0\tilde\chi_j^+h_n^+} \ =
\ N_{ik}^* \, V_{jl}^* \, b_{\tilde\psi_k^0\tilde\psi_l^+h_n^+}~,~~~~
b_{\tilde\chi_i^0\tilde\chi_j^+h_n^+} \ =\ N_{ik} \, U_{jl} \,
a_{\tilde\psi_k^0\tilde\psi_l^+h_n^+}~.
\end{equation}
To find the couplings to charged-Higgs mass eigenstates, we rotate
both $a$- and $b$-couplings by the angle $\beta$,
\begin{equation}
\label{NCH B}
\left(\begin{array}{c} a_{\tilde\chi^0\tilde\chi^+G^+} \\[2mm]
a_{\tilde\chi^0\tilde\chi^+H^+}
\end{array}\right) \ =\ \left(\begin{array}{cc}c_\beta & s_\beta\\[2mm]
-s_\beta & c_\beta\end{array}\right)\left(\begin{array}{c}
a_{\tilde\chi^0\tilde\chi^+h_1^+} \\[2mm]
a_{\tilde\chi^0\tilde\chi^+h_2^+}\end{array}\right)~.
\end{equation}

\vspace{.5cm} {\noindent\large\bf Gluino}
\vspace{.5cm}

The gluino self-energy appears in Refs.~\cite{MV,PP2,Kras,Donini}.
The physical gluino mass satisfies
\begin{equation}
m_{\tilde g}\ =\ M_3(Q)\ -\ {\cal R}e\,\Sigma_{\tilde g}(m_{\tilde
g}^2)\ ,
\end{equation}
where
\begin{eqnarray}
\Sigma_{\tilde g}(p^2) &=& {g_3^2\over16\pi^2}\Biggl\{ -\ m_{\tilde g}
\left(15 + 9\ln{Q^2\over m^2_{\tilde g}}\right)\ +\ \sum_q\sum_{i=1}^2
m_{\tilde g}B_1(m_q,m_{\tilde q_i})\nonumber\\ && \qquad +\ \sum_q m_q
s_{2\theta_q} \biggl[ B_0(m_q, m_{\tilde q_1})- B_0(m_q,m_{\tilde
q_2})\biggr]\Biggr\}\ ,
\end{eqnarray}
where $Q$ is the renormalization scale.

\vspace{.5cm}
\noindent{\large\bf Squarks and sleptons}
\vspace{.5cm}

We find the sfermion masses by taking the real part of the poles of
the propagator matrix
\begin{equation}
{\rm Det}\biggl[~p_i^2~-~{\cal M}^2_{\tilde f}(p_i^2)~\biggr]\ =\ 0~,
\qquad m_{\tilde f_i}^2 = {\cal R}e(p_i^2)\ ,
\end{equation}
where
\begin{equation}
{\cal M}_{\tilde f}^2(p^2)\ =\ \left(\begin{array}{cc} M_{\tilde
f_L\tilde f_L}^2-\Pi_{\tilde f_L\tilde f_L}(p^2) & \qquad
M_{\tilde f_L\tilde f_R}^2-\Pi_{\tilde f_L\tilde f_R}(p^2)
\\[2mm] M_{\tilde f_R\tilde f_L}^2-\Pi_{\tilde f_R\tilde
f_L}(p^2) & \qquad M_{\tilde f_R\tilde f_R}^2-\Pi_{\tilde
f_R\tilde f_R}(p^2)
\end{array}\right)~.
\end{equation}
The matrix formalism allows us to determine the one-loop
masses and mixing angles for arbitrary tree-level parameters.
In this expression,
the $M_{\tilde f_i\tilde f_j}^2,~(i,j=L,R)$ are the
\mbox{\footnotesize$\overline{\rm DR}~$} tree-level mass matrix
entries given in Eqs.~(\ref{sqm u},\ \ref{sqm d}): all the entries
contain running \mbox{\footnotesize$\overline{\rm DR}~$} parameters at
a common scale $Q$.  In particular, the
\mbox{\footnotesize$\overline{\rm DR}~$} tree-level matrix contains
corrections from the replacements $M_Z^2\rightarrow\hat M_Z^2 = M_Z^2
+ {\cal R}e\,\Pi^T_{ZZ}(M_Z^2)$ and $m_f\rightarrow\hat m_f = m_f +
{\cal R}e\,\Sigma_f(m_f^2)$.  (The arguments of these self-energy
functions are external momenta, {\em not} the scale $Q$.) The
$\Pi_{\tilde f_i\tilde f_j},~(i,j=L,R)$ are the sfermion self-energy
functions evaluated at the scale $Q$. Of course, for the first two
generations of sfermions, both the tree-level and one-loop
contributions to the off-diagonal elements of the mass matrices are
negligible. Note $\Pi_{\tilde f_R\tilde f_L}\ne
\Pi^*_{\tilde f_L\tilde f_R}$ because of the absorptive part,
which contributes to the mass-squared at ${\cal O}(\alpha^2)$.

For a $\tilde t_L$ squark we have
\begin{eqnarray}
&& 16\pi^2~\Pi_{\tilde t_L\tilde t_L}(p^2)\nonumber\\ &=&
{4g_3^2\over3} \biggl[\, 2G(m_{\tilde g},m_t) +c_t^2 F(m_{{\tilde
t}_1},0) + s_t^2 F(m_{{\tilde t}_2},0) +c_t^2 A_0(m_{{\tilde t}_1}) +
s_t^2 A_0(m_{{\tilde t}_2}) \,\biggr]\nonumber\\ &+& \lambda_t^2
\biggl( s_t^2 A_0(m_{{\tilde t}_1}) + c_t^2 A_0(m_{{\tilde t}_2})
\biggr)\ + \ \lambda_b^2 \biggl( s_b^2 A_0(m_{{\tilde
b}_1})+c_b^2A_0(m_{{\tilde b}_2}) \biggr) \nonumber\\ &+&
{1\over2}\sum_{n=1}^4 \biggl(\lambda_t^2\,D_{nu} - {g^2
g_{t_L}\over2\hat c^2} C_n\biggr) A_0(m_{H^0_n}) \ +\ \sum_{n=3}^4
\biggl(\lambda_b^2D_{nu}+g^2\left( {g_{t_L}\over2\hat
c^2}-I_3^t\right)C_n\biggr) A_0(m_{H^+_{n-2}}) \nonumber\\ &+&
\sum_{n=1}^4\sum_{i=1}^2 (\lambda_{H^0_n{\tilde t}_L{\tilde t}_i})^2
B_0(m_{H^0_n},m_{{\tilde t}_i}) \ +\ \sum_{i,n=1}^2
(\lambda_{H^+_n{\tilde t}_L{\tilde b}_i})^2 B_0(m_{{\tilde
b}_i},m_{H^+_n}) \nonumber\\ &+& \frac{4g^2}{\hat c^2} (g_{t_L})^2
A_0(M_Z) + 2 g^2 A_0(M_W)\ +\ (e_te)^2 \biggl( c_t^2 F(m_{{\tilde
t}_1},0) + s_t^2 F(m_{{\tilde t}_2},0) \biggr) \nonumber \\ &+&
\frac{g^2}{\hat c^2} (g_{t_L})^2 \biggl[\, c_t^2 F(m_{{\tilde
t}_1},M_Z) + s_t^2 F(m_{{\tilde t}_2},M_Z) \,\biggr] \ +\ {g^2\over2}
\biggl[\, c_b^2 F(m_{{\tilde b}_1},M_W) + s_b^2 F(m_{{\tilde
b}_2},M_W) \,\biggr] \nonumber \\ &+& ~{g^2\over4} \biggl[\, c_t^2
A_0(m_{{\tilde t}_1}) + s_t^2 A_0(m_{{\tilde t}_2}) + 2\biggl( c_b^2
A_0(m_{{\tilde b}_1}) + s_b^2 A_0(m_{{\tilde b}_2}) \biggr) \,\biggr]
\nonumber \\ &+& g^2\sum_{f} N_c^f I_3^t I_3^f \biggl( c_f^2
A_0(m_{{\tilde f}_1}) +s_f^2 A_0(m_{{\tilde f}_2}) \biggr) \ +
\ {g'^2\over4} (Y_{t_L})^2 \biggl( c_t^2 A_0(m_{{\tilde t}_1}) + s_t^2
A_0(m_{{\tilde t}_2}) \biggr) \nonumber\\ &+& {g'^2\over4} Y_{t_L}
\sum_{f} N_c^f \biggl[\, Y_{f_L} \biggl( c_f^2 A_0(m_{{\tilde f}_1})
+s_f^2 A_0(m_{{\tilde f}_2}) \biggr) \ +\ Y_{f_R} \biggl( s_f^2
A_0(m_{{\tilde f}_1}) +c_f^2 A_0(m_{{\tilde f}_2})
\biggr)\,\biggr]\nonumber\\ &+& \sum_{i=1}^4 \biggl[\, f_{it\tilde
t_{LL}} G(m_{\tilde\chi_i^0},m_t) - 2\,g_{it\tilde t_{LL}}\,
m_{\tilde\chi_i^0}m_t B_0(m_{\tilde\chi_i^0},m_t) \,\biggr]
\nonumber\\ &+& \sum_{i=1}^2 \biggl[\, f_{ib\tilde t_{LL}}
G(m_{\tilde\chi_i^+},m_b) - 2\,g_{ib\tilde t_{LL}}\,
m_{\tilde\chi_i^+}m_b B_0(m_{\tilde\chi_i^+},m_b)\biggr]~,
\end{eqnarray}
and similarly for a ${\tilde t}_R$ squark,
\begin{eqnarray}
&& 16\pi^2~\Pi_{\tilde t_R\tilde t_R}(p^2) \nonumber\\ &=&
{4g_3^2\over3} \biggl[\, 2G(m_{\tilde g},m_t) +s_t^2 F(m_{{\tilde
t}_1},0) + c_t^2 F(m_{{\tilde t}_2},0) +s_t^2 A_0(m_{{\tilde t}_1}) +
c_t^2 A_0(m_{{\tilde t}_2}) \,\biggr] \nonumber \\ &+& \lambda_t^2
\biggl( c_t^2 A_0(m_{{\tilde t}_1}) + s_t^2 A_0(m_{{\tilde t}_2}) + c_b^2
A_0(m_{{\tilde b}_1}) + s_b^2 A_0(m_{{\tilde b}_2}) \biggr) \nonumber\\ &+&
{1\over2}\sum_{n=1}^4 \biggl( \lambda_t^2\,D_{nu}
-{g^2g_{t_R}\over2\hat c^2} C_n \biggr) A_0(m_{H^0_n}) \ +\ \sum_{n=3}^4
\biggl(\lambda_t^2D_{nd}+{g^2g_{t_R}\over2\hat c^2}
C_n\biggr)A_0(m_{H^+_{n-2}})\nonumber\\ &+& \sum_{n=1}^4\sum_{i=1}^2
(\lambda_{H^0_n{\tilde t}_R{\tilde t}_i})^2 B_0(m_{H^0_n},m_{{\tilde
t}_i}) \ +\ \sum_{i,n=1}^2 (\lambda_{H^+_n{\tilde t}_R{\tilde b}_i})^2
B_0(m_{{\tilde b}_i},m_{H^+_n}) \nonumber\\ &+& \frac{4g^2}{\hat c^2}
(g_{t_R})^2 A_0(M_Z) +(e_te)^2 \biggl( s_t^2 F(m_{{\tilde t}_1},0) +
c_t^2 F(m_{{\tilde t}_2},0) \biggr) \nonumber \\ &+& \frac{g^2}{\hat
c^2} (g_{t_R})^2 \biggl[\, s_t^2 F(m_{{\tilde t}_1},M_Z) + c_t^2
F(m_{{\tilde t}_2},M_Z) \,\biggr] \nonumber \\ &+& {g'^2\over4}
(Y_{t_R})^2 \biggl( s_t^2 A_0(m_{{\tilde t}_1}) + c_t^2 A_0(m_{{\tilde
t}_2}) \biggr) \nonumber\\ &+& {g'^2\over4} Y_{t_R} \sum_{f} N_c^f
\biggl[\, Y_{f_L} \biggl( c_f^2 A_0(m_{{\tilde f}_1}) +s_f^2
A_0(m_{{\tilde f}_2}) \biggr) \ +\ Y_{f_R} \biggl( s_f^2 A_0(m_{{\tilde
f}_1}) +c_f^2 A_0(m_{{\tilde f}_2}) \biggr) \,\biggr]\nonumber\\ &+&
\sum_{i=1}^4 \biggl[\, f_{it\tilde t_{RR}} G(m_{\tilde\chi_i^0},m_t) -
2\,g_{it\tilde t_{RR}}\, m_{\tilde\chi_i^0}m_t
B_0(m_{\tilde\chi_i^0},m_t) \,\biggr] \nonumber\\ &+& \sum_{i=1}^2
\biggl[\, f_{ib\tilde t_{RR}} G(m_{\tilde\chi_i^+},m_b) -
2\,g_{ib\tilde t_{RR}}\, m_{\tilde\chi_i^+}m_b
B_0(m_{\tilde\chi_i^+},m_b)\biggr]~.
\end{eqnarray}
The off-diagonal self-energy is
\begin{eqnarray}
&& 16\pi^2~\Pi_{\tilde t_L\tilde t_R}(p^2) \nonumber\\ &=& \frac{4
g_3^2}{3} \biggl[\, 4 m_{\tilde g} m_t B_0(m_{\tilde g},m_t) +s_t c_t
\biggl( F(m_{{\tilde t}_1},0) - F(m_{{\tilde t}_2},0) - A_0(m_{{\tilde
t}_1}) + A_0(m_{{\tilde t}_2}) \biggr) \,\biggr] \nonumber \\ &+&
\sum_{n=1}^4\sum_{i=1}^2 \lambda_{H^0_n\tilde t_L\tilde t_i}
\lambda_{H^0_n\tilde t_R\tilde t_i} B_0(m_{H^0_n},m_{{\tilde t}_i})
\ +\ \sum_{i,n=1}^2 \lambda_{H^+_n\tilde t_L\tilde b_i}
\lambda_{H^+_n\tilde t_R\tilde b_i} B_0(m_{\tilde
b_i},m_{H^+_n})\nonumber\\ &+& {\lambda_t\over2} \sum_{f_u} N_c^f
\lambda_u s_{2\theta_u} \biggl( A_0(m_{\tilde u_1}) - A_0(m_{\tilde u_2})
\biggr) \ +\ {g'^2\over4} Y_{t_L} Y_{t_R} s_t c_t \biggl( A_0(m_{{\tilde
t}_1}) - A_0(m_{{\tilde t}_2}) \biggr) \nonumber \\ &+& (e_te)^2 s_t c_t
\biggl( F(m_{{\tilde t}_1},0) - F(m_{{\tilde t}_2},0) \biggr) \ -
\ \frac{g^2}{\hat c^2} g_{t_L} g_{t_R} s_t c_t \biggl( F(m_{{\tilde
t}_1},M_Z) - F(m_{{\tilde t}_2},M_Z) \biggr) \nonumber \\ &+&
\sum_{i=1}^4 \biggl[\, f_{it\tilde t_{LR}} G(m_{\tilde\chi_i^0},m_t) -
2\,g_{it\tilde t_{LR}}\, m_{\tilde\chi_i^0}m_t
B_0(m_{\tilde\chi_i^0},m_t) \,\biggr] \nonumber\\ &+& \sum_{i=1}^2
\biggl[\, f_{ib\tilde t_{LR}} G(m_{\tilde\chi_i^+},m_b) -
2\,g_{ib\tilde t_{LR}}\, m_{\tilde\chi_i^+}m_b
B_0(m_{\tilde\chi_i^+},m_b)\biggr]~.
\end{eqnarray}
Inside the sum $\sum_f$, the sub- or superscript $f$ refers to
(s)quarks and (s)leptons, and in the sum $\sum_{f_u}$, the sub- or
superscript $u$ refers to up-type (s)quarks and (s)leptons.  The $g_f$
are defined in Eq.~(\ref{glgr}).  The electric charges $e_f$,
hypercharges $Y_f$ and third component of isospin $I_3^f$ are given in
Eq.~(\ref{table1}).  These results are equivalent to those of
Ref.~\cite{Donini} in the limit $g,\ g',$ and $\lambda_b \rightarrow
0$.

For the $\tilde\chi f\tilde f$ couplings, we have defined
\begin{equation}
f_{it\tilde t_{LR}} \ =\ a_{\tilde\chi_i^0t\tilde
t_L}^*\,a_{\tilde\chi_i^0t\tilde t_R} + b_{\tilde\chi_i^0t\tilde
t_L}^*\,b_{\tilde\chi_i^0t\tilde t_R}~,~~~~ g_{it\tilde t_{LR}} \ =
\ b^*_{\tilde\chi_i^0t\tilde t_L}\,a_{\tilde\chi_i^0t\tilde t_R}
+ a^*_{\tilde\chi_i^0t\tilde t_L}\,b_{\tilde\chi_i^0t\tilde t_R}~,
\end{equation}
\begin{equation}
f_{ib\tilde t_{LR}} \ =\ a_{\tilde\chi_i^+b\tilde
t_L}^*\,a_{\tilde\chi_i^+b\tilde t_R} + b_{\tilde\chi_i^+b\tilde
t_L}^*\,b_{\tilde\chi_i^+b\tilde t_R}~,~~~~ g_{ib\tilde t_{LR}} \ =
\ b^*_{\tilde\chi_i^+b\tilde t_L}\,a_{\tilde\chi_i^+b\tilde t_R}
+ a^*_{\tilde\chi_i^+b\tilde t_L}\,b_{\tilde\chi_i^+b\tilde t_R}~,
\end{equation}
with analogous definitions for the $LL$ and $RR$ couplings.  The
$\tilde\chi f\tilde f$ couplings are listed in
Eqs.~(\ref{NQS}--\ref{CQS}).

The Higgs bosons $H^0_n$ refer to $H,\ h,\ G^0$, and $A$, and
$H_n^+$ refer to $H^+,\ G^+$. The $H_n$-$H_n$-sfermion-sfermion
couplings involve $C_n$ and $D_{nf}$, and are given
in the following table,
\begin{equation}
\begin{array}{cccc}
n&~~~C_n&~~~D_{nu}&~~~D_{nd} \\ [1mm] \hline \\
1&~~~-c_{2\alpha}&~~~s_\alpha^2&~~~c_\alpha^2 \\ [2mm]
2&~~~~c_{2\alpha}&~~~c_\alpha^2&~~~s_\alpha^2 \\ [2mm]
3&~~~-c_{2\beta}&~~~s_\beta^2&~~~c_\beta^2 \\[2mm]
4&~~~~c_{2\beta}&~~~c_\beta^2&~~~s_\beta^2 \\ [2mm] \hline
\end{array}
\end{equation}

We write the Feynman rules associated with the
CP-even-Higgs-sfermion-sfermion vertices as $-i\lambda$, and list the
couplings $\lambda_{s\tilde f\tilde f}$ in the following table,
\vspace{.4cm}
\begin{equation}
\begin{array}{c|cc}
& \qquad\qquad\qquad s_1 \qquad\qquad\qquad & \qquad\qquad\qquad s_2
\qquad\qquad\qquad \\ \hline\\ \tilde u_L\tilde u_L & {gM_Z\over \hat
c}g_{u_L}c_\beta & -{gM_Z\over \hat
c}g_{u_L}s_\beta+\sqrt2\lambda_um_u \\[2mm] \tilde u_R\tilde u_R &
{gM_Z\over \hat c}g_{u_R}c_\beta & -{gM_Z\over \hat
c}g_{u_R}s_\beta+\sqrt2\lambda_um_u \\[2mm] \tilde u_L\tilde u_R &
{\lambda_u\over\sqrt2}\mu & {\lambda_u\over\sqrt2}A_u \\[2mm]\hline\\
\tilde d_L\tilde d_L & {gM_Z\over \hat
c}g_{d_L}c_\beta+\sqrt2\lambda_dm_d & -{gM_Z\over\hat
c}g_{d_L}s_\beta\\[2mm] \tilde d_R\tilde d_R & {gM_Z\over \hat
c}g_{d_R}c_\beta+\sqrt2\lambda_dm_d & -{gM_Z\over\hat
c}g_{d_R}s_\beta\\[2mm] \tilde d_L\tilde d_R &
{\lambda_d\over\sqrt2}A_d & {\lambda_d\over\sqrt2}\mu \\[2mm]
\end{array}\label{H1H2}
\end{equation}
We find the couplings in the $\tilde f_{1,2}$ sfermion basis via
\begin{equation}
\label{sf rot}
\left( \begin{array}{cc} \lambda_{s_n\tilde f_1\tilde f_1}&~
\lambda_{s_n\tilde f_1\tilde f_2} \\[2mm] \lambda_{s_n\tilde f_2\tilde
f_1}&~ \lambda_{s_n\tilde f_2\tilde f_2}
\end{array}  \right) \ =
\ \left( \begin{array}{cc} {}~c_f & ~s_f \\ [2mm] -s_f & ~c_f
\end{array}  \right)
\left( \begin{array}{cc} \lambda_{s_n\tilde f_L\tilde f_L}&~
\lambda_{s_n\tilde f_L\tilde f_R} \\[2mm] \lambda_{s_n\tilde f_R\tilde
f_L}&~ \lambda_{s_n\tilde f_R\tilde f_R}
\end{array}\right)
\left( \begin{array}{cc} c_f & ~-s_f \\ [2mm] s_f & ~c_f
\end{array}  \right)\ ;
\end{equation}
we obtain the couplings in the mixed $\tilde f_{L,R}\tilde f_{1,2}$
basis by omitting the left-most matrix on the right hand side of the
above equation.  The couplings to the CP-even Higgs-boson eigenstates
$(H,\,h)$ are obtained from the couplings to $(s_1,\,s_2)$ using the
rotation
\begin{equation}
 \label{a rot} \left(\begin{array}{c} \lambda_{H\tilde f_i\tilde f_j}
\\[2mm] \lambda_{h\tilde f_i\tilde f_j}
\end{array}\right) \ =\ \left(\begin{array}{cc}c_\alpha & s_\alpha \\[2mm]
-s_\alpha & c_\alpha\end{array}\right)\left(\begin{array}{c}
\lambda_{s_1\tilde f_i\tilde f_j} \\[2mm]\lambda_{s_2\tilde f_i\tilde
f_j}
\end{array}\right)~.
\end{equation}

The couplings $\lambda_{G^0\tilde f_i\tilde f_j}$ and
$\lambda_{A\tilde f_i\tilde f_j}$ vanish for $i=j$, while for $i\ne j$
they satisfy $\lambda_{A\tilde f_i\tilde f_j}=-\lambda_{A\tilde
f_j\tilde f_i}$.  We write the Feynman rules for these couplings for
incoming $\tilde f_L$ as $\lambda$. They are
\vspace{2mm}
\begin{equation}
\begin{array}{c|cc}
       & \qquad\qquad\qquad G^0 \qquad\qquad\qquad &
 \qquad\qquad\qquad A \qquad\qquad\qquad \\ \hline\\ \tilde u_L\tilde
 u_R & {\lambda_u\over\sqrt2}\left(\mu c_\beta + A_us_\beta\right) &
 -{\lambda_u\over\sqrt2}\left(\mu s_\beta - A_uc_\beta\right) \\[2mm]
 \tilde d_L\tilde d_R & -{\lambda_d\over\sqrt2}\left(\mu s_\beta +
 A_dc_\beta\right) & -{\lambda_d\over\sqrt2}\left(\mu c_\beta -
 A_ds_\beta\right) \\[2mm]
\end{array}\label{AG}
\end{equation}
We obtain these couplings in the $\tilde f_{L,R}\tilde f_{1,2}$ basis
by a rotation as described after Eq.~(\ref{sf rot}).

We also write the Feynman rules for the
charged-Higgs-sfermion-sfermion vertices in the form $-i\lambda$.  The
couplings $\lambda_{G^+\tilde f_i\tilde f'_j}$ and $\lambda_{H^+\tilde
f_i\tilde f'_j}$ are
\begin{equation}
\begin{array}{c|cc}
       & G^+ & H^+ \\ \hline\\ \tilde u_L\tilde d_L &
-{gM_W\over\sqrt2}c_{2\beta} -\lambda_um_us_\beta+\lambda_dm_dc_\beta
\qquad & {gM_W\over\sqrt2}s_{2\beta}
-\lambda_um_uc_\beta-\lambda_dm_ds_\beta\\[2mm] \tilde u_R\tilde d_R &
0 & -\lambda_um_dc_\beta-\lambda_dm_us_\beta \\[2mm] \tilde u_L\tilde
d_R & \lambda_d\left(\mu s_\beta + A_dc_\beta\right) \qquad &
\lambda_d\left(\mu c_\beta - A_ds_\beta\right) \\[2mm] \tilde
u_R\tilde d_L & -\lambda_u\left(\mu c_\beta + A_us_\beta\right) \qquad
& \lambda_u\left(\mu s_\beta - A_uc_\beta\right) \\[2mm]
\end{array}\label{H+G+}
\end{equation}
These couplings are obtained in the $\tilde f_{1,2}$ basis via
\begin{equation}
\label{ud rot}\left(\begin{array}{cc}
\lambda_{H^+\tilde u_1\tilde d_1} & \lambda_{H^+\tilde u_1\tilde
d_2}\\[2mm] \lambda_{H^+\tilde u_2\tilde d_1} & \lambda_{H^+\tilde
u_2\tilde d_2}
\end{array}\right)\ =\ \left(
\begin{array}{cc}c_u & s_u\\[2mm]-s_u & c_u\end{array}\right)\left(
\begin{array}{cc}
\lambda_{H^+\tilde u_L\tilde d_L} & \lambda_{H^+\tilde u_L\tilde
d_R}\\[2mm] \lambda_{H^+\tilde u_R\tilde d_L} & \lambda_{H^+\tilde
u_R\tilde d_R}
\end{array}\right)\left(
\begin{array}{cc}c_d & -s_d\\[2mm]s_d & c_d\end{array}\right)~.
\end{equation}
We obtain the mixed sfermion basis couplings to $\tilde u_{L,R}\tilde
d_{1,2} \ (\tilde u_{1,2}\tilde d_{L,R})$ by leaving off the left-most
(right-most) matrix on the right hand side of the above equation.

The expressions for $\Pi_{\tilde b_i\tilde b_j}$ are obtained from
$\Pi_{\tilde t_i\tilde t_j}$ by interchanging the indices
$t\leftrightarrow b$, replacing $u \rightarrow d$, and substituting
$c_\beta\leftrightarrow s_\beta$. The self-energy of a charged slepton
(sneutrino) is given by a formula similar to that for a $b$-squark
($t$-squark), with the $SU(3)$ correction set to zero and with the
appropriate $SU(2)\times U(1)$ quantum-number substitutions.

\vspace{.5cm}
\noindent{\large\bf Higgs bosons}
\vspace{.5cm}

The full one-loop MSSM Higgs-boson self-energies appear in
Refs.~\cite{CPR}. Corrections to the Higgs boson
masses are the subject of Refs.~\cite{H+,Higgs mass}.
We discuss the relations between the self-energies and the
pole masses of the Higgs bosons in Appendix E. Here we list
the self-energies.

The Higgs-boson contributions to the Higgs-boson self-energies involve
the trilinear and quartic couplings, which we denote
$\lambda_{H_n^0H_m^0s_k}$, $\lambda_{H_n^+H_m^-s_k}$, and
$\lambda_{H_nH_nH_mH_m}$, $\lambda_{H_nH_nH^+_mH^-_m}$, where the
$H^0_n$ refer to the $H,\ h,\ G^0$, and $A$ Higgs bosons, and the
$H^+_n$ refer to the $G^+$ and $H^+$ Higgses. The $G^0$ and $G^+$ are
the neutral and charged Goldstone bosons, which in the
't~Hooft-Feynman gauge have masses $M_Z$ and $M_W$, respectively.

For the two CP-even Higgs bosons, we have
%
\begin{eqnarray}
16\pi^2~\Pi_{s_1s_1}(p^2)&=& \sum_{f_d}N_c^f\lambda_d^2 \Biggl[\, (p^2
- 4 m_d^2)B_0(m_d,m_d) - 2A_0(m_d) + A_0(m_{\tilde d_1})+A_0(m_{\tilde
d_2})\, \Biggr]\nonumber\\ &+& \sum_f{N_c^f g^2\over2 \hat
c^2}\Biggl[\, g_{f_L} \biggl(c_f^2 A_0(m_{\tilde f_1})+ s_f^2
A_0(m_{\tilde f_2})\biggr) \ +\ g_{f_R} \biggl(s_f^2 A_0(m_{\tilde f_1}) +
c_f^2 A_0(m_{\tilde f_2})\biggr)\,\Biggr]\nonumber\\ &+&
\sum_f\sum_{i,j=1}^2N_c^f\lambda_{s_1\tilde f_i\tilde f_j}^2
B_0(m_{\tilde f_i},m_{\tilde f_j}) \nonumber \\ &+&
{g^2\over4}\Biggl\{\,s^2_\beta\biggl[\,2F(m_{H^+},M_W)+{F(m_A,M_Z)
\over\hat c^2}\,\biggr] \nonumber \\ &&\qquad\qquad +
\ c^2_\beta\biggl[\,2F(M_W,M_W)+{F(M_Z,M_Z)\over \hat
c^2}\,\biggr]\,\Biggr\} \nonumber\\ &+& {7\over4}g^2c^2_\beta\biggl[\,
2 M^2_W B_0(M_W,M_W) +{M^2_Z B_0(M_Z,M_Z)\over\hat c^2}\,\biggr]
\nonumber\\ &+& g^2\biggl[\,2A_0(M_W) + {A_0(M_Z)\over\hat c^2}\,\biggr]
\nonumber\\ &+& {1\over2}\sum_{n=1}^4\biggl[\,\sum_{m=1}^4
\lambda_{H^0_nH^0_ms_1}^2B_0(m_{H^0_n},m_{H^0_m}) \ +
\ \lambda_{H^0_nH^0_ns_1s_1}A_0(m_{H^0_n})\,\biggr]\nonumber\\ &+&
\sum_{n=1}^2\biggl[\,\sum_{m=1}^2
\lambda^2_{H_n^+H_m^-s_1}B_0(m_{H_n^+},m_{H_m^+}) \ +
\ \lambda_{H^+_nH^-_ns_1s_1}A_0(m_{H^+_n})\,\biggr]\nonumber\\ &+&
{1\over2}\sum_{i,j=1}^4\biggl[\,f_{ijs_{11}}^0
G(m_{\tilde\chi^0_i},m_{\tilde\chi^0_j}) \ -
\ 2\,g_{ijs_{11}}^0\,m_{\tilde\chi^0_i}
m_{\tilde\chi^0_j}B_0(m_{\tilde\chi^0_i},m_{\tilde\chi^0_j})\,\biggr]
\nonumber\\ &+& \sum_{i,j=1}^2\biggl[\,f_{ijs_{11}}^+
G(m_{\tilde\chi^+_i},m_{\tilde\chi^+_j}) \ -\ 2\,g_{ijs_{11}}^+\,
m_{\tilde\chi^+_i} m_{\tilde\chi^+_j}
B_0(m_{\tilde\chi^+_i},m_{\tilde\chi^+_j})\, \biggr]~,
\label{s1s1}
\end{eqnarray}
%
\begin{eqnarray}
16\pi^2~\Pi_{s_2s_2}(p^2)&=&\sum_{f_u}N_c^f\lambda_u^2 \Biggl[\, (p^2
- 4 m_u^2)B_0(m_u,m_u) -2A_0(m_u) + A_0(m_{\tilde u_1})+ A_0(m_{\tilde
u_2})\,\Biggr] \nonumber\\ &-& \sum_f{N_c^f g^2\over2 \hat
c^2}\Biggl[\, g_{f_L} \biggl(c_f^2 A_0(m_{\tilde f_1}) + s_f^2
A_0(m_{\tilde f_2})\biggr)\ +\ g_{f_R} \biggl(s_f^2 A_0(m_{\tilde f_1}) +
c_f^2 A_0(m_{\tilde f_2})\biggr)\,\Biggr]\nonumber\\ &+&
\sum_f\sum_{i,j=1}^2N_c^f\lambda_{s_2\tilde f_i\tilde f_j}^2
B_0(m_{\tilde f_i},m_{\tilde f_j}) \nonumber\\ &+&
{g^2\over4}\Biggl\{\,c^2_\beta\biggl[\,2F(m_{H^+},M_W)
+{F(m_A,M_Z)\over\hat c^2}\,\biggr] \nonumber\\ && \qquad\qquad+
\ s^2_\beta\biggl[\,2F(M_W,M_W)+ {F(M_Z,M_Z)\over\hat
c^2}\,\biggr]\,\Biggr\} \nonumber\\ &+&
{7\over4}g^2s^2_\beta\biggl[\,2M^2_WB_0(M_W,M_W)
+{M^2_ZB_0(M_Z,M_Z)\over\hat c^2}\,\biggr] \nonumber\\ &+&
g^2\biggl[\,2A_0(M_W)+{A_0(M_Z)\over\hat c^2}\,\biggr] \nonumber\\ &+&
{1\over2}\sum_{n=1}^4\biggl[\,\sum_{m=1}^4 \lambda_{H^0_nH^0_ms_2}^2
B_0(m_{H_n^0},m_{H^0_m})+\lambda_{H^0_nH^0_ns_2s_2}A_0(m_{H_n^0})\,\biggr]
\nonumber\\ &+& \sum_{n=1}^2\biggl[\,\sum_{m=1}^2
\lambda^2_{H_n^+H_m^-s_2}B_0(m_{H_n^+},m_{H_m^+}) \ +
\ \lambda_{H^+_nH^-_ns_2s_2}A_0(m_{H^+_n})\,\biggr]\nonumber\\ &+&
{1\over2}\sum_{i,j=1}^4\biggl[\,f_{ijs_{22}}^0
G(m_{\tilde\chi^0_i},m_{\tilde\chi^0_j}) \ -\ 2\,g_{ijs_{22}}^0\,
m_{\tilde\chi^0_i} m_{\tilde\chi^0_j}
B_0(m_{\tilde\chi^0_i},m_{\tilde\chi^0_j})\,\biggr] \nonumber\\ &+&
\sum_{i,j=1}^2\biggl[\,f_{ijs_{22}}^+
G(m_{\tilde\chi^+_i},m_{\tilde\chi^+_j}) \ -
\ 2\,g_{ijs_{22}}^+m\,_{\tilde\chi^+_i}m_{\tilde\chi^+_j}
B_0(m_{\tilde\chi^+_i},m_{\tilde\chi^+_j})\,\biggr]~,
\label{s2s2}
\end{eqnarray}
%
\begin{eqnarray}
16\pi^2~\Pi_{s_1s_2}(p^2) &=& \sum_f\sum_{i,j=1}^2N_c^f
\lambda_{s_1\tilde f_i\tilde f_j}\lambda_{s_2\tilde f_i\tilde f_j}
B_0(m_{\tilde f_i},m_{\tilde f_j}) \qquad\qquad\nonumber\\ &+&
{1\over4}g^2 s_\beta c_\beta\Biggl\{\, 2F(M_W,M_W)\ -\ 2F(m_{H^+},M_W)
\nonumber\\ && \qquad\qquad +\ {F(M_Z,M_Z)-F(m_A,M_Z)\over\hat c^2}
\nonumber\\ && \qquad\qquad +\ 7\biggl[\, 2M^2_W B_0(M_W,M_W) +{M^2_Z
B_0(M_Z,M_Z)\over\hat c^2}\,\biggr]\,\Biggr\} \nonumber \\ &+&
{1\over2}\sum_{n=1}^4\biggl[\,\sum_{m=1}^4
\lambda_{H^0_nH^0_ms_1}\lambda_{H^0_nH^0_ms_2}B_0(m_{H^0_n},m_{H^0_m})
\ +\ \lambda_{H^0_nH^0_ns_1s_2}A_0(m_{H_n^0})\,\biggr]\nonumber\\ &+&
\sum_{n=1}^2\biggl[\,\sum_{m=1}^2
\lambda_{H_n^+H_m^-s_1}\lambda_{H_n^+H_m^-s_2}B_0(m_{H_n^+},m_{H_m^+})
\ +\ \lambda_{H^+_nH^-_ns_1s_2}A_0(m_{H^+_n})\,\biggr]\nonumber\\ &+&
{1\over2}\sum_{i,j=1}^4\biggl[\,f_{ijs_{12}}^0
G(m_{\tilde\chi^0_i},m_{\tilde\chi^0_j}) \ -\ 2\,g_{ijs_{12}}^0\,
m_{\tilde\chi^0_i} m_{\tilde\chi^0_j}
B_0(m_{\tilde\chi^0_i},m_{\tilde\chi^0_j})\,\biggr] \nonumber\\ &+&
\sum_{i,j=1}^2\biggl[\,f_{ijs_{12}}^+
G(m_{\tilde\chi^+_i},m_{\tilde\chi^+_j}) \ -
\ 2\,g_{ijs_{12}}^+\,m_{\tilde\chi^+_i}m_{\tilde\chi^+_j}
B_0(m_{\tilde\chi^+_i},m_{\tilde\chi^+_j})\,\biggr]~,
\label{s1s2}
\end{eqnarray}
where $N_c^f$ is the number of colors, which is 3 if $f$ is a (s)quark
and 1 if $f$ is a (s)lepton.  The neutral current couplings $g_f$ are
defined in Eq.~(\ref{glgr}).

The $\tilde\chi_i\tilde\chi_j$-Higgs couplings $f^+_{ijs_{kl}},
\ g^+_{ijs_{kl}}, \ f^0_{ijs_{kl}}$, and $g^0_{ijs_{kl}}$ are defined by
\begin{equation}
f_{ijs_{kl}} \ =\ a^*_{\tilde\chi_i\tilde\chi_js_k}\,
a_{\tilde\chi_i\tilde\chi_js_l}\,+\,
b^*_{\tilde\chi_i\tilde\chi_js_k}\,
b_{\tilde\chi_i\tilde\chi_js_l}~,~~~~ g_{ijs_{kl}} \ =
\ b^*_{\tilde\chi_i\tilde\chi_js_k}\,a_{\tilde\chi_i\tilde\chi_js_l}
+ a^*_{\tilde\chi_i\tilde\chi_js_k}\,b_{\tilde\chi_i\tilde\chi_js_l}~,
\end{equation}
and the $a_{\tilde\chi_i\tilde\chi_js_k}$ and
$b_{\tilde\chi_i\tilde\chi_js_k}$ couplings are defined in
Eqs.~(\ref{NNS}), (\ref{NNS N}-38).  The Higgs-squark-squark
couplings $\lambda_{H\tilde f_j\tilde f_k}$ and $\lambda_{h\tilde
f_j\tilde f_k}$ are given in Eqs.~(\ref{H1H2}--\ref{sf rot}).

We write the Feynman rules for the relevant quartic Higgs couplings as
$-i\lambda$, and define $\lambda_{H_nH_nH_mH_m} = g^2/(4\hat
c^2)\overline \lambda_{H_nH_nH_mH_m}$. We list the necessary
$\overline\lambda_{H_nH_nH_mH_m}$ couplings in the following two
tables,
\vspace{4mm}
\begin{equation}
\begin{array}{c|ccc}
& \qquad\qquad s_1s_1 \qquad\qquad & \qquad\qquad s_2s_2 \qquad\qquad&
\qquad\qquad s_1s_2 \qquad\qquad\\ \hline\\ HH &
3c_\alpha^2-s_\alpha^2 & 3s_\alpha^2-c_\alpha^2 & -s_{2\alpha} \\[2mm]
hh & 3s_\alpha^2-c_\alpha^2 & 3c_\alpha^2-s_\alpha^2 & ~\,s_{2\alpha}
\\[2mm] G^0G^0 & ~\,c_{2\beta} & -c_{2\beta} & 0 \\[2mm] AA &
-c_{2\beta} & ~\,c_{2\beta} & 0 \\[2mm] G^+G^- & \hat c^2+\hat
s^2c_{2\beta} & \hat c^2-\hat s^2c_{2\beta} & -\hat c^2s_{2\beta}
\\[2mm] H^+H^- & \hat c^2-\hat s^2c_{2\beta} & \hat c^2+\hat
s^2c_{2\beta} &\,\hat c^2s_{2\beta}\\
\end{array}
\end{equation}
\vspace{5mm}
\begin{equation}
\begin{array}{c|cc}
& \qquad\qquad\qquad AA \qquad\qquad\qquad & \qquad\qquad\qquad H^+H^-
\qquad\qquad\qquad \\ \hline\\ G^0G^0 & 3s_{2\beta}^2-1 & \hat
c^2(1+s_{2\beta}^2)-\hat s^2c_{2\beta}^2 \\ [2mm] AA & 3c_{2\beta}^2 &
c_{2\beta}^2 \\ [2mm] G^+G^- & \hat c^2(1+s_{2\beta}^2)-\hat
s^2c_{2\beta}^2 & 2s_{2\beta}^2-1 \\ [2mm] H^+H^- & c_{2\beta}^2 &
2c_{2\beta}^2 \\ [2mm]
\end{array}
\end{equation}
For the couplings involving $(s_1,\,s_2)$, we obtain the corresponding
couplings in the $(H,\,h)$ eigenstate basis by the following rotations
\begin{equation}
\left(\begin{array}{cc} \lambda_{H_nH_nHH} & \lambda_{H_nH_nHh}\\[2mm]
\lambda_{H_nH_nHh} & \lambda_{H_nH_nhh}\end{array}\right)\ =\ \left(
\begin{array}{cc}c_\alpha & s_\alpha\\[2mm]-s_\alpha
& c_\alpha\end{array}\right)\left(
\begin{array}{cc} \lambda_{H_nH_ns_1s_1} & \lambda_{H_nH_ns_1s_2}\\[2mm]
\lambda_{H_nH_ns_1s_2} &
\lambda_{H_nH_ns_2s_2}\end{array}\right)\left(
\begin{array}{cc}c_\alpha & -s_\alpha \\[2mm]
s_\alpha & c_\alpha\end{array}\right)~.
\end{equation}

We write the Feynman rules for the trilinear Higgs-boson couplings as
$-i\lambda$, and define $\lambda_{H_nH_ms_l} = gM_Z/(2\hat
c)\overline\lambda_{H_nH_ms_l}$. We list the
$\overline\lambda_{H_nH_ms_l}$ in the following two tables,
\begin{equation}
\begin{array}{c|ccc}
 &\qquad\qquad\quad HH\qquad\qquad\quad &\qquad\qquad\quad
hh\qquad\qquad\quad &\qquad\qquad\quad Hh\qquad\qquad\quad \\ \hline\\
s_1 & c_\beta(3c_\alpha^2-s_\alpha^2)-s_\beta s_{2\alpha} &
c_\beta(3s_\alpha^2-c_\alpha^2)+s_\beta s_{2\alpha} & -2c_\beta
s_{2\alpha}-s_\beta c_{2\alpha} \\ [2mm] s_2 &
s_\beta(3s_\alpha^2-c_\alpha^2)-c_\beta s_{2\alpha} &
s_\beta(3c_\alpha^2-s_\alpha^2)+c_\beta s_{2\alpha} & 2s_\beta
s_{2\alpha}-c_\beta c_{2\alpha}\\[2mm] \end{array}
\label{sHH}
\end{equation}
\vspace{2mm}
\begin{equation}
\begin{array}{c|cccccc}
 & \quad ~G^0G^0 \quad& \quad ~AA\quad & \quad ~G^0A\quad & \quad
{}~G^+G^-\quad & \quad H^+H^- \quad & \quad G^+H^- \quad\\ \hline\\
s_1 & ~c_{2\beta}c_\beta & -c_{2\beta}c_\beta & -s_{2\beta}c_\beta &
{}~c_{2\beta}c_\beta & -c_{2\beta}c_\beta+2\hat c^2c_\beta &
-s_{2\beta}c_\beta+\hat c^2s_\beta \\ [2mm] s_2 & -c_{2\beta}s_\beta &
{}~c_{2\beta}s_\beta & ~s_{2\beta}s_\beta & -c_{2\beta}s_\beta &
{}~c_{2\beta}s_\beta+2\hat c^2s_\beta & ~s_{2\beta}s_\beta-\hat
c^2c_\beta\\[2mm]
\end{array}\label{sAA}
\end{equation}
\noindent To obtain the couplings involving $(s_1,\,s_2)$ in the
$(H,\, h)$ eigenstate basis, we rotate the $(s_1,\, s_2)$ couplings by
the angle $\alpha$, as in Eq.~(\ref{a rot}).

The CP-odd Higgs boson $A$ and charged Higgs $H^+$ self-energies are
%
\begin{eqnarray}
16\pi^2~\Pi_{AA}(p^2) &=& \Biggl\{\,
c_\beta^2\sum_{f_u}N_c^f\lambda_u^2 \biggl[\,p^2 B_0(m_u,m_u)\ -
\ 2A_0(m_u)\,\biggr]\nonumber\\ &+&
\sum_{f_u}N_c^f\left(\lambda_u^2c_\beta^2 - {g^2\over\hat
c^2}I_3^ug^u_Lc_{2\beta} \right)\biggl[\,c_u^2A_0(m_{\tilde u_1})+
s_u^2A_0(m_{\tilde u_2})\,\biggr]\nonumber\\ &+&
\sum_{f_u}N^f_c\left(\lambda_u^2c^2_\beta - {g^2\over\hat
c^2}I_3^ug^u_Rc_{2\beta} \right)\biggl[\,s_u^2A_0(m_{\tilde u_1})+
c^2_uA_0(m_{\tilde u_2})\,\biggr] \nonumber\\ &+&
\sum_{f_u}\sum_{i,j=1}^2 N_c^f\lambda^2_{A\tilde u_i\tilde u_j}
B_0(m_{\tilde u_i},m_{\tilde u_j})\ +\ \left(
\begin{array}{c} u\rightarrow d\\[1mm] c_\beta\leftrightarrow s_\beta
\end{array}\right)\, \Biggr\}\nonumber\\
&+& {g^2\over4}\biggl[\,2F(m_{H^+},M_W)\ +
\ {s^2_{\alpha\beta}\over\hat c^2}F(m_H,M_Z) \ +
\ {c^2_{\alpha\beta}\over \hat c^2}F(m_h,M_Z)\,\biggr] \nonumber\\ &+&
{1\over2}\sum_{n=1}^4\biggl[\,\sum_{m=1}^4\lambda_{AH^0_nH^0_m}^2
B_0(m_{H^0_n},m_{H^0_m}) \ +
\ \lambda_{AAH^0_nH^0_n}A_0(m_{H^0_n})\,\biggr] \nonumber\\ &+&
{g^2M_W^2\over2}B_0(M_W,m_{H^+}) \ +\ \sum_{n=1}^2
\lambda_{AAH^+_nH^-_n}A_0(m_{H^+_n}) \nonumber\\ &+&
g^2\biggl[\,2A_0(M_W)+{A_0(M_Z)\over\hat c^2}\,\biggr] \nonumber\\ &+&
{1\over2}\sum_{i,j=1}^4\biggl[\,f_{ijA}^0
G(m_{\tilde\chi^0_i},m_{\tilde\chi^0_j}) \ -\ 2\,g_{ijA}^0\,
m_{\tilde\chi^0_i} m_{\tilde\chi^0_j}
B_0(m_{\tilde\chi^0_i},m_{\tilde\chi^0_j})\,\biggr] \nonumber\\ &+&
\sum_{i,j=1}^2\biggl[\,f_{ijA}^+
G(m_{\tilde\chi^+_i},m_{\tilde\chi^+_j}) \ -\ 2\,g_{ijA}^+\,
m_{\tilde\chi^+_i} m_{\tilde\chi^+_j}
B_0(m_{\tilde\chi^+_i},m_{\tilde\chi^+_j})\,\biggr]~,
\label{pia}
\end{eqnarray}
%
\begin{eqnarray}
16\pi^2~\Pi_{H^+H^-}(p^2) &=& \sum_{f_u/f_d}N_c^f
\biggl[\,(\lambda^2_uc^2_\beta + \lambda^2_ds^2_\beta)G(m_u,m_d) \ -
\ 2\lambda_u\lambda_dm_um_ds_{2\beta}B_0(m_u,m_d)\,\biggr] \nonumber\\
&+& \sum_{f_u/f_d}\sum_{i,j=1}^2 N_c^f\lambda_{H^+\tilde u_i\tilde
d_j}^2B_0(m_{\tilde u_i},m_{\tilde d_j}) \nonumber \\ &+& \Biggl\{\,
\sum_{f_u}N_c^f\left(\lambda_d^2s^2_\beta - {g^2\over\hat
c^2}I_3^ug^u_Lc_{2\beta} + {g^2\over2}c_{2\beta}
\right)\biggl[\,c_u^2A_0(m_{\tilde u_1})+s_u^2A_0(m_{\tilde u_2})\,\biggr]
\nonumber\\ &+& \sum_{f_u}N_c^f\left(\lambda^2_uc^2_\beta -
{g^2\over\hat c^2}I_3^ug^u_Rc_{2\beta}
\right)\biggl[\,s_u^2A_0(m_{\tilde u_1})+c_u^2A_0(m_{\tilde u_2})\,\biggr]
\ +\ \left(\begin{array}{c} u\leftrightarrow d\\[1mm]
s_\beta\leftrightarrow c_\beta
\end{array}\right)\, \Biggr\}\nonumber\\
&+& {g^2\over 4}\biggl[\,s^2_{\alpha\beta}F(m_H,M_W)\ +
\ c^2_{\alpha\beta}F(m_h,M_W) \ +\ F(m_A,M_W)\nonumber\\ &&
\qquad\qquad+\ {c^2_{2\hat\theta_W}\over\hat
c^2}F(m_{H^+},M_Z)\,\biggr] \nonumber \\ &+&
e^2F(m_{H^+},0)+2g^2A_0(M_W) \ +\ {g^2c^2_{2\hat\theta_W}\over\hat
c^2}A_0(M_Z) \nonumber\\ &+& \sum_{n,m=1}^2
\lambda_{H^+H^0_nH^-_m}^2B_0(m_{H^0_n},m_{H^-_m})\nonumber\\ &+&
{g^2M_W^2\over4}B_0(M_W,m_A) \ +\ {1\over2} \sum_{n=1}^4
\lambda_{H^+H^-H^0_nH^0_n} \, A_0(m_{H^0_n}) \nonumber\\ &+&
\sum_{n=1}^2\lambda_{H^+H^-H^+_nH^-_n}A_0(m_{H^+_n})\nonumber\\ &+&
\sum_{i=1}^2\sum_{j=1}^4\biggl[\,f_{ijH^+}
G(m_{\tilde\chi^+_i},m_{\tilde\chi^0_j}) \ -\ 2\,g_{ijH^+}\,
m_{\tilde\chi^+_i} m_{\tilde\chi^0_j}
B_0(m_{\tilde\chi^+_i},m_{\tilde\chi^0_j})\,\biggr]~,
\label{piH+}
\end{eqnarray}
where $c_{\alpha\beta}~(s_{\alpha\beta})$ denotes
$\cos(\alpha-\beta)~(\sin(\alpha-\beta))$.  The $g_{f_L},~g_{f_R}$ are
defined in Eq.~(\ref{glgr}), and the $I_3^f$ are listed in the table
of Eq.~(\ref{table1}).  $N_c^f$ denotes the number of colors, which is
3 for a (s)quark.

The $\tilde\chi_i\tilde\chi_jA$ couplings $f^+_{ijA},\ g^+_{ijA},
\ f^0_{ijA}$, and $g^0_{ijA}$ are defined by
\begin{equation}
f_{ijA} \ =\ \left|a_{\tilde\chi_i\tilde\chi_jA}\right|^2\,+\,
\left|b_{\tilde\chi_i\tilde\chi_jA}\right|^2~,~~~~ g_{ijA} \ =\ 2\,{\cal
R}e\,\left(b^*_{\tilde\chi_i\tilde\chi_jA} \,
a_{\tilde\chi_i\tilde\chi_jA}\right)~,
\end{equation}
and similarly for the $f_{ijH^+},\ g_{ijH^+}$ couplings.  The
$a_{\tilde\chi_i\tilde\chi_jA},~b_{\tilde\chi_i\tilde\chi_jA}$
couplings are given in Eqs.~(\ref{NNP}--\ref{CCH U}), and
those of the charged Higgs are listed in Eqs.~(\ref{NCH}--\ref{NCH
B}). The Higgs-sfermion-sfermion couplings $\lambda_{A\tilde
f_j\tilde f_k}$ and $\lambda_{H^+\tilde f_j\tilde f_k}$ are given in
Eqs.~(\ref{AG}--\ref{ud rot}).  The $\tilde f_{L,R}$ basis couplings
$\lambda_{A\tilde f\tilde f}$ of Eq.~(\ref{AG}) also apply in the
$\tilde f_{1,2}$ basis.

\vspace{.7cm} {\noindent\Large\bf Appendix E: One-loop Higgs boson
masses}
\vspace{.7cm} \setcounter{equation}{0}
\renewcommand{\theequation}{E.\arabic{equation}}

In this appendix we will present the formalism necessary to obtain
accurate Higgs-boson masses at the one-loop level.  The tadpole
diagrams play an important role in determining the masses.  The
one-loop tadpole contributions are listed in Refs.~\cite{CPR,BBO}.
At any given order in perturbation theory, minimizing the scalar
potential is equivalent to requiring that the tadpoles vanish. At tree
level, we have $T_1=T_2=0$, with the tadpoles given by the
\mbox{\footnotesize$\overline{\rm DR}~$} relations
\begin{eqnarray}
&& {T_1\over v_1}\ =\ {1\over2}\hat M_Z^2c_{2\beta}\ +\ m_{H_1}^2\ +
\ \mu^2 \ +\ B\mu\tan\beta\ ,\label{T1}\\ && {T_2\over v_2}\ =\ -
\ {1\over2}\hat M_Z^2c_{2\beta}\ +\ m_{H_2}^2\ +\ \mu^2 \ +
\ B\mu\cot\beta\ ,\label{T2}
\end{eqnarray}
where the Higgs-sector soft supersymmetry-breaking potential is
\begin{equation}
V_{\rm soft}\ =\ \ m_{H_1}^2\left|H_1\right|^2 \ +
\ m_{H_2}^2\left|H_2\right|^2\ +\ \left(B\mu \epsilon_{ij}H_1^iH_2^j +
\rm h.c.\right)\ .
\end{equation}

At one-loop level, the total (tree-level plus one-loop) tadpole must
vanish, so $T_1-t_1=0,\ T_2-t_2=0$, with
\begin{eqnarray}
16\pi^2~{t_1\over v_1} &=& -\ \sum_{f_d}2 N_c^f \lambda_d^2 A_0(m_d)
\ +\ \sum_f\sum_{i=1}^2 N_c^f {g\lambda_{s_1\tilde f_i\tilde
f_i}\over2M_Wc_\beta} A_0(m_{\tilde f_i}) \nonumber\\ &-&
{g^2c_{2\beta}\over8\hat c^2}\biggl(A_0(m_A) + 2A_0(m_{H^+})\biggr) \ +
\ {g^2\over2} A_0(m_{H^+}) \nonumber\\ &+& {g^2\over8\hat
c^2}\biggl(3s_\alpha^2-c_\alpha^2+s_{2\alpha}\tan\beta\biggr)A_0(m_h)
\ +\ {g^2\over8\hat c^2}\biggl(3c_\alpha^2-s_\alpha^2-s_{2\alpha}
\tan\beta\biggr)A_0(m_H)\nonumber\\
&-& \sum_{i=1}^4g^2{m_{\tilde\chi_i^0}\over M_Wc_\beta} {\cal R}e \,
\biggl[\,N_{i3}(N_{i2}-N_{i1}\tan\hat\theta_W) \,
\biggr]A_0(m_{\tilde\chi_i^0}) \nonumber\\
&-& \sum_{i=1}^2\sqrt{2}g^2{m_{\tilde\chi_i^+}\over M_Wc_\beta}
{\cal R}e\,\biggl[\,V_{i1}U_{i2}\,\biggr]A_0(m_{\tilde\chi_i^+})
\nonumber\\ &+& {3g^2\over4}\biggl(2A_0(M_W) + {A_0(M_Z)\over\hat c^2}
\biggr) \ +\ {g^2c_{2\beta}\over8\hat c^2}
\biggl(2A_0(M_W)+A_0(M_Z)\biggr)\ ,
\label{t1}
\end{eqnarray}
and
\begin{eqnarray}
16\pi^2~{t_2\over v_2} &=& -\ \sum_{f_u}2 N_c^f \lambda_u^2 A_0(m_u)
\ +\ \sum_f\sum_{i=1}^2 N_c^f {g\lambda_{s_2\tilde f_i\tilde
f_i}\over2M_Ws_\beta}A_0(m_{\tilde f_i}) \nonumber\\ &+&
{g^2c_{2\beta}\over8\hat c^2}\biggl(A_0(m_A) + 2A_0(m_{H^+})\biggr) \ +
\ {g^2\over2} A_0(m_{H^+}) \nonumber\\ &+& {g^2\over8\hat
c^2}\biggl(3c_\alpha^2-s_\alpha^2+s_{2\alpha}\cot\beta\biggr)A_0(m_h)
\ +\ {g^2\over8\hat c^2}\biggl(3s_\alpha^2 - c_\alpha^2 -
s_{2\alpha}\cot\beta\biggr)A_0(m_H)\nonumber\\
&+& \sum_{i=1}^4g^2{m_{\tilde\chi_i^0}\over M_Ws_\beta} {\cal R}e \,
\biggl[\,N_{i4} (N_{i2}-N_{i1} \tan\hat\theta_W) \, \biggr]
A_0(m_{\tilde\chi_i^0}) \nonumber\\
&-& \sum_{i=1}^2\sqrt{2}g^2{m_{\tilde\chi_i^+}\over
M_Ws_\beta} {\cal R}e\,\biggl[\,V_{i2}U_{i1}\,\biggr]
A_0(m_{\tilde\chi_i^+})\nonumber\\
&+& {3g^2\over4}\biggl(2A_0(M_W)+{A_0(M_Z)\over\hat c^2}\biggr) \ -
\ {g^2c_{2\beta}\over8\hat c^2}\biggl(2A_0(M_W)+A_0(M_Z)\biggr)\ ,
\label{t2}
\end{eqnarray}
where $N_c^f$ is the number of colors, 3 if $f$ is a (s)quark and 1
otherwise.  The $A_0$ function is given in Eq.~(\ref{A}); $\hat c$
denotes $\cos\hat\theta_W$ and $c_\beta=\cos\beta$, etc.  The matrices
$N,\ U$, and $V$ are described in Appendix A and the couplings
$\lambda_{s_1\tilde f_i\tilde f_j}, \ \lambda_{s_2\tilde f_i\tilde
f_j}$ are given by Eqs.~(\ref{H1H2},\ \ref{sf rot}).

The \mbox{\footnotesize$\overline{\rm DR}~$} (tree-level) CP-odd Higgs
mass is given by $\hat m_A^2 = -B\mu(\tan\beta+\cot\beta)$, and
Eqs.~(\ref{T1},\ \ref{T2}) allow us to solve for the
\mbox{\footnotesize$\overline{\rm DR}~$} parameter, $\mu^2,$ and the
pole mass\footnote{In case $m_A$ is very close to $M_Z$, there is an
additional $\cal O(\alpha)$ correction to $m_A$ from the off-diagonal
element of the CP-odd mass matrix.}, $m_A$,
\begin{eqnarray}
\mu^2 &=& {1\over2}\Biggl[\, \tan2\beta\biggl(\overline
m_{H_2}^2\tan\beta-\overline m_{H_1}^2\cot\beta\biggr) \ -\ M_Z^2\ -
\ {\cal R}\!e\,\Pi_{ZZ}^T(M_Z^2)\, \Biggr]\ ,\nonumber\\ m_A^2 &=&
{1\over c_{2\beta}}\biggl(\overline m_{H_2}^2-\overline
m_{H_1}^2\biggr) \ -\ M_Z^2 \ -\ {\cal R}e\,\Pi_{ZZ}^T(M_Z^2) \ -
\ {\cal R}e\,\Pi_{AA}(m_A^2) \ +\ b_A\ ,
\end{eqnarray}
where $\overline m_{H_1}^2 = m_{H_1}^2 - t_1/v_1, \ \overline
m_{H_2}^2 = m_{H_2}^2 - t_2/v_2$.  The self-energies $\Pi_{ZZ}^T$
and $\Pi_{AA}$ are given in Eqs.~(\ref{piz}) and (\ref{pia}),
respectively, and $b_A = s_\beta^2\, t_1/v_1 +
c_\beta^2\,t_2/v_2$.

Having determined the physical CP-odd Higgs-boson mass $m_A$, we are
in a position to compute the remaining Higgs masses.  The physical
mass for the charged Higgs boson $H^+$ is
\begin{equation}
m^2_{H^+}\ =\ m_A^2\ +\ M_W^2\ +\ {\cal R}e\biggl[\Pi_{AA}(m^2_A) -
\Pi_{H^+H^-}(m_{H^+}^2)+\Pi^T_{WW}(M^2_W)\biggr]~,
\end{equation}
where the $W$-boson self-energy is given in Eq.~(\ref{piw}), and the
charged-Higgs self-energy in Eq.~(\ref{piH+}).

The CP-even Higgs-boson masses are obtained from the real part of the
poles of the propagator matrix,
\begin{equation}
{\rm Det}\biggl[\,~p_i^2\,{\bf1}~-~{\cal M}^2_s(p_i^2)~\,\biggr]\ =
~0~, \qquad m_i^2 = {\cal R}e(p_i^2)~,
\end{equation}
where the matrix ${\cal M}^2_s(p^2)$ is
\begin{equation}
{\cal M}^2_s(p^2) \ =\ \left(\begin{array}{cc} \hat M_Z^2c_\beta^2 +
\hat m_A^2s_\beta^2 -\,\Pi_{s_1s_1}(p^2) + t_1/v_1 & -(\hat
M_Z^2 + \hat m_A^2)s_\beta c_\beta -\,\Pi_{s_1s_2}(p^2)\\[2mm]
-(\hat M_Z^2 + \hat m_A^2)s_\beta c_\beta -\,\Pi_{s_2s_1}(p^2)
& \hat M_Z^2s_\beta^2 + \hat m_A^2c_\beta^2 -\,\Pi_{s_2s_2}(p^2)
+ t_2/v_2
\end{array}\right)\ .
\end{equation}
In this expression, $\hat M_Z^2$ and $\hat m_A^2$ are the $Z$- and
$A$-boson \mbox{\footnotesize$\overline{\rm DR}~$} masses ($\hat M_Z^2
= M_Z^2 + {\cal R}e\, \Pi^T_{ZZ}(M_Z^2),$ $\hat m_A^2 = m_A^2 + {\cal
R}e\, \Pi_{AA}(m_A^2) -b_A$).  The self-energies $\Pi_{s_is_j}$ are
given in Eqs.~(\ref{s1s1}--\ref{s1s2}).  At one loop, the angle
$\alpha$ diagonalizes the matrix ${\cal M}_s^2(p^2)$ for some choice
of momentum $p^2$; we choose $p^2=m_h^2$.

\end{document}